\documentclass[aip,jcp,reprint,noshowkeys,superscriptaddress]{revtex4-1}
\usepackage{graphicx,dcolumn,bm,xcolor,multirow,amscd,amsmath,amssymb,amsfonts,physics,longtable,wrapfig,bbold,siunitx,xspace}
\usepackage[version=4]{mhchem}

\usepackage[utf8]{inputenc}
\usepackage[T1]{fontenc}

\usepackage{hyperref}
\hypersetup{
    colorlinks,
    linkcolor={red!50!black},
    citecolor={red!70!black},
    urlcolor={red!80!black}
}

\usepackage{listings}
\definecolor{codegreen}{rgb}{0.58,0.4,0.2}
\definecolor{codegray}{rgb}{0.5,0.5,0.5}
\definecolor{codepurple}{rgb}{0.25,0.35,0.55}
\definecolor{codeblue}{rgb}{0.30,0.60,0.8}
\definecolor{backcolour}{rgb}{0.98,0.98,0.98}
\definecolor{mygray}{rgb}{0.5,0.5,0.5}

\definecolor{sqred}{rgb}{0.85,0.1,0.1}
\definecolor{sqgreen}{rgb}{0.25,0.65,0.15}
\definecolor{sqorange}{rgb}{0.90,0.50,0.15}
\definecolor{sqblue}{rgb}{0.10,0.3,0.60}

\lstdefinestyle{mystyle}{
    backgroundcolor=\color{backcolour},
    commentstyle=\color{codegreen},
    keywordstyle=\color{codeblue},
    numberstyle=\tiny\color{codegray},
    stringstyle=\color{codepurple},
    basicstyle=\ttfamily\footnotesize,
    breakatwhitespace=false,
    breaklines=true,
    captionpos=b,
    keepspaces=true,
    numbers=left,
    numbersep=5pt,
    numberstyle=\ttfamily\tiny\color{mygray},
    showspaces=false,
    showstringspaces=false,
    showtabs=false,
    tabsize=2
  }

  \newcolumntype{d}{D{.}{.}{-1}}

\newcommand{\ie}{\textit{i.e.}}

\newcommand{\alert}[1]{\textcolor{black}{#1}}
\usepackage[normalem]{ulem}

\newcommand{\SupMat}{\textcolor{blue}{supplementary material}}

\newcommand{\mc}{\multicolumn}

\newcommand{\tabc}[1]{\multicolumn{1}{c}{#1}}

\newcommand{\br}{\boldsymbol{r}}
\newcommand{\bx}{\boldsymbol{x}}

\newcommand{\GW}{\text{$GW$}}	
\newcommand{\GT}{\text{$GT$}}	
\newcommand{\GF}{\text{GF}}	
\newcommand{\HF}{\text{HF}}
\newcommand{\ph}{\text{ph}}
\newcommand{\pp}{\text{pp}}
\newcommand{\hh}{\text{hh}}

\newcommand{\BSE}{\text{BSE}}
\newcommand{\dBSE}{\text{dBSE}}

\newcommand{\e}{\epsilon}
\newcommand{\Om}{\Omega}
\newcommand{\Sig}{\Sigma}
\newcommand{\MO}[1]{\psi_{#1}}
\newcommand{\ERI}[2]{\mel{#1}{}{#2}}
\newcommand{\sERI}[2]{M_{#1}^{#2}}

\newcommand{\bO}{\boldsymbol{0}}
\newcommand{\bId}{\boldsymbol{1}}
\newcommand{\bI}{\boldsymbol{I}}
\newcommand{\bJ}{\boldsymbol{J}}
\newcommand{\bK}{\boldsymbol{K}}
\newcommand{\bL}{\boldsymbol{L}}
\newcommand{\bH}{\boldsymbol{H}}

\newcommand{\bOm}{\boldsymbol{\Omega}}
\newcommand{\bXi}{\boldsymbol{\Xi}}
\newcommand{\bA}{\boldsymbol{A}}
\newcommand{\bB}{\boldsymbol{B}}
\newcommand{\bC}{\boldsymbol{C}}
\newcommand{\bD}{\boldsymbol{D}}
\newcommand{\bV}{\boldsymbol{V}}
\newcommand{\bX}{\boldsymbol{X}}
\newcommand{\bY}{\boldsymbol{Y}}
\newcommand{\bc}{\boldsymbol{c}}

\newcommand{\LCPQ}{Laboratoire de Chimie et Physique Quantiques (UMR 5626), Universit\'e de Toulouse, CNRS, UPS, France}

\begin{document}	

\title{Connections and performances of Green's function methods for charged and neutral excitations}

\author{Enzo \surname{Monino}}
	\email{enzo.monino@irsamc.ups-tlse.fr}
	\affiliation{\LCPQ}
\author{Pierre-Fran\c{c}ois \surname{Loos}}
	\email{loos@irsamc.ups-tlse.fr}
	\affiliation{\LCPQ}

\begin{abstract}
In recent years, Green's function methods have garnered considerable interest due to their ability to target both charged and neutral excitations. Among them, the well-established $GW$ approximation provides accurate ionization potentials and electron affinities and can be extended to neutral excitations using the Bethe-Salpeter equation (BSE) formalism. Here, we investigate the connections between various Green's function methods and evaluate their performance for charged and neutral excitations. Comparisons with other widely-known second-order wave function methods are also reported. Additionally, we calculate the singlet-triplet gap of cycl[3,3,3]azine, a model molecular emitter for thermally activated delayed fluorescence, which has the particularity of having an inverted gap thanks to a substantial contribution from the double excitations. We demonstrate that, within the $GW$ approximation, a second-order BSE kernel with dynamical correction is required to predict this distinctive characteristic.
\end{abstract}

\maketitle

\section{Green's function methods}
Recent developments and investigations in Green's function approaches have generated significant interest within the electronic structure community, \cite{CsanakBook,FetterBook,Martin_2016} especially in quantum chemistry. \cite{Bruneval_2016,Golze_2019,Blase_2018,Blase_2020} 
The pillar of Green's function many-body perturbation theory is the one-body Green's function (or electron propagator). \cite{Schirmer_2018}
It has the ability to provide the charged excitations (\ie, ionization potentials and electron affinities) of the system in a single calculation as measured in direct or inverse photoemission spectroscopy. 
This avoids using state-specific methods where one has to perform separate calculations on the neutral and ionized species. \cite{Das_1973,Dalgaard_1978,Lengsfield_1980,Bauschlicher_1980a,Bauschlicher_1980b,Werner_1981,Golab_1983,Ziegler_1977,Kowalczyk_2011}

Obviously, the exact one-body Green's function $G$ is in general unknown but its mean-field Hartree-Fock (HF) version $G_{\HF}$ can be linked to the exact one, via a Dyson equation involving a key quantity known as the self-energy $\Sigma$, which includes correlation effects:
\begin{equation}
\label{eq:Dyson}
	G(12) = G_{\HF}(12)+\int G_{\HF}(13)\Sigma(34)G(42)d3d4
\end{equation}
Here, $1 \equiv (\bx_1, t_1)$ is a composite coordinate gathering spin-space and time variables. 

The HF one-body Green's function is given by
\begin{equation}
\label{eq:1-GF}
	G_{\HF}(\bx_1,\bx_2;\omega) 
	= \sum_i \frac{\MO{i}(\bx_1)\MO{i}(\bx_2)}{\omega-\e_{i}^{\HF} - i\eta} 
	+ \sum_a \frac{\MO{a}(\bx_1)\MO{a}(\bx_2)}{\omega-\e_{a}^{\HF} + i\eta}
\end{equation}
where $\MO{p}(\bx)$ is the $p$th HF spinorbital and $\e_{p}^{\HF}$ its corresponding energy, while $\eta$ is a positive infinitesimal that we shall set to zero in the remaining of this paper. 
Throughout this article, we assume real orbitals and energies.
The indices $p$, $q$, $r$, and $s$ are general spinorbitals, $i$, $j$, $k$, and $l$ denote occupied spinorbitals, $a$, $b$, $c$, and $d$ are vacant spinorbitals, while $m$ and $n$ label single excitations/deexcitations and double electron attachments/detachments, respectively.
Here, we systematically consider a HF starting point but the present analysis can be straightforwardly extended to a Kohn-Sham starting point. 

Approximations to the self-energy, such as $GW$, \cite{Aryasetiawan_1998,Onida_2002,Reining_2017,Golze_2019,Bruneval_2021} are needed to solve the Dyson equation defined in Eq.~\eqref{eq:Dyson}.
Even though the $GW$ approximation has proven to produce accurate charged excitations in solids \cite{Strinati_1980,Strinati_1982a,Strinati_1982b,Hybertsen_1985,Hybertsen_1986,Godby_1986,Godby_1987,Godby_1987a,Godby_1988,Blase_1995}  and molecules, \cite{Rohlfing_1999a,Horst_1999,Puschnig_2002,Tiago_2003,Rocca_2010,Boulanger_2014,Jacquemin_2015a,Bruneval_2015,Jacquemin_2015b,Hirose_2015,Jacquemin_2017a,Jacquemin_2017b,Rangel_2017,Krause_2017,Gui_2018,Blase_2018,Liu_2020,Li_2017,Li_2019,Li_2020,Li_2021,Blase_2020,Holzer_2018a,Holzer_2018b,Loos_2020e,Loos_2021,McKeon_2022,vanSetten_2018,Jin_2019a,Jin_2019b,Golze_2018,Golze_2020,Li_2022a,Li_2022b,Li_2022c,Forster_2022a} it is not the only approximation to the self-energy. 
Indeed, other approximations exist such as the second-order Green's function (GF2), \cite{Casida_1989,Casida_1991,Ortiz_2013,Phillips_2014,Phillips_2015,Rusakov_2014,Rusakov_2016,Hirata_2015,Hirata_2017,Backhouse_2021,Backhouse_2020b,Backhouse_2020a,Pokhilko_2021a,Pokhilko_2021b,Pokhilko_2022} also known as second Born in the condensed matter community, \cite{Martin_2016,Stefanucci_2013} the $T$-matrix \cite{Liebsch_1981,Bickers_1989a,Bickers_1991,Katsnelson_1999,Katsnelson_2002,Zhukov_2005,vonFriesen_2010,Romaniello_2012,Gukelberger_2015,Muller_2019,Friedrich_2019,Biswas_2021,Zhang_2017,Li_2021b,Loos_2022} (or Bethe-Goldstone approximation \cite{Bethe_1957,Baym_1961,Baym_1962,Danielewicz_1984a,Danielewicz_1984b}).
Going beyond these approximations has been shown to be rather challenging \alert{from a computational point of view. \cite{Baym_1961,Baym_1962,DeDominicis_1964a,DeDominicis_1964b,Bickers_1989a,Bickers_1989b,Bickers_1991,Hedin_1999,Bickers_2004,Shirley_1996,DelSol_1994,Schindlmayr_1998,Morris_2007,Shishkin_2007b,Romaniello_2009a,Romaniello_2012,Gruneis_2014,Hung_2017,Maggio_2017b,Mejuto-Zaera_2022,Ren_2015,Wang_2021,Wang_2022,Forster_2022b,Vlcek_2019,Pavlyukh_2020} Moreover, the overall accuracy does not always improve. \cite{Lewis_2019a,Bruneval_2021}}
Here, for the sake of simplicity, we consider only one-shot schemes where one does not self-consistently update the self-energy,
\cite{Strinati_1980,Hybertsen_1985a,Godby_1988,Linden_1988,Northrup_1991,Blase_1994,Rohlfing_1995} 
but the same analysis can be performed in the case of (partially) self-consistent schemes.  \cite{Hybertsen_1986,Shishkin_2007a,Blase_2011,Faber_2011,Rangel_2016,Gui_2018,Faleev_2004,vanSchilfgaarde_2006,Kotani_2007,Ke_2011,Kaplan_2016}
\alert{Note that by only considering one-shot schemes, we neglect all the diagrammatic contributions stemming from the self-consistency.}

Another attractive point concerning Green's function-based techniques is the Bethe-Salpeter equation (BSE) formalism \cite{Salpeter_1951,Strinati_1988,Blase_2018,Blase_2020} that allows access to the neutral (\ie, optical) excitations of a given system. 
BSE relies on the two-body Green's function $G_2$ (or polarization propagator) via its link with the two-body correlation function
\begin{equation}
	iL(12;1'2') = -G_2(12;1'2') + G(11')G(22')
\end{equation}
that also satisfies a Dyson equation
\begin{multline}
	L(12,1'2') = L_0(12,1'2') + \\
	\int L_0(14,1'3) \Xi(35,46)L(62,52')d3d4d5d6
\end{multline}
with 
\begin{equation}
	iL_0(12;1'2') = G(12')G(21')
\end{equation}
and where
\begin{equation}
\label{eq:kernel}
	\Xi(13,24)=i\fdv{\Sigma(12)}{G(43)}
\end{equation}
is the so-called BSE kernel.
As we shall discuss later, like its time-dependent density-functional theory (TD-DFT) cousin, \cite{Runge_1984,Casida_1995,Petersilka_1996,UlrichBook} BSE can be written in the form of Casida-like equations. 

Because the BSE kernel is the functional derivative of $\Sigma$ with respect to $G$, one can readily see from Eq.~\eqref{eq:kernel} that $\Xi$ strongly depends on the choice of approximate self-energy. 
The most popular BSE kernel is based on the $GW$ approximation and leans on the dynamically-screened Coulomb potential $W$ to provide rather accurate neutral excitations for molecular systems. \cite{Strinati_1988}
However, one can also rely on kernels based on the $T$-matrix \cite{Loos_2022} or GF2. \cite{Zhang_2013,Rebolini_2016,Dou_2022}

Undoubtedly, the self-energy and kernel approximations discussed earlier possess inherent limitations, both stemming from their intrinsic nature and the typical methods employed to solve these equations, as well as potential additional approximations involved.
For example, unphysical discontinuities in energy surfaces have been recently discovered and studied in the $GW$ approximation \cite{Loos_2018b,Veril_2018,Loos_2020e,Berger_2021,DiSabatino_2021} but similar observations can be made with the other approximations. The issue can be traced down to the multiple solution character of the quasiparticle equation. \cite{Veril_2018} 
This problem of discontinuities can be partially addressed by using linearization of the quasiparticle equation, but irregularities (or ``bumps'') remain, for example, in potential energy surfaces. \cite{Veril_2018} 
One can deal with this issue by using a static Coulomb-hole plus screened-exchange (COHSEX),\cite{Hedin_1965,Hybertsen_1986,Hedin_1999,Bruneval_2006,Berger_2021} by adopting a fully self-consistent scheme,\cite{Koval_2014,Wilhelm_2018,Stan_2006,Stan_2009,Rostgaard_2010,Caruso_2012,Caruso_2013a,Caruso_2013b,Caruso_2013,DiSabatino_2021,Scott_2023} or via regularization techniques. \cite{Monino_2022,Marie_2023}

Moreover, the BSE is considered, in general, within the so-called static approximation where the dynamical (\ie, frequency-dependent) BSE kernel is approximated by its static limit. 
By doing so, the static BSE scheme, \cite{Ankudinov_2003,Romaniello_2009b,Sangalli_2011,Loos_2020h,Authier_2020} like the adiabatic approximation of TD-DFT, \cite{Levine_2006,Tozer_2000,Elliott_2011,Maitra_2012,Maitra_2016} does not permit the description of double and higher excitations. 
The endeavor to go beyond the static approximation was first addressed by Strinati for core excitons in semiconductors.\cite{Strinati_1988,Strinati_1982a,Strinati_1982b,Strinati_1984} 
Then, using first-order perturbation theory, Rohlfing and co-workers have developed a way to take into account dynamical effects via the plasmon-pole approximation combined with the Tamm-Dancoff approximation (TDA). \cite{Rohlfing_2000,Ma_2009a,Ma_2009b,Kaczmarski_2010,Baumeier_2012a,Baumeier_2012b,Rohlfing_2012,Lettmann_2019}
Recently, Loos and Blase \cite{Loos_2020h} proposed a dynamical scheme similar to Rohlfing's that goes beyond the plasmon-pole approximation where the dynamical screening of the Coulomb interaction is computed exactly within the random-phase approximation (RPA). \cite{Bohm_1951,Pines_1952,Bohm_1953,Ren_2012,Chen_2017}  

Unfortunately, even though this dynamical scheme allows to dynamically correct the single excitations obtained from the static approach, it does not permit access to double excitations. 
A way to obtain these higher solutions is to resort to the spin-flip formalism where one considers a higher spin state as a reference.\cite{Krylov_2001a,Monino_2021} 
Note, however, that the spin-flip formalism does not give access to all double excitations and is hampered by spin contamination.\cite{Casanova_2020,Monino_2022b}

Recently, Backhouse and Booth have introduced an upfolding version of the non-linear GF2 equations which provides a linear eigenvalue problem of larger dimension \cite{Backhouse_2020a} (see also Ref.~\onlinecite{Schirmer_2018}).
Bintrim and Berkelbach have extended it to the non-linear $GW$ equations. \cite{Bintrim_2021a} 
This linear eigenvalue problem allowed us to understand the role of intruder states in the origin of the energy surface discontinuities, \cite{Monino_2022} as well as the connections between Green's function methods and coupled-cluster theory. \cite{Lange_2018,Quintero_2022,Tolle_2023}
More importantly, when combined with other computational techniques, the upfolding framework provides a way to significantly lower the computational scaling of these approaches. \cite{Backhouse_2020a,Backhouse_2020b,Bintrim_2021a,Backhouse_2021,Scott_2023,Bintrim_2021a}
The same concept was also applied to the dynamical BSE eigenvalue problem built from the $GW$ kernel in order to go beyond the static approximation. \cite{Bintrim_2022} 
This upfolding approach produces a linear eigenvalue problem in an expanded space of single and double excitations, hence direct access to doubly-excited states with, however, limited success in terms of accuracy.

In this work, we investigate both charged and neutral excitations in Secs.~\ref{sec:CE} and \ref{sec:NE}, respectively.
We begin by reviewing the equations associated with the GF2 (Sec.~\ref{sec:Sig_GF2}), $GW$ (Sec.~\ref{sec:Sig_GW}), and $T$-matrix  (Sec.~\ref{sec:Sig_GT}) self-energies in various forms.
Subsequently, we present and study various static and dynamic BSE kernels based on the GF2 (Sec.~\ref{sec:Xi_GF2}), $GW$, (Secs.~\ref{sec:Xi_GW} and \ref{sec:Theta_GW}), and $T$-matrix (Sec.~\ref{sec:Xi_GT}) approximations, elucidating their interconnections and similarities with other theories.
Computational details are provided in Sec.~\ref{sec:comp_det}.
In Sec.~\ref{sec:res_CEs}, we assess the accuracy of the three different self-energies in calculating the principal ionization potentials of a subset of atoms and molecules taken from the $GW100$ dataset. \cite{vanSetten_2015}
Section \ref{sec:res_NEs} reports the computation of neutral excitations for another set of molecules using four different kernels within the static approximation. 
Additionally, we evaluate dynamical corrections through perturbation theory.
Our analysis considers various types of excited states, predominantly valence and Rydberg states, and investigates the performance of these kernels based on the specific type of states. 
Finally, in Sec.~\ref{sec:2T}, we compute, at various levels of theory, the gap between the first singlet and triplet excited states of cycl[3,3,3]azine, a model light-emitting diode (OLED) emitter for thermally activated delayed fluorescence (TADF), which has the particularity of having an inverted singlet-triplet gap.
Our conclusions are presented in Sec.~\ref{sec:ccl}. 
Atomic units are consistently employed throughout.

\section{Charged excitations}
\label{sec:CE}

\subsection{Quasiparticle equation}
\label{sec:QP}
Within the one-shot scheme, in order to obtain the quasiparticle energies and the corresponding satellites, one solve, for each spinorbital $p$ and assuming real values of the frequency $\omega$, the following (non-linear) quasiparticle equation
\begin{equation}
\label{eq:qp_eq}
	\e_{p}^{\HF} + \Sig_{pp}(\omega) - \omega = 0
\end{equation}
where $\Sig_{pp}(\omega)$ is a diagonal element of the correlation part of the self-energy. 
Due to the fact that one is usually interested in the quasiparticle solution, Eq.~\eqref{eq:qp_eq} is often linearized around $\omega = \e_{p}^{\HF}$, \ie, 
\begin{equation}
	\Sig_{pp}(\omega) 
	\approx \Sig_{pp}(\e_{p}^{\HF}) 
	+ (\omega - \e_{p}^{\HF}) \eval{\pdv{\Sig_{pp}(\omega)}{\omega}}_{\omega = \e_{p}^{\HF}}
\end{equation}
which yields 
\begin{equation}
\label{eq:lin_qp_eq}
	\e_{p} =  \e_{p}^{\HF} + Z_p \Sig_{pp}(\e_{p}^{\HF}) 
\end{equation}
where 
\begin{equation}
	Z_p = \qty [1 - \eval{\pdv{\Sig_{pp}(\omega)}{\omega}}_{\omega = \e_{p}^{\HF}}]^{-1}
\end{equation}
is a renormalization factor ($0 \leq Z_p \leq 1$) which represents the spectral weight of the quasiparticle solution.

The non-linear quasiparticle equation \eqref{eq:qp_eq} can be \textit{exactly} transformed into a larger linear problem via the upfolding process mentioned earlier where the 2h1p and 2p1h sectors are upfolded from the 1h and 1p sectors. \cite{Backhouse_2020a,Backhouse_2020b,Bintrim_2021a,Backhouse_2021,Monino_2022,Riva_2022,Quintero_2022,Tolle_2023,Marie_2023}
For each orbital $p$, this yields a linear eigenvalue problem of the form
\begin{equation}
	\bH_{p} \cdot \bc_{\nu} = \e_{\nu} \bc_{\nu}
\end{equation}
where $\nu$ runs over all solutions, quasiparticle and satellites, and with \cite{Tolle_2023}
\begin{equation}
\label{eq:Hp}
	\bH_{p} = 
	\begin{pmatrix}
		\e_{p}^{\HF}		&	\bV_{p}^{\text{2h1p}}	&	\bV_{p}^{\text{2p1h}}
		\\
		\qty(\bV_{p}^{\text{2h1p}})^{\dag}	&	\bC^{\text{2h1p}}			&	\bO
		\\
		\qty(\bV_{p}^{\text{2p1h}})^{\dag}	&	\bO				&	\bC^{\text{2p1h}}	
	\end{pmatrix}
\end{equation}

The diagonalization of $\bH_{p}$ is equivalent to solving the quasiparticle equation \eqref{eq:qp_eq}.
This can be further illustrated by expanding the secular equation associated with Eq.~\eqref{eq:Hp}
\begin{equation}
	\det[ \bH_{p} - \omega \bId ] = 0 
\end{equation}
and comparing it with Eq.~\eqref{eq:qp_eq} by setting
\begin{equation}
\label{eq:self-energy}
\begin{split}
	\Sig_{pp}(\omega) 
	& = \bV_{p}^{\text{2h1p}} \cdot \qty(\omega \bId - \bC^{\text{2h1p}} )^{-1} \cdot \qty(\bV_{p}^{\text{2h1p}})^{\dag}
	\\
	& + \bV_{p}^{\text{2p1h}} \cdot \qty(\omega \bId - \bC^{\text{2p1h}} )^{-1} \cdot \qty(\bV_{p}^{\text{2p1h}})^{\dag} 
\end{split}
\end{equation}
where $\bId$ is the identity matrix. 

It can be readily seen from Eq.~\eqref{eq:Hp} that the hole (h) and particle (p) sectors are potentially coupled.
This coupling, which is absent in coupled-cluster theory, \cite{Lange_2018,Rishi_2020,Quintero_2022} is critical for generating effective higher-order diagrams in Green's function methods. \cite{Schirmer_2018}

In the present work, we look at various approximations for the dynamical self-energy $\Sig_{pp}(\omega)$ and it obviously leads to different expressions for the blocks $\bC^{\text{2h1p}}$, $\bC^{\text{2p1h}}$, $\bV_{p}^{\text{2h1p}}$, and $\bV_{p}^{\text{2p1h}}$. 
In the following, for each approximation, we provide the expression for the self-energy and the different blocks. 

\subsection{GF2 self-energy}
\label{sec:Sig_GF2}
Within the {\GF2} approximation, one only takes into account the direct and exchange second-order diagrams, \cite{MattuckBook} and the self-energy is given by \cite{Cederbaum_1977,Oddershede_1984,SzaboBook} 
\begin{equation}
\label{eq:SigGF2}
\begin{split}
	\Sigma^{\GF2}(12) 
	& = G(12) \int v(13)G(34)G(43)v(42)d3d4 
	\\
	& - \int G(13)v(14)G(34)G(42)v(32)d3d4
\end{split}
\end{equation}
where $v(12)=\abs{\br_1 -\br_2}^{-1}$ is the bare Coulomb operator.

In the spinorbital basis, the self-energy is constituted by a hole and a particle term as follows
\begin{equation}
\label{eq:SigCGF2}
\begin{split}
	\Sig_{pq}^{\GF2}(\omega) 
	& = \frac{1}{2}\sum_{ija} \frac{\ERI{pa}{ij} \ERI{qa}{ij}}{\omega + \e_{a}^{\HF} - \e_{i}^{\HF} - \e_{j}^{\HF}}
	\\
	& + \frac{1}{2}\sum_{iab} \frac{\ERI{pi}{ab} \ERI{qi}{ab}}{\omega + \e_{i}^{\HF} - \e_{a}^{\HF} - \e_{b}^{\HF}}
\end{split}
\end{equation}
with $\ERI{pq}{rs} = \braket*{pq}{rs}- \braket*{pq}{sr}$ the antisymmetrized two-electron integrals written in Dirac's notation, \ie, 
\begin{equation}
	\braket*{pq}{rs} = \iint \MO{p}(\bx_1) \MO{q}(\bx_2) v(12) \MO{r}(\bx_1) \MO{s}(\bx_2) d\bx_1 d\bx_2
\end{equation}
As mentioned above, one can rely on an equivalent linear eigenvalue problem [see Eq.~\eqref{eq:Hp}] where the diagonal blocks are given by \cite{Schirmer_2018,Backhouse_2020a} 
\begin{subequations}
\begin{align}
	C^\text{2h1p}_{ija,klc} & = \qty(- \e_{a}^{\HF} + \e_{i}^{\HF} + \e_{j}^{\HF}) \delta_{ik} \delta_{jl} \delta_{ac} 
	\\
	C^\text{2p1h}_{iab,kcd} & = \qty(- \e_{i}^{\HF} + \e_{a}^{\HF} + \e_{b}^{\HF})\delta_{ik} \delta_{ac} \delta_{bd}
\end{align}
\end{subequations}
and the corresponding coupling blocks read
\begin{align}
	V^\text{2h1p}_{p,klc} & = \frac{\ERI{pc}{kl}}{\sqrt{2}} 
	&
	V^\text{2p1h}_{p,kcd} & = \frac{\ERI{pk}{dc}}{\sqrt{2}} 
\end{align}
Using Eq.~\eqref{eq:self-energy} we can see that one easily retrieves the self-energy expression in Eq.~\eqref{eq:SigCGF2}.
As already discussed in the literature, \cite{Schirmer_2018,Backhouse_2021} it is worth mentioning that we recover the same secular equations as the second-order algebraic-diagrammatic construction [ADC(2)] treatment of the electron propagator in its Dyson form. \cite{Schirmer_1982,Schirmer_1983,Schirmer_1984,Schirmer_2018,Dreuw_2015}

\subsection{$GW$ self-energy}
\label{sec:Sig_GW}

$GW$ is an approximation to Hedin's equations, a set of exact coupled integro-differential equations. \cite{Hedin_1965} 
Diagrammatically, $GW$ takes into account all the direct ring diagrams via a resummation technique \cite{MattuckBook} and is adequate in the high-density regime where correlation is weak. \cite{Gell-Mann_1957,Nozieres_1958}
Therefore, $GW$ includes the second-order direct term contained in GF2 but lacks its second-order exchange counterpart.
The $GW$ approximation is a relatively low computational cost method \cite{Foerster_2011,Liu_2016,Wilhelm_2018,Forster_2021,Duchemin_2019,Duchemin_2020,Duchemin_2021,Romanova_2022,Weng_2021,Romanova_2020,Brooks_2020,Vlcek_2019,Vlcek_2017} that relies on the dynamically screened Coulomb potential $W$ \alert{that is usually} computed at the direct (\ie, without exchange) particle-hole RPA (ph-RPA) level. 
In solids and large molecular systems, screening is usually significant, and the (frequency-dependent) screened Coulomb interaction is noticeably weaker than the (static) bare one.
\alert{One can also include so-called internal vertex corrections for the calculation of the polarizability.
In this case, one talks about test charge-test charge (tc-tc) polarizabilities, and various choices are possible. \cite{Cunningham_2018,Dadkhah_2023,Grzeszczyk_2023,Shishkin_2007b,Chen_2015,Caruso_2016,Lewis_2019a}}

The ph-RPA equations take the form of a non-Hermitian eigenvalue problem written in the basis of single excitations and deexcitations: 
\begin{multline}
\label{eq:phRPA}
	\begin{pmatrix}
		\bA^{\ph} & \bB^{\ph} 
		\\
		- \bB^{\ph} &  -\bA^{\ph}
	\end{pmatrix}
	\cdot
	\begin{pmatrix}
		\bX^{\ph} & \bY^{\ph}
		\\
		\bY^{\ph} & \bX^{\ph}
	\end{pmatrix}
	\\
	=
	\begin{pmatrix}
		\bX^{\ph} & \bY^{\ph}
		\\
		\bY^{\ph} & \bX^{\ph}
	\end{pmatrix}
	\cdot
	\begin{pmatrix}
		\bOm^{\ph} & \bO
		\\
		\bO & -\bOm^{\ph}
	\end{pmatrix}
\end{multline} 
with 
\begin{subequations}
\begin{align}
\label{eq:A_phRPA}
	A_{ia,jb}^\ph & = (\e_{a}^{\HF} - \e_{i}^{\HF}) \delta_{ij} \delta_{ab} + \braket{ib}{aj} 
	\\
\label{eq:B_phRPA}	
	B_{ia,jb}^\ph & = \braket{ij}{ab} 
\end{align}
\end{subequations}
In the absence of instabilities, \ie, when $\bA^{\ph} - \bB^{\ph}$ is positive definite, the ph-RPA problem reduces to a Hermitian problem of half the size.
If one includes exchange in Eq.~\eqref{eq:A_phRPA} and \eqref{eq:B_phRPA}, one ends up with RPA with exchange (RPAx) which is equivalent to time-dependent HF (TDHF). 
Note that TDHF within the TDA, where one removes the coupling between excitations and deexcitations, \ie, $\bB=\bO$, is equivalent to configuration interaction with singles (CIS). \cite{Dreuw_2005} 

Within the $GW$ approximation, the self-energy is defined by the following simple expression:
\begin{equation}
\label{eq:SigGW}
	\Sigma^{GW}(12) = i G(12)\alert{W_\text{c}}(12)
\end{equation}
which clearly justifies the name of this approximation. 
\alert{Here $W_\text{c} = W - v$ is the correlation part of the screened Coulomb interaction.}
In the spinorbital basis, the self-energy reads 
\begin{equation}
\label{eq:SigCGW}
	\Sig_{pq}^{\GW}(\omega) 
	= \sum_{im} \frac{\sERI{pi,m}{\ph}\sERI{qi,m}{\ph}}{\omega - \e_{i}^{\HF} + \Om_{m}^{\ph}}
	+ \sum_{am} \frac{\sERI{pa,m}{\ph}\sERI{qa,m}{\ph}}{\omega - \e_{a}^{\HF} - \Om_{m}^{\ph}}
\end{equation}
where the screened two-electron integrals are given by 
\begin{equation}
\label{eq:sERI_RPA}
	\sERI{pq,m}{\ph} = \sum_{ia} \braket*{pi}{qa} \qty(\bX^{\ph} + \bY^{\ph} )_{ia,m}
\end{equation}

As shown by Bintrim and Berkelbach \cite{Bintrim_2021a}, and more recently by T\"olle and Chan \cite{Tolle_2023} (who have been able to eschew the use of the TDA), the blocks $\bC^{\text{2h1p}}$ and $\bC^{\text{2p1h}}$ defined in Eq.~\eqref{eq:Hp} are diagonal with elements 
\begin{align}
	C^\text{2h1p}_{im,im} & = \e_{i}^{\HF} - \Om_{m}^{\ph}
	&
	C^\text{2p1h}_{am,am} & = \e_{a}^{\HF} + \Om_{m}^{\ph}
\end{align}
and the coupling blocks read
\begin{align}
	V^\text{2h1p}_{p,im} & = \sERI{pi,m}{\ph} 
	&
	V^\text{2p1h}_{p,am} & = \sERI{pa,m}{\ph} 
\end{align}
where $\sERI{pq,m}{\ph}$ are the screened integrals of Eq.~\eqref{eq:sERI_RPA}. 
Using the expressions of the different blocks one can, via the inverse process, obtain the expression of the self-energy as described in Eq.~\eqref{eq:self-energy} and recover Eq.~\eqref{eq:SigCGW}.

Note that an attempt of decoupling the 2h1p and 2p1h spaces within $GW$, as it is done in non-Dyson ADC, \cite{Schirmer_1998} has been made but with very mitigated results. \cite{Bintrim_2021a} 

\subsection{$T$-matrix self-energy}
\label{sec:Sig_GT}
While $GW$ depends on the dynamically screened Coulomb potential $W$, the $T$-matrix approximation relies on the so-called $T$-matrix, which, diagrammatically, corresponds to a resummation of a different class of diagrams known as ladder diagrams. \cite{MattuckBook}
Unlike the two-point quantity $W$, the four-point $T$-matrix is spin-dependent, and mixes the singlet and triplet spin channels in the computation of the self-energy.
The $T$-matrix approximation, which contains both second-order diagrams as well as additional higher-order ladder diagrams, is usually preferred to $GW$ when the screening is weak or, in other words, in the low-density regime.
 
While $W$ is computed using ph-RPA, the $T$-matrix is computed using the particle-particle (pp) RPA (pp-RPA) problem which is a non-Hermitian eigenvalue problem expressed in the basis of double electron attachments and double electron detachments: \cite{Schuck_Book,vanAggelen_2013,Peng_2013,Scuseria_2013,Yang_2013,Yang_2013b,vanAggelen_2014,Yang_2014a,Zhang_2015,Zhang_2016,Bannwarth_2020} 
\begin{multline}
\label{eq:ppRPA}
	\begin{pmatrix}
		\bC^{\pp} & \bB^{\pp/\hh}
		\\
		-\qty(\bB^{\pp/\hh})^{\dag} &  -\bD^{\hh}
	\end{pmatrix}
	\cdot
	\begin{pmatrix}
		\bX^{\pp} & \bY^{\hh}
		\\
		\bY^{\pp} & \bX^{\hh}
	\end{pmatrix}
	\\
	=
	\begin{pmatrix}
		\bOm^{\pp} & \bO 
		\\
		\bO & \bOm^{\hh}
	\end{pmatrix}
	\cdot
	\begin{pmatrix}
		\bX^{\pp} & \bY^{\hh}
		\\
		\bY^{\pp} & \bX^{\hh}
	\end{pmatrix}
\end{multline} 
where
\begin{subequations}
\begin{align}
	C_{ab,cd}^{\pp} 
	& = (\e_{a}^{\HF} + \e_{b}^{\HF}) \delta_{ac} \delta_{bd} + \ERI{ab}{cd}
	\\
	B_{ab,ij}^{\pp/\hh} 
	& = \ERI{ab}{ij}
	\\
	D_{ij,kl}^{\hh} 
	& = -(\e_{i}^{\HF} + \e_{j}^{\HF}) \delta_{ik} \delta_{jl} + \ERI{ij}{kl}
\end{align}
\end{subequations}
with the following index restrictions $a<b$, $c<d$, $i<j$, and $k<l$. 

Within the $T$-matrix approximation, \alert{the correlation part of the self-energy is}
\begin{equation}
\label{eq:SigGT}
	\Sigma^{GT}(12) = i \int G(43)\alert{T_\text{c}}(13,24)d3d4
\end{equation}
\alert{(where $\alert{T_\text{c}}$ is the correlation part of the $T$ matrix)} and the elements of the self-energy in the spinorbital basis are explicitly given by \cite{Romaniello_2012,Zhang_2017}
\begin{equation}
\label{eq:SigCGT}
	\Sig_{pq}^{\GT}(\omega) 
	= \sum_{in} \frac{\sERI{pi,n}{\pp}\sERI{qi,n}{\pp}}{\omega + \e_{i}^{\HF} - \Om_{n}^{\pp}}
	+ \sum_{an} \frac{\sERI{pa,n}{\hh}\sERI{qa,n}{\hh}}{\omega + \e_{a}^{\HF} - \Om_{n}^{\hh}}
\end{equation}
where the pp and hh versions of the screened two-electron integrals read
\begin{subequations}
\begin{align}
\label{eq:chi_RPA1}
	\sERI{pq,n}{\pp} 
	& = \sum_{c<d} \ERI{pq}{cd} X_{cd,n}^{\pp} 
	+ \sum_{k<l} \ERI{pq}{kl} Y_{kl,n}^{\pp}
	\\
\label{eq:chi_RPA2}
	\sERI{pq,n}{\hh} 
	& = \sum_{c<d} \ERI{pq}{cd} X_{cd,n}^{\hh} 
	+ \sum_{k<l} \ERI{pq}{kl} Y_{kl,n}^{\hh} 
\end{align}
\end{subequations}

Following the upfolding process of Bintrim and Berkelbach \cite{Bintrim_2021a} together with the generalization of T\"olle and Chan, \cite{Tolle_2023} in the case of the $T$-matrix, we have the following diagonal elements for the blocks $\bC^{\text{2h1p}}$ and $\bC^{\text{2p1h}}$ of Eq.~\eqref{eq:Hp}
\begin{align}
	C^\text{2h1p}_{an,an} & = - \e_{a}^{\HF} + \Om_{n}^{\hh}
	&
	C^\text{2p1h}_{in,in} & = - \e_{i}^{\HF} + \Om_{n}^{\pp}
\end{align}
and the corresponding coupling blocks are
\begin{align}
	V^\text{2h1p}_{p,an} & = \sERI{pa,n}{\hh}
	&
	V^\text{2p1h}_{p,in} & = \sERI{pi,n}{\pp}
\end{align}
where the screened integrals are given by Eqs.~\eqref{eq:chi_RPA1} and \eqref{eq:chi_RPA2}. 

\section{Neutral excitations}
\label{sec:NE}

\subsection{Bethe-Salpeter equation}
\label{sec:BSE}
Within the BSE formalism, one must solve, in the general setting, a non-linear eigenvalue problem of the form 
\begin{multline}
\label{eq:BSE}
    \begin{pmatrix}
    	\bA^{\BSE}(\Om_{\nu}^{\BSE}) & \bB^{\BSE}(\Om_{\nu}^{\BSE}) 
		\\
	    -\bB^{\BSE}(-\Om_{\nu}^{\BSE}) & -\bA^{\BSE}(-\Om_{\nu}^{\BSE})
    \end{pmatrix}
	\cdot 
    \begin{pmatrix}
	    \bX_{\nu}^{\BSE} 
	    \\
	    \bY_{\nu}^{\BSE}
    \end{pmatrix}
	\\
	=
	\Om_{\nu}^{\BSE}
	\begin{pmatrix}
		\bX_{\nu}^{\BSE} 
		\\
		\bY_{\nu}^{\BSE}
	\end{pmatrix}
\end{multline}
where the (anti)resonant block $\pm\bA^{\BSE}(\omega)$ and the coupling blocks $\pm\bB^{\BSE}(\omega)$ are dynamical quantities and the index $v$ runs over single, double, and potentially higher excitations.
Of course, their expressions depend on the type of quasiparticles and the kernel that one considers but they have the following generic expressions
\begin{subequations}
\begin{align}
	A_{ia,jb}^{\BSE}(\omega) 
	& = A_{ia,jb} + \Xi_{ia,jb}(\omega)
	\\
	B_{ia,jb}^{\BSE}(\omega) 
	& = B_{ia,jb} + \Xi_{ia,bj}(\omega)
\end{align}
\end{subequations}
with the following static parts
\begin{subequations}
\begin{align}
	A_{ia,jb}
	& = (\e_{a} - \e_{i}) \delta_{ij} \delta_{ab} + \ERI{ib}{aj}
	\\
	B_{ia,jb}
	& = \ERI{ij}{ab}
\end{align}
\end{subequations}
where the $\e_{p}$'s are quasiparticle energies and $\Xi_{pq,rs}(\omega)$ is an element of the dynamical correlation kernel computed at a given level of theory. 
Note that, although these matrices are built in the single excitation and deexcitation manifolds, thanks to the frequency dependence of these quantities, one can potentially access higher excitations.

Again, one can enforce the TDA to obtain a simpler non-linear system
\begin{equation} 
\label{eq:BSE_TDA}
	\bA^{\BSE}(\Om_{\nu}^{\BSE}) \cdot \bX_{\nu}^{\BSE} = \Om_{\nu}^{\BSE} \bX_{\nu}^{\BSE}
\end{equation}
Below, we present three different ways of tackling the BSE problem.

First, one can enforce the so-called static approximation where one sets 
\begin{subequations}
\begin{align}
	A_{ia,jb}^{\BSE}
	& = A_{ia,jb} + \Xi_{ia,jb}
	\\
	B_{ia,jb}^{\BSE}
	& = B_{ia,jb} + \Xi_{ia,bj}
\end{align}
\end{subequations}
to get
\begin{equation}
\label{eq:BSEstat}
	\begin{pmatrix}
		\bA^{\BSE}		&	\bB^{\BSE} 
		\\
		-\bB^{\BSE}	&	-\bA^{\BSE}
	\end{pmatrix}
	\cdot 
	\begin{pmatrix}
		\bX_{m}^{\BSE} 
		\\
		\bY_{m}^{\BSE}
	\end{pmatrix}
	 =
	\Om_{m}^{\BSE}
	\begin{pmatrix}
		\bX_{m}^{\BSE} 
		\\
		\bY_{m}^{\BSE}
	\end{pmatrix}
\end{equation}
In this case, because of the frequency-independent nature of the static kernel's elements $\Xi_{pq,rs}$, one only accesses single excitations.

Second, one can go beyond the static approximation by using a renormalized first-order perturbative correction to the static BSE excitation energies.  
(We refer the interested reader to Ref.~\onlinecite{Loos_2020h} for a detailed discussion. Here, we only provide the main equations.)
This dynamical correction to the static BSE kernel (labeled dBSE in the following) allows us to recover additional relaxation effects coming from higher excitations.  

The dBSE excitation energies are then obtained via
\begin{equation}
	\Om_{m}^{\dBSE} = \Om_{m}^{\BSE} + \zeta_{m} \Tilde{\Om}_{m}^{\BSE}
\end{equation}
where the $\Om_{m}^{\BSE}$'s are the static (zeroth-order) BSE excitation energies defined in Eq.~\eqref{eq:BSEstat} and 
\begin{equation}
\label{eq:Om1-TDA}
	\Tilde{\Om}_{m}^{\BSE} = (\bX_{m}^{\BSE})^{\dag} \cdot \Delta \bXi(\Om_{m}^{\BSE}) \cdot \bX_{m}^{\BSE}
\end{equation}
are first-order corrections obtained within the dynamical TDA (\ie, as commonly done, only the resonant block is corrected for dynamical effects) with the renormalization factor
\begin{equation}
\label{eq:Z}
	\zeta_{m} = \qty[ 1 - (\bX_{m}^{\BSE})^{\dag} \cdot \eval{ \pdv{\Delta \bXi(\omega)}{\omega} }_{\omega = \Om_{m}^{\BSE}} \cdot \bX_{m}^{\BSE} ]^{-1}
\end{equation}
The generic expression for $\Delta \bXi(\omega)$ is
\begin{equation}
	\Delta \Xi_{ia,jb}(\omega) = \Tilde{\Xi}_{ia,jb}(\omega) - \Xi_{ia,jb}
\end{equation}
where $\Tilde{\Xi}_{ia,jb}(\omega)$ is an element of the so-called effective dynamical kernel.
Unlike in the quasiparticle case (see Sec.~\ref{sec:QP}), this renormalization factor $\zeta_{m}$ is not restricted between 0 and 1. 
However, it has been found to be close to unity in most cases which indicates the satisfactory convergence properties of the perturbative series. \cite{Loos_2020h,Monino_2021}

Third, within the TDA, it is also possible to transform the non-linear eigenvalue problem \eqref{eq:BSE_TDA} into a larger linear problem via an upfolding process where the 2h2p sector is upfolded from the 1h1p sector. 
The structure (and the dimension) of this matrix $\Tilde{\bH}$ depends on the nature and origin of the kernel. 
In the following, we present four different kernels and, for each of them, we provide the corresponding working equations.   

\subsection{Second-order GF2 kernel}
\label{sec:Xi_GF2}
We first discuss the BSE correlation kernel based on the GF2 self-energy considered in Eq.~\eqref{eq:SigGF2}:

\begin{equation}
	\Xi^{\GF2}(35,46) = i \fdv{\Sigma^{\GF2}(34)}{G(65)} 
\end{equation}
To avoid lengthy derivations and expressions, we refer the interested reader to the work of Zhang \textit{et al.} (and particularly to the supplementary material) for the full derivation of the GF2 kernel. \cite{Zhang_2013}
Additional details and complements can be found in the work of Rebolini and Toulouse. \cite{Rebolini_2016,Rebolini_PhD}

At the BSE@GF2 level, we have 
\begin{subequations}
\begin{align}
	A_{ia,jb}^{\GF2}(\omega) 
	& = A_{ia,jb}^{\GF2} + \Xi^{\GF2}_{ia,jb}(\omega)
	\\
	B_{ia,jb}^{\GF2}(\omega) 
	& = B_{ia,jb}^{\GF2} + \Xi^{\GF2}_{ia,bj}(\omega)
\end{align}
\end{subequations}
with
\begin{subequations}
\begin{align}
	A_{ia,jb}^{\GF2}
	& =  (\e_{a}^{\GF2} - \e_{i}^{\GF2}) \delta_{ij} \delta_{ab} + \ERI{ib}{aj}
	\\
	B_{ia,jb}^{\GF2}
	& = \ERI{ij}{ab}
\end{align}
\end{subequations}
and the elements of the second-order (with respect to the Coulomb interaction) static kernel for the (anti)resonant and coupling blocks are given by the following expression:
\begin{equation}
\label{eq:XiGF2}
\begin{split}
	\Xi^{\GF2}_{pq,rs} 
	& = \sum_{kc} \frac{\ERI{rc}{pk}\ERI{kq}{cs}}{ \e_{c}^{\GF2} - \e_{k}^{\GF2}} 
	+ \sum_{kc} \frac{\ERI{rk}{pc}\ERI{cq}{ks}}{ \e_{c}^{\GF2} - \e_{k}^{\GF2}} 
	\\
	& + \frac{1}{2} \sum_{kl} \frac{\ERI{qr}{kl}\ERI{lk}{sp}}{ \e_{k}^{\GF2} + \e_{l}^{\GF2}} 
	+ \frac{1}{2} \sum_{cd} \frac{\ERI{qr}{cd}\ERI{dc}{sp}}{ \e_{c}^{\GF2} + \e_{d}^{\GF2}} 
\end{split}
\end{equation}

Going beyond the static approximation, the elements of the dynamical kernel for the resonant block are 
\begin{equation}
\label{eq:dynamical_BSE2_GF2}
\begin{split}
	\Tilde{\Xi}^{\GF2}_{ia,jb}(\omega) 
	= & -\sum_{kc} \frac{\ERI{jc}{ik} \ERI{ka}{cb}}{\omega - (\e_{b}^{\GF2} + \e_{c}^{\GF2} - \e_{i}^{\GF2} - \e_{k}^{\GF2})} 
	\\
	& - \sum_{kc} \frac{\ERI{jk}{ic} \ERI{ca}{kb}}{\omega - (\e_{a}^{\GF2} + \e_{c}^{\GF2} - \e_{j}^{\GF2} - \e_{k}^{\GF2})} 
	\\
	& + \frac{1}{2} \sum_{kl} \frac{\ERI{aj}{kl} \ERI{lk}{bi}}{\omega - (\e_{a}^{\GF2} + \e_{b}^{\GF2} - \e_{k}^{\GF2} - \e_{l}^{\GF2})} 
	\\
	& + \frac{1}{2} \sum_{cd} \frac{\ERI{aj}{cd} \ERI{dc}{bi}}{\omega - (\e_{c}^{\GF2} + \e_{d}^{\GF2} - \e_{i}^{\GF2} - \e_{j}^{\GF2})} 
\end{split}
\end{equation}
The first two terms in Eqs.~\eqref{eq:XiGF2} and \eqref{eq:dynamical_BSE2_GF2} are ph and hp terms, while the third and fourth ones correspond to hh and pp terms, respectively.

Within the TDA, the upfolding process leads to the following linear eigenvalue problem 
\begin{equation}
\label{eq:H}
	\tilde{\bH}^{\GF2} = 
	\begin{pmatrix}
		\bA^{\GF2}			&	\bI+\bK		&	\bJ			&	\bK				
		\\
		\qty(\bJ+\bL)^{\dag}	&	\bC^{\GF2}	&	\bO			&	\bO							
		\\
		\bK^{\dag}			&	\bO			&	\bC^{\GF2}	&	\bO			
		\\								
		\bJ^{\dag}			&	\bO			&	\bO			&	\bC^{\GF2}	
	\end{pmatrix}
\end{equation}
where, the block $\bC^{\GF2}$ is diagonal with elements 
\begin{equation}
	C_{ijab,ijab}^{\GF2} = \e_{a}^{\GF2} + \e_{b}^{\GF2} - \e_{i}^{\GF2} - \e_{j}^{\GF2}
\end{equation}
while the various coupling terms read
\begin{subequations}
\begin{align}
	I_{ia,klcd} & = + \frac{\delta_{ac}}{\sqrt{2}} \ERI{di}{kl} 
	\\
	J_{ia,klcd} & = - \frac{\delta_{ad}}{\sqrt{2}} \ERI{ci}{kl}
	\\
	K_{ia,klcd} & = + \frac{\delta_{il}}{\sqrt{2}} \ERI{dc}{ak}
	\\
	L_{ia,klcd} & = - \frac{\delta_{ik}}{\sqrt{2}} \ERI{dc}{al}
\end{align}
\end{subequations}
Note that, in this case, the upfolded matrix is non-Hermitian and contains three blocks of double excitations (\ie, 2h2p configurations).
Therefore, spurious (\ie, non-physical) solutions are expected to appear due to the redundancy of the basis set. \cite{Romaniello_2009b,Sangalli_2011,Authier_2020}
By downfolding the three subspaces of double excitations onto the space of single excitations, \ie, 
\begin{equation}
\begin{split}
	\Tilde{\bXi}^{\GF2}(\omega) 
	& = \qty(\bI+\bK) \cdot \qty( \omega \bId -  \bC^{\GF2})^{-1} \cdot \qty(\bJ+\bL)^{\dag} 
	\\
	& + \bJ \cdot \qty( \omega \bId -  \bC^{\GF2})^{-1} \cdot \bK^{\dag} 
	\\
	& + \bK \cdot \qty( \omega \bId -  \bC^{\GF2})^{-1} \cdot \bJ^{\dag} 
\end{split}
\end{equation}
one recovers exactly the dynamical kernel defined in Eq.~\eqref{eq:dynamical_BSE2_GF2}.

Following Bintrim and Berkelbach, \cite{Bintrim_2022} we attempt to symmetrize $\Tilde{\bH}^{\GF2}$ and remove the redundant sets of 2h2p configurations by simply defining
\begin{equation}
\label{eq:BarH_GF2}
	\Bar{\bH}^{\GF2} = 
	\begin{pmatrix}
		\bA^{\HF}		&	\bV			
		\\
		\bV^{\dag}	&	\bC^{\HF}							
	\end{pmatrix}
\end{equation}
with $\bV = (\bI + \bJ + \bK + \bL)/\sqrt{2}$, and
\begin{subequations}
\begin{align}
	A_{ia,jb}^{\HF}
	& =  (\e_{a}^{\HF} - \e_{i}^{\HF}) \delta_{ij} \delta_{ab} + \ERI{ib}{aj}
	\\
	C_{ijab,ijab}^{\HF} 
	& =\e_{a}^{\HF} + \e_{b}^{\HF} - \e_{i}^{\HF} - \e_{j}^{\HF}
\end{align}
\end{subequations}
In this case, we obtain the dynamical kernel
\begin{equation}
	\Bar{\bXi}(\omega) = \bV \cdot \qty(\omega \bId - \bC^{\HF})^{-1} \cdot \bV^\dag
\end{equation}
with
\begin{equation}
\label{eq:bXi_GF2}
\begin{split}
	\Bar{\Xi}_{ia,jb}^{\GF2}(\omega) & =
	\frac{\delta_{ab}}{2} \sum_{klc} \frac{\ERI{kl}{ic} \ERI{kl}{jc} }{\omega - \qty(\e_{a}^{\HF} + \e_{c}^{\HF} - \e_{k}^{\HF} - \e_{l}^{\HF})} 
	\\
	& + \frac{\delta_{ij}}{2} \sum_{kcd} \frac{\ERI{ak}{cd} \ERI{bk}{cd} }{\omega - \qty(\e_{c}^{\HF} + \e_{d}^{\HF} - \e_{k}^{\HF} - \e_{i}^{\HF})} 
	\\
	& -  \sum_{kc} \frac{\ERI{jc}{ik} \ERI{ka}{cb} }{\omega - \qty(\e_{b}^{\HF} + \e_{c}^{\HF} - \e_{k}^{\HF} - \e_{i}^{\HF})}
	\\
	& -  \sum_{kc} \frac{\ERI{jk}{ic} \ERI{ca}{kb} }{\omega - \qty(\e_{a}^{\HF} + \e_{c}^{\HF} - \e_{k}^{\HF} - \e_{j}^{\HF})}
	\\
	& + \frac{1}{2} \sum_{kl} \frac{\ERI{aj}{kl} \ERI{lk}{bi} }{\omega - \qty(\e_{a}^{\HF} + \e_{b}^{\HF} - \e_{k}^{\HF} - \e_{l}^{\HF})} 
	\\
	& + \frac{1}{2} \sum_{cd} \frac{\ERI{aj}{cd} \ERI{dc}{bi} }{\omega - \qty(\e_{c}^{\HF} + \e_{d}^{\HF} - \e_{i}^{\HF} - \e_{j}^{\HF})} 
\end{split}
\end{equation}
where one can see that we recover the four terms of the original dynamical kernel \eqref{eq:dynamical_BSE2_GF2} with two additional self-energy terms that correspond to partial renormalization of the particle and hole sectors of the self-energy (forward time-ordered diagrams).
Therefore, in order to avoid double counting, one must use HF orbital energies instead of GF2 quasiparticle energies in Eq.~\eqref{eq:BarH_GF2}.
However, the particle (hole) propagator is only renormalized by the 2p1h (2h1p) configurations and not the 2h1p (2p1h) configurations. 

The expression \eqref{eq:bXi_GF2} has a strong connection with the ADC(2) method for the polarization propagator (\ie, for neutral excitations). \cite{Schirmer_1982,Schirmer_2004,Wormit_2014,Wormit_PhD,Dreuw_2015}
Indeed, the blocks $\bC^{\HF}$ and $\bV$ in ADC(2) have identical expressions.
However, in ADC(2), the 1h1p block has three additional second-order static terms.
By replacing $\bA^{\HF}$ by $\bA^{\HF} + \Bar{\bA}^{\GF2}$ in Eq.~\eqref{eq:BarH_GF2} with
\begin{widetext}
\begin{equation}
\begin{split}
	\Bar{A}_{ia,jb}^{\GF2}
	& = \frac{\delta_{ij}}{4} \sum_{klc}
	\qty[\frac{\ERI{ac}{kl} \ERI{kl}{bc}}{\e_{a}^{\HF} - \e_{k}^{\HF} + \e_{c}^{\HF} - \e_{l}^{\HF}} 
	+ \frac{\ERI{ac}{kl} \ERI{kl}{bc}}{\e_{b}^{\HF} - \e_{k}^{\HF} + \e_{c}^{\HF} - \e_{l}^{\HF}}]
	\\
	& + \frac{\delta_{ab}}{4} \sum_{kcd}\qty[\frac{\ERI{cd}{ik} \ERI{jk}{cd}}{\e_{c}^{\HF} - \e_{i}^{\HF} + \e_{d}^{\HF} - \e_{k}^{\HF}} 
	+ \frac{\ERI{cd}{ik} \ERI{jk}{cd}}{\e_{c}^{\HF} - \e_{j}^{\HF} + \e_{d}^{\HF} - \e_{k}^{\HF}}]
	\\
	& -\frac{1}{2} \sum_{kc}\qty[\frac{\ERI{ac}{ik} \ERI{jk}{bc}}{\e_{a}^{\HF} - \e_{i}^{\HF} + \e_{c}^{\HF} - \e_{k}^{\HF}} 
	+ \frac{\ERI{ac}{ik} \ERI{jk}{bc}}{\e_{b}^{\HF} - \e_{j}^{\HF} + \e_{c}^{\HF} - \e_{k}^{\HF}}]
\end{split}
\end{equation}
\end{widetext}
one ends up with exactly the ADC(2) secular equations.
Although static, the first two terms are particularly crucial as they complete the renormalization of the HF orbital energies via the introduction of the missing backward time-ordered diagrams.

\subsection{First-order $GW$ kernel}
\label{sec:Xi_GW}
Within the $GW$ approximation, using the self-energy defined in Eq.~\eqref{eq:SigGW}, the BSE kernel reads \cite{Albrecht_1998,Benedict_1998,Onida_2002,Marini_2003}
\begin{equation}
\label{eq:BSE2_GW}
\begin{split}
	i\fdv{\Sigma^{GW}(34)}{G(65)} 
	& = -\fdv{(G(34)\alert{W_\text{c}(34)})}{G(65)} 
	\\
	& = -\alert{W_\text{c}(34)}\fdv{G(34)}{G(65)} - G(34) \fdv{\alert{W_\text{c}(34)}}{G(65)}
	\\
	& = \Xi^{\GW}(35,46) + \Theta^{\GW}(35,46)
\end{split}
\end{equation}
and it is common practice to neglect $\Theta^{\GW}$. \cite{Hanke_1980,Strinati_1982a,Strinati_1982b,Strinati_1984,Strinati_1988}
(We shall come back to this point later on.) 
Thus, one gets the following static kernel elements \cite{Strinati_1988}
\begin{equation}
	\Xi_{pq,rs}^{\GW} =  2 \sum_m \frac{\sERI{pr,m}{\ph} \sERI{qs,m}{\ph}}{\Om_{m}^{\ph}}
\end{equation}

As for the GF2 case, it is possible to go beyond the static approximation by taking into account the dynamical structure of $W$, that is, \cite{Monino_2021}
\begin{equation}
\begin{split}
\label{eq:W}
	\Tilde{\Xi}_{ia,jb}^{GW}(\omega) = 
	 - \sum_m  \frac{\sERI{ij,m}{\ph} \sERI{ab,m}{\ph}}{\omega - (\e_{b}^{\GW} - \e_{i}^{\GW} +  \Om_{m}^{\ph})} 
	\\
	- \sum_m  \frac{\sERI{ij,m}{\ph} \sERI{ab,m}{\ph}}{\omega -  (\e_{a}^{\GW} - \e_{j}^{\GW} + \Om_{m}^{\ph})} 
\end{split}
\end{equation}
By removing the screening effects from $GW$, \ie, by performing the following substitutions, $\Om_{m}^{\ph} \to  \e_{a}^{\HF} - \e_{i}^{\HF}$ and $\sERI{pq,m}{\ph} \to \braket{pi}{qa}$, one recovers the two ph terms of Eq.~\eqref{eq:dynamical_BSE2_GF2}, without, of course, the exchange part.

As shown in Ref.~\onlinecite{Bintrim_2022}, at the BSE@$GW$ level, the upfolding process yields 
\begin{equation}
\label{eq:upfolded_BSE_GW}
	\Tilde{\bH}^{\GW} = 
	\begin{pmatrix}
		\bA^{\GW}			&	\bJ^{\ph}			&	\bK^{\ph}
		\\
		\qty(\bK^{\ph})^{\dag}	&	\bC^{\GW}	&	\bO
		\\
		\qty(\bJ^{\ph})^{\dag}	&	\bO			&	\bC^{\GW}	
	\end{pmatrix}
\end{equation}
with the usual static expression for the 1h1p part
\begin{equation}
	A_{ia,jb}^{\GW} =  (\e_{a}^{\GW} - \e_{i}^{\GW}) \delta_{ij} \delta_{ab} + \ERI{ib}{aj}
\end{equation}
a diagonal 2h2p block with elements
\begin{equation}
	C_{iam,iam}^{\GW} = \Om_{m}^{\ph} + \e_{a}^{\GW} - \e_{i}^{\GW} 
\end{equation}
and coupling blocks that read
\begin{subequations}
\begin{align}
	J_{ia,kcm}^{\ph} & =
	- \delta_{ac} \sERI{ik,m}{\ph}
	\\
	K_{ia,kcm}^{\ph} & =
	+ \delta_{ik} \sERI{ac,m}{\ph}
\end{align}
\end{subequations}
Again, $\Tilde{\bH}^{\GW}$ is a non-Hermitian matrix with two sets of double excitations.
By downfolding we get 
\begin{equation}
\begin{split}
	\Tilde{\bXi}^{\GW}(\omega) 
	& = \bJ^{\ph} \cdot \qty( \omega \bId -  \bC^{\GW})^{-1} \cdot \qty(\bK^{\ph})^{\dag} 
	\\
	& + \bK^{\ph} \cdot \qty( \omega \bId -  \bC^{\GW})^{-1} \cdot \qty(\bJ^{\ph})^{\dag} 
\end{split}
\end{equation}
which gives us back the dynamical kernel \eqref{eq:W}.
As proposed by Bintrim and Berkelbach, \cite{Bintrim_2022} one can also symmetrize $\Tilde{\bH}^{\GW}$ and remove the additional 2h2p block by defining
\begin{equation}
\label{eq:BarH_GW}
	\Bar{\bH}^{\GW} = 
	\begin{pmatrix}
		\bA^{\HF}	+ \Bar{\bA}^{\GW}	&	\bJ^{\ph} + \bK^{\ph}	\\
		\\
		\qty(\bJ^{\ph} + \bK^{\ph})^{\dag}	&	\Bar{\bC}^{\GW}	\\
	\end{pmatrix}
\end{equation}
with
\begin{equation}
	\bar{C}_{iam,iam}^{\GW} = \Om_{m}^{\ph} + \e_{a}^{\HF} - \e_{i}^{\HF} 
\end{equation}
but, again, the resulting dynamical kernel
\begin{equation}
	\Bar{\bXi}^{\GW}(\omega) 
	= \qty(\bJ^{\ph} + \bK^{\ph}) \cdot \qty( \omega \bId - \Bar{\bC}^{\GW})^{-1} \cdot \qty(\bJ^{\ph} + \bK^{\ph})^{\dag}
\end{equation}
contains additional self-energy terms:
\begin{equation}
\begin{split}
	\Bar{\Xi}_{ia,jb}^{GW}(\omega) 
	& = \delta_{ab} \sum_{km} \frac{\sERI{ik,m}{\ph} \sERI{jk,m}{\ph}}{\omega - (\e_{a}^{\HF} - \e_{k}^{\HF} +  \Om_{m}^{\ph})} 
	\\
	& + \delta_{ij} \sum_{cm} \frac{\sERI{ac,m}{\ph} \sERI{bc,m}{\ph}}{\omega -  (\e_{c}^{\HF} - \e_{i}^{\HF} + \Om_{m}^{\ph})} 
	\\
	& - \sum_m  \frac{\sERI{ij,m}{\ph} \sERI{ab,m}{\ph}}{\omega - (\e_{b}^{\HF} - \e_{i}^{\HF} +  \Om_{m}^{\ph})} 
	\\
	& - \sum_m  \frac{\sERI{ij,m}{\ph} \sERI{ab,m}{\ph}}{\omega -  (\e_{a}^{\HF} - \e_{j}^{\HF} + \Om_{m}^{\ph})} 
\end{split}
\end{equation}
It has been found to severely affect the excitation energies due to the lack of backward time-ordered diagrams in the self-energy. \cite{Bintrim_2022}
Hence, inspired by the ADC(2) expression, one could consider adding the missing self-energy terms (which correspond to the inclusion of the backward time-ordered diagrams) by defining the elements of $\Bar{\bA}^{\GW}$ in Eq.~\eqref{eq:BarH_GW} as 
\begin{equation}
\begin{split}
	\Bar{A}_{ia,jb}^{\GW} 
	& = \frac{\delta_{ij}}{2} \sum_{km} \qty[\frac{\sERI{ak,m}{\ph} \sERI{bk,m}{\ph}}{\e_{a}^{\HF} - \e_{k}^{\HF} + \Om_{m}^{\ph}} 
	+ \frac{\sERI{ak,m}{\ph} \sERI{bk,m}{\ph}}{\e_{b}^{\HF} - \e_{k}^{\HF} + \Om_{m}^{\ph} }]
	\\
	& - \frac{\delta_{ab}}{2} \sum_{cm}\qty[\frac{\sERI{ic,m}{\ph} \sERI{jc,m}{\ph}}{ \e_{i}^{\HF} - \e_{c}^{\HF} - \Om_{m}^{\ph}} 
	+ \frac{\sERI{ic,m}{\ph} \sERI{jc,m}{\ph}}{\e_{j}^{\HF} - \e_{c}^{\HF} - \Om_{m}^{\ph} }]
\end{split}
\end{equation}
One can then solely rely on HF orbital energies in the previous expressions, instead of the $GW$ quasiparticles. \cite{Bintrim_2022}
The study of the performance of this new scheme is left for future work.

\subsection{Second-order $GW$ kernel}
\label{sec:Theta_GW}
As mentioned above, it is customary to neglect the functional derivative $\fdv*{\alert{W_\text{c}}}{G}$ in the expression of the $GW$ kernel [see Eq.~\eqref{eq:BSE2_GW}]. 
However, a second-order $GW$ kernel, $\Theta^{\GW}$, that takes into account this additional term has been recently derived by Yamada \textit{et al.} \cite{Yamada_2022} and tested on the Thiel benchmark set \cite{Schreiber_2008,Silva-Junior_2010,Silva-Junior_2010b,Silva-Junior_2010c} within the plasmon-pole approximation. 
In the following, we refer to this scheme as BSE2@$GW$.

The second-order $GW$ kernel is naturally divided into two terms as follows: 
\begin{equation}
\begin{split}
	\Theta^{\GW}(35,46) 
	& = iG(35)G(64)W(34)W(56) 
	\\
	& + iG(35)G(64)W(36)W(54)
\end{split}
\end{equation}
Contrary to $\Xi^{\GW}$ which corresponds to the screening of the exchange term, the two additional second-order terms included in $\Theta^{\GW}$ screen the direct term.
As a consequence, BSE2@{\GW} only alters the excitation energies of the singlet excited states, while triplet states remain unaffected by this second-order correction.

In the spinorbital basis, we obtain the following static kernel elements:
\begin{equation}
\begin{split}
	\Theta^{\GW}_{pq,rs}
	& = \sum_{kc} \frac{W_{rk,pc}W_{qc,sk}}{\e_{c}^{\GW} - \e_{k}^{\GW}} 
	+ \sum_{kc} \frac{W_{rc,pk}W_{qk,sc}}{\e_{c}^{\GW} - \e_{k}^{\GW}} 
	\\
	& + \sum_{kl} \frac{W_{qr,kl}W_{kl,ps}}{\e_{k}^{\GW} + \e_{l}^{\GW}} 
	- \sum_{cd} \frac{W_{qr,cd}W_{cd,ps}}{\e_{c}^{\GW} + \e_{d}^{\GW}} 
\end{split}
\end{equation}
where
\begin{equation}
	W_{pq,rs} = - \braket{pq}{rs} + \Xi_{pq,rs}^{GW}
\end{equation}
are the elements of the dynamically-screened Coulomb potential in its static limit, while the elements of the dynamical kernel for the resonant block are \cite{Yamada_2022}
\begin{equation}
\begin{split}
	\Tilde{\Theta}^{\GW}_{ia,jb}(\omega) =
	& - \sum_{kc} \frac{W_{ac,bk} W_{jk,ic}}{ \omega - \qty(\e_{a}^{GW} + \e_{c}^{GW} - \e_{k}^{GW} - \e_{j}^{GW})}
	\\
	& - \sum_{kc} \frac{W_{ak,bc} W_{ki,cj}}{ \omega - \qty(\e_{c}^{GW} + \e_{b}^{GW} - \e_{i}^{GW} - \e_{k}^{GW})} 
	\\
	& + \sum_{cd} \frac{W_{aj,cd} W_{cd,ib}}{ \omega - \qty(\e_{c}^{GW} + \e_{d}^{GW} - \e_{j}^{GW} - \e_{i}^{GW})} 
	\\
	& + \sum_{kl} \frac{W_{aj,kl} W_{kl,ib}}{ \omega - \qty(\e_{a}^{GW} + \e_{b}^{GW} - \e_{k}^{GW} - \e_{l}^{GW})} 
\end{split}
\end{equation}
where one readily sees that hp, ph, pp, and hh contributions are included at the BSE2@$GW$ level.
As for the GF2 kernel (see Sec.~\ref{sec:Xi_GF2}), one can easily derive an upfolded version of this second-order kernel.

\subsection{First-order $T$-matrix kernel}
\label{sec:Xi_GT}
Another possible BSE kernel can be constructed using the $T$-matrix self-energy [see Eq.~\eqref{eq:SigGT}]. 
A detailed study of this kernel is performed in Ref.~\onlinecite{Loos_2022}.
Following a similar derivation as the {\GW} kernel, one gets at the BSE@$GT$ level
\begin{equation}
\begin{split}
	\Xi^{GT}(35,46) 
	& =  i\fdv{\Sigma^{GT}(34)}{G(65)} 
	= -\fdv{(G(87)\alert{T_\text{c}}(37,48))}{G(65)} 
	\\
	& = -\alert{T_\text{c}}(37,48)\fdv{G(87)}{G(65)} - G(87) \fdv{\alert{T_\text{c}}(37,48)}{G(65)}
	\\
	& = -\alert{T_\text{c}}(35,46)
\end{split}
\end{equation}
where again we neglect the functional derivative $\fdv*{\alert{T_\text{c}}}{G}$. 
(To be best of our knowledge, a second-order expression of the $T$-matrix kernel has not yet been derived.)

The elements of the static $T$-matrix kernel are given by \cite{Zhang_2017}
\begin{equation}
	\Xi_{pq,rs}^{\GT}
	= - \sum_n \frac{\sERI{pq,n}{\pp}\sERI{rs,n}{\pp}}{\Om_{n}^{\pp}} 
	+ \sum_n \frac{\sERI{pq,n}{\hh}\sERI{rs,n}{\hh}}{\Om_{n}^{\hh}}
\end{equation}
where the expressions for the screened integrals have already been established in Sec.~\ref{sec:Sig_GT}.

Going beyond the static approximation, one gets the dynamical $T$-matrix kernel \cite{Loos_2022}
\begin{equation}
\label{eq:T}
\begin{split}
	\Tilde{\Xi}_{ia,jb}^{\GT}(\omega)
	& = \sum_n \frac{\sERI{aj,n}{\pp}\sERI{bi,n}{\pp}}{\omega - (\Om_{n}^{\pp} - \e_{i}^{\GT} - \e_{j}^{\GT})} 
	\\
	& + \sum_n \frac{\sERI{aj,n}{\hh}\sERI{bi,n}{\hh}}{\omega - (\e_{a}^{\GT} + \e_{b}^{\GT} - \Om_{n}^{\hh})}
\end{split}
\end{equation}
It is interesting to note that, by removing the resummation effect of the $T$-matrix, \ie, by performing the following substitutions, $\Om_{n}^{\pp} \to  \e_{a}^{\HF} + \e_{b}^{\HF}$, $\Om_{n}^{\hh} \to \e_{i}^{\HF} + \e_{j}^{\HF}$, $\sERI{pq,m}{\pp} \to \ERI{pq}{cd}$, and $\sERI{pq,m}{\hh} \to \ERI{pq}{ij}$, one recovers both the direct and exchange parts of the pp and hh terms from Eq.~\eqref{eq:dynamical_BSE2_GF2}.

The upfolding process gives us 
\begin{equation}
\label{eq:upfolded_BSE_GT}
	\Tilde{\bH}^{\GT} = 
	\begin{pmatrix}
		\bA^{\GT}	&	 \bK^{\pp}		&	 \bI^{\hh}
		\\
		\qty(\bL^{\pp})^{\dag}	&	\bC^{\pp}	&	\bO
		\\
		\qty(\bJ^{\hh})^{\dag}	&	\bO			&	\bC^{\hh}	
	\end{pmatrix}
\end{equation}
with 
\begin{equation}
	A_{ia,jb}^{\GT} =  (\e_{a}^{\GT} - \e_{i}^{\GT}) \delta_{ij} \delta_{ab} + \ERI{ib}{aj}
\end{equation}
and the following expressions for the diagonal blocks $\bC^{\pp}$ and $\bC^{\hh}$
\begin{subequations}
\begin{align}
	C_{ijn,ijn}^{\pp} 
	& = \Om_{n}^{\pp} -\e_{i}^{\GT} - \e_{j}^{\GT} 
		\\
	C_{abn,abn}^{\hh} 
	& = \e_{a}^{\GT} + \e_{b}^{\GT} - \Om_{n}^{\hh}
\end{align}
\end{subequations}
while the coupling blocks read
\begin{subequations}
\begin{align}
	I_{ia,cdn}^{\hh} & = \delta_{ac} \sERI{di,n}{\hh}
	&
	J_{ia,cdn}^{\hh} & = \delta_{ad} \sERI{ci,n}{\hh}
	\\
	K_{ia,kln}^{\pp} & = \delta_{il} \sERI{ak,n}{\pp}
	&
	L_{ia,kln}^{\pp} & = \delta_{ik} \sERI{al,n}{\pp}
\end{align}
\end{subequations}
By downfolding Eq.~\eqref{eq:upfolded_BSE_GT}, we obtain 
\begin{equation}
\begin{split}
	\Tilde{\bXi}^{\GT}(\omega) 
	& = \bK^{\pp} \cdot \qty( \omega \bId - \bC^{\pp} )^{-1} \cdot \qty(\bL^{\pp})^{\dag} 
	\\
	& + \bI^{\hh} \cdot \qty( \omega \bId - \bC^{\hh} )^{-1} \cdot \qty(\bJ^{\hh})^{\dag} 
\end{split}
\end{equation}
which gives back the dynamical kernel \eqref{eq:T}.
Symmetrizing Eq.~\eqref{eq:upfolded_BSE_GT} has been revealed to be challenging, and we have not found any satisfying form.

\section{Computational Details}
\label{sec:comp_det}
All systems investigated in this study possess a closed-shell singlet ground state, and thus we employ the restricted formalism exclusively. 
As mentioned earlier, we initiate all calculations from HF orbitals and energies. 
We focus on two sets of atoms and molecules: one set pertains to charged excitations, where we solely consider the principal ionization potentials (IPs), while the other set concerns neutral excitations, where we compute singlet and triplet vertical excitation energies.
In all calculations, the positive infinitesimal $\eta$ is set to zero.

The first set comprises 20 atoms and molecules from the $GW100$ test set, \cite{vanSetten_2015} denoted as $GW20$, previously explored in Refs.~\onlinecite{Lewis_2019a,Loos_2020}. 
We adopt the geometries for the $GW20$ set from Ref.~\onlinecite{vanSetten_2015}. 
Calculations of IPs are performed using three different schemes: GF2, $GW$, and $GT$. 
All occupied and virtual orbitals are corrected.
For each scheme, we compute the linearized solution of the quasiparticle equation by solving Eq.~\eqref{eq:lin_qp_eq} and the dynamical solution by employing Newton's method starting from the linearized solution. 
The results presented in the {\SupMat} indicate that the linearization procedure has minimal impact on the $GW$ and $GT$ quasiparticles energies, while it improves the accuracy of GF2. 
Consequently, all quasiparticle energies are obtained via linearization of quasiparticle equation [see Eq.~\eqref{eq:lin_qp_eq}]. 
It is important to note that the $GW$ and $GT$ calculations are carried out without the TDA for the calculation of $W$ and $T$, respectively. 
As reference data, we rely on CCSD(T) IPs computed in the same basis.

The second set comprises 7 molecules as considered in Ref.~\onlinecite{Loos_2020h}. 
The corresponding geometries are extracted from the same work. 
Singlet and triplet transition energies are computed using the aug-cc-pVTZ basis via BSE utilizing the quasiparticle energies and kernels from the three different approximations under consideration (GF2, $GW$, and $GT$). 
For each scheme (BSE@GF2, BSE@$GW$, BSE2@$GW$, and BSE@$GT$), we also incorporate their respective dynamical corrections, named dBSE@GF2, dBSE@$GW$, dBSE2@$GW$, and dBSE@$GT$. 
To facilitate comparison, we also perform TDHF and CIS calculations. 
Our results are benchmarked against the theoretical best estimates (TBEs) from Ref.~\onlinecite{Loos_2020h}, from which we also extract transition energies computed using various second-order methods: CIS(D), \cite{Head-Gordon_1994,Head-Gordon_1995} ADC(2), \cite{Trofimov_1997,Dreuw_2015} CC2, \cite{Christiansen_1995a} and EOM-CCSD. \cite{Purvis_1982,Stanton_1993}

\alert{Here, we detail the strategy for the computation of neutral excitations within the BSE formalism. First, we solve the linearized quasiparticle equation [see Eq.~\eqref{eq:lin_qp_eq}] to obtain the quasiparticle solutions. Then, for the static calculations, we solve the eigenvalue problem of Eq.~\eqref{eq:BSEstat} using the various static kernels. Finally, for the dynamic corrections, we compute the corrected excitation energies using Eqs.~\eqref{eq:Om1-TDA} and \eqref{eq:Z} and the corresponding dynamical kernels.}

Various statistical quantities with respect to the reference values [CCSD(T) for IPs and TBEs for transition energies] are reported: mean absolute error (MAE), mean signed error (MSE), root-mean-square error (RMSE), and maximum error (Max). 
All static and dynamic BSE calculations, as well as CIS and TDHF calculations, are performed using the freely available software \textsc{quack}, which can be found on \textsc{github}. \cite{QuAcK}
 
\section{Results and discussion}

\subsection{Ionization potentials}
\label{sec:res_CEs}

\begin{table}
\caption{Principal IPs (in \si{\eV}) of the $GW20$ set computed with various approximations using the cc-pVTZ basis.}
\label{Tab:IPs}
	\begin{ruledtabular}
		\begin{tabular}{lddddd}
Mol. & \tabc{HF} &\tabc{GF2} & \tabc{$GW$} & \tabc{$GT$} & \tabc{$\Delta$CCSD(T)} \\
		\hline
\ce{He} & 24.97 & 24.54 & 24.58 & 24.77 & 24.53 \\
\ce{Ne} & 23.01 & 20.13 & 21.40 & 21.02 & 21.30 \\
\ce{H2} & 16.17 & 16.31 & 16.49 & 16.26 & 16.40 \\
\ce{Li2} & 4.95 & 5.19 & 5.35 & 5.04 & 5.23 \\
\ce{LiH} & 8.20 & 7.99 & 8.16 & 8.14 & 7.99 \\
\ce{HF} & 17.53 & 14.72 & 16.18 & 15.63 & 15.98 \\
\ce{Ar} & 16.06 & 15.39 & 15.70 & 15.49 & 15.53 \\
\ce{H2O} & 13.75 & 11.52 & 12.81 & 12.24 & 12.53 \\
\ce{LiF} & 12.92 & 9.81 & 11.38 & 10.95 & 11.39 \\
\ce{HCl} & 12.95 & 12.40 & 12.75 & 12.48 & 12.59 \\
\ce{BeO} & 10.50 & 8.38 & 9.78 & 9.21 & 9.98 \\
\ce{CO} & 15.35 & 14.17 & 15.03 & 14.44 & 14.21 \\
\ce{N2} & 17.23 & 15.09 & 17.09 & 15.70 & 15.49 \\
\ce{CH4} & 14.84 & 14.11 & 14.75 & 14.28 & 14.38 \\
\ce{BH3} & 13.56 & 13.25 & 13.65 & 13.30 & 13.28 \\
\ce{NH3} & 11.61 & 10.18 & 11.15 & 10.62 & 10.78 \\
\ce{BF} & 11.00 & 11.02 & 11.29 & 10.92 & 11.09 \\
\ce{BN} & 11.52 & 10.99 & 11.70 & 11.12 & 11.99 \\
\ce{SH2} & 10.46 & 10.15 & 10.46 & 10.15 & 10.32 \\
\ce{F2} & 18.09 & 14.26 & 16.31 & 15.38 & 15.68 \\
\hline
MAE & 0.81 & 0.56 & 0.28 & 0.26 & \\
MSE & 0.70 & -0.55 & 0.23 & -0.18 & \\
RMSE & 1.04 & 0.80 & 0.36 & 0.34 & \\
Max & 2.41 & 1.60 & 0.85 & 0.87 & \\
		\end{tabular}
	\end{ruledtabular}
\end{table}

The IPs of the $GW20$ set using the different approximations of the self-energy are reported in Table \ref{Tab:IPs}, where we also report the HF values.
It clearly shows the superiority of the $GW$ approximation for the calculation of IPs compared to the GF2 approximation. 
Indeed, we can see that the different statistical errors associated with $GW$ (MAE and MSE of \SI{0.28}{\eV} and \SI{0.23}{\eV}, respectively) are much smaller than the ones of GF2  (MAE and MSE of \SI{0.56}{\eV} and \SI{-0.55}{\eV}, respectively). 
For example, we have a maximum error of \SI{1.60}{eV} for GF2 whereas $GW$ has a maximum error of \SI{0.85}{eV}. 
We can note that the $GT$ approximation (MAE and MSE of \SI{0.26}{\eV} and \SI{-0.18}{\eV}, respectively) presents a similar MAE and maximum error as $GW$, while its MSE has also a similar magnitude but opposite sign. 
An analogous conclusion was reached by Zhang and coworkers \cite{Zhang_2017} for larger systems. 
\alert{Note also that similar trends on ionization potentials were found by Bruneval \textit{et al.} \cite{Bruneval_2021}}

\subsection{Vertical transition energies}
\label{sec:res_NEs}

\begin{squeezetable}
\begin{table*}
\caption{Singlet excitation energies (in \si{\eV}) of various molecules computed using the aug-cc-pVTZ basis set at different levels of theory. The dynamically-corrected BSE transition energies (dBSE) are reported in parentheses. CT stands for charge transfer.
The statistical descriptors associated with the errors with respect to the reference values are also reported for the entire dataset and separately for valence (Val.) and Rydberg (Ryd.) excited states.}
\label{Tab:singlets}
	\begin{ruledtabular}
		\begin{tabular}{lcddddddddddddddd}

					Mol.			& \mc{1}{c}{Nature} & \mc{1}{c}{CIS} & \mc{1}{c}{TDHF} 
					& \mc{2}{c}{BSE@{\GF2}}  & \mc{2}{c}{BSE@{\GW}}
					&\mc{2}{c}{BSE2@{\GW}} 
				& \mc{2}{c}{BSE@{\GT}} 
& \mc{1}{c}{CIS(D)}	& \mc{1}{c}{ADC(2)} & \mc{1}{c}{CC2} & \mc{1}{c}{CCSD} & \mc{1}{c}{TBE} \\ 
		
		\hline

\ce{HCl} & \text{CT} & 8.32 & 8.27 & 8.17 & (7.99) & 8.30 & (8.19) & 8.48 & (8.36) & \
7.56 & (7.52) & 6.07 & 7.97 & 7.96 & 7.91 & 7.84 \\
\ce{H2O} & Ryd. & 8.69 & 8.64 & 7.13 & (7.01) & 8.09 & (8.01) & 8.24 & (8.14) & \
7.12 & (7.08) & 7.62 & 7.18 & 7.23 & 7.60 & 7.17 \\
 & Ryd. & 10.36 & 10.31 & 8.71 & (8.66) & 9.80 & (9.72) & 9.91 & (9.84) & \
8.88 & (8.84) & 9.41 & 8.84 & 8.89 & 9.36 & 8.92 \\
 & Ryd. & 10.96 & 10.93 & 9.49 & (9.36) & 10.42 & (10.35) & 10.53 & (10.45) & \
9.55 & (9.51) & 9.99 & 9.52 & 9.58 & 9.96 & 9.52 \\
\ce{N2} & Val. & 9.95 & 9.70 & 9.83 & (9.28) & 10.42 & (9.99) & 11.28 & (10.74) & \
7.89 & (7.78) & 9.66 & 9.48 & 9.44 & 9.41 & 9.34 \\
 & Val. & 8.43 & 7.86 & 10.72 & (9.69) & 10.11 & (9.66) & 11.35 & (10.70) & \
8.18 & (8.02) & 10.31 & 10.26 & 10.32 & 10.00 & 9.88 \\
 & Val. & 8.98 & 8.68 & 11.28 & (10.34) & 10.75 & (10.33) & 11.45 & (10.86) & \
8.47 & (8.36) & 10.85 & 10.79 & 10.86 & 10.44 & 10.29 \\
 & Ryd. & 14.48 & 14.46 & 12.30 & (12.29) & 13.60 & (13.57) & 13.61 & (13.57) \
& 12.71 & (12.68) & 13.67 & 12.99 & 12.83 & 13.15 & 12.98 \\
 & Ryd. & 14.95 & 14.87 & 14.19 & (14.07) & 13.98 & (13.94) & 14.08 & (14.03) \
& 13.69 & (13.66) & 13.64 & 13.32 & 13.15 & 13.43 & 13.03 \\
 & Ryd. & 14.42 & 13.98 & 12.84 & (12.84) & 13.98 & (13.91) & 14.13 & (14.08) \
& 13.16 & (13.11) & 13.75 & 13.07 & 12.89 & 13.26 & 13.09 \\
 & Ryd. & 13.56 & 13.54 & 12.99 & (12.96) & 14.24 & (14.21) & 14.30 & (14.27) \
& 13.54 & (13.47) & 14.52 & 14.00 & 13.96 & 13.67 & 13.46 \\
\ce{CO} & Val. & 9.00 & 8.72 & 9.40 & (8.84) & 9.54 & (9.20) & 10.15 & (9.74) & \
7.63 & (7.53) & 8.78 & 8.69 & 8.64 & 8.59 & 8.49 \\
 & Val. & 9.61 & 9.25 & 10.11 & (9.43) & 10.25 & (9.91) & 11.27 & (10.79) & \
8.62 & (8.52) & 10.13 & 10.03 & 10.30 & 9.99 & 9.92 \\
 & Val. & 10.02 & 9.82 & 10.39 & (9.83) & 10.72 & (10.40) & 11.23 & (10.77) & \
8.80 & (8.72) & 10.41 & 10.30 & 10.60 & 10.12 & 10.06 \\
 & Ryd. & 12.12 & 12.08 & 11.04 & (11.00) & 11.88 & (11.85) & 11.86 & (11.83) \
& 11.16 & (11.13) & 11.48 & 11.32 & 11.11 & 11.22 & 10.95 \\
 & Ryd. & 12.72 & 12.71 & 11.72 & (11.65) & 12.39 & (12.37) & 12.45 & (12.42) \
& 11.81 & (11.80) & 11.71 & 11.83 & 11.63 & 11.75 & 11.52 \\
 & Ryd. & 12.82 & 12.81 & 11.69 & (11.62) & 12.37 & (12.32) & 12.46 & (12.41) \
& 11.68 & (11.67) & 12.06 & 12.03 & 11.83 & 11.96 & 11.72 \\
\ce{C2H2} & Val. & 6.27 & 5.90 & 7.95 & (7.33) & 7.37 & (7.05) & 8.09 & (7.67) & \
5.72 & (5.63) & 7.28 & 7.24 & 7.26 & 7.15 & 7.10 \\
 & Val. & 6.61 & 6.42 & 8.15 & (7.59) & 7.74 & (7.46) &  8.17 & (7.81) & \
5.94&  (5.87) & 7.62 & 7.56 & 7.59 & 7.48 & 7.44 \\
\ce{C2H4} & Ryd. & 7.15 & 7.13 & 7.41 & (7.31) & 7.64 & (7.62) & 7.69 & (7.66) & \
7.01 & (6.98) & 7.35 & 7.34 & 7.29 & 7.42 & 7.39 \\
 & Val. & 7.72 & 7.37 & 8.36 & (8.11) & 8.19 & (8.04) & 8.34 & (8.31) & \
7.02 & (6.97) & 7.95 & 7.91 & 7.92 & 8.02 & 7.93 \\
 & Ryd. & 7.74 & 7.73 & 8.04 & (7.97) & 8.29 & (8.26) & 8.35 & (8.35) & \
7.64 & (7.61) & 8.01 & 7.99 & 7.95 & 8.08 & 8.08 \\
\ce{CH2O} & Val. & 4.57 & 4.39 & 4.82 & (4.26) & 5.03 & (4.68) & 5.66 & (5.17) & \
2.78 & (2.68) & 4.04 & 3.92 & 4.07 & 4.01 & 3.98 \\
 & Ryd. & 8.59 & 8.59 & 6.36 & (6.40) & 7.87 & (7.85) & 7.87 & (7.88) & \
7.11 & (7.09) & 6.64 & 6.50 & 6.56 & 7.23 & 7.23 \\
 & Ryd. & 9.41 & 9.40 & 7.50 & (7.45) & 8.76 & (8.72) & 8.83 & (8.79) & \
7.87 & (7.85) & 7.56 & 7.53 & 7.57 & 8.12 & 8.13 \\
 & Ryd. & 9.53 & 9.58 & 7.39 & (7.41 )& 8.85 & (8.84) & 8.85 & (8.86) & \
8.12 & (8.11) & 8.16 & 7.47 & 7.52 & 8.21 & 8.23 \\
 & Ryd. & 10.02 & 10.02 & 7.40 & (7.37) & 8.87 & (8.85) & 8.92 & (8.89) & \
8.00 & (7.99) & 8.04 & 7.99 & 8.04 & 8.65 & 8.67 \\
 & Val. & 9.82 & 9.57 & 10.00 & (9.34) & 10.19 & (9.77) & 11.00 & (10.48) & \
7.54 & (7.44) & 9.38 & 9.17 & 9.32 & 9.28 & 9.22 \\
 & Val. & 9.72 & 9.21 & 9.95 & (9.82) & 10.06 & (9.82) & 10.39 & (10.14) & \
8.38 & (8.31) & 9.08 & 9.46 & 9.54 & 9.67 & 9.43 \\
\hline
MAE  &  & 0.92 & 0.94 & 0.52 & (0.35)  & 0.64 & (0.50) & 0.96 & (0.76) & 0.69  & (0.74)  & 0.43 & 0.24 & 0.25 & 0.15 &  \\
MSE  &  & 0.54 & 0.38 & 0.15 & (-0.13) & 0.64 & (0.48) & 0.96 & (0.76) & -0.60 & (-0.66) & 0.14 & 0.02 & 0.03 & 0.14 &  \\
RMSE &  & 1.06 & 1.09 & 0.63 & (0.47)  & 0.71 & (0.58) & 1.06 & (0.82) & 0.92  & (0.98)  & 0.55 & 0.33 & 0.33 & 0.20 &  \\
Max  &  & 1.92 & 2.02 & 1.27 & (1.30)  & 1.08 & (0.91) & 1.94 & (1.40) & 1.82  &  (1.93) & 1.77 & 0.76 & 0.71 & 0.44 &  \\
\hline
MAE  & Val. & 0.63 & 0.74 & 0.66 & (0.23) & 0.61 & (0.32) & 1.27 & (0.84) & 1.34  & (1.44)  & 0.26 & 0.17 & 0.23 & 0.09 \\
MSE  & Val. & -0.20 & -0.52 & 0.66 & (0.06) & 0.61 & (0.27) & 1.27 & (0.84) & -1.34 & (-1.44) & 0.20 & 0.14 & 0.23 & 0.09 \\
RMSE & Val. & 0.75 & 0.94 & 0.70 & (0.26) & 0.69 & (0.41) & 1.35 & (0.91) & 1.37  & (1.47)  & 0.30 & 0.22 & 0.30 & 0.11 \\
Max  & Val. & 1.45 & 2.02 & 0.99 & (0.49) & 1.08 & (0.71) & 1.94 & (1.40) & 1.82  & (1.93)  & 0.56 & 0.50 & 0.57 & 0.24 \\
\hline
MAE  & Ryd. & 1.16 & 1.12 & 0.43  & (0.45)  & 0.68 & (0.64) & 0.75 & (0.71) & 0.23  & (0.23)  & 0.47 & 0.30  & 0.27  & 0.19 \\
MSE  & Ryd. & 1.09 & 1.04 & -0.24 & (-0.30) & 0.68 & (0.64) & 0.75 & (0.71) & -0.07 & (-0.09) & 0.22 & -0.07 & -0.13 & 0.19 \\
RMSE & Ryd. & 1.26 & 1.22 & 0.59  & (0.59)  & 0.73 & (0.69) & 0.80 & (0.76) & 0.31  & (0.31)  & 0.54 & 0.41  & 0.36  & 0.25 \\
Max  & Ryd. & 1.92 & 1.84 & 1.27  & (1.30)  & 0.95 & (0.91) & 1.07 & (1.00) & 0.67  & (0.68)  & 1.06 & 0.76  & 0.71  & 0.44 \\
		\end{tabular}
	\end{ruledtabular}
\end{table*}
\end{squeezetable}

\begin{squeezetable}
\begin{table*}
\caption{Triplet excitation energies (in \si{\eV}) of various molecules computed using the aug-cc-pVTZ basis set at different levels of theory. The dynamically-corrected BSE transition energies (dBSE) are reported in parentheses.
The statistical descriptors associated with the errors with respect to the reference values are also reported for the entire dataset and separately for valence (Val.) and Rydberg (Ryd.) excited states.}
\label{Tab:triplets}
	\begin{ruledtabular}
\begin{tabular}{lcddddddddddddd}

		Mol.			& \mc{1}{c}{Nature} & \mc{1}{c}{CIS} & \mc{1}{c}{TDHF} 
& \mc{2}{c}{BSE@{\GF2}} & \mc{2}{c}{BSE@{\GW}}
				& \mc{2}{c}{BSE@{\GT}}
& \mc{1}{c}{CIS(D)}	& \mc{1}{c}{ADC(2)} & \mc{1}{c}{CC2} & \mc{1}{c}{CCSD} & \mc{1}{c}{TBE} \\ 			
		\hline

\ce{H2O} & Ryd. & 8.00 & 7.88 & 7.02 & (6.80) & 7.62 & (7.48) & \
6.60 & (6.54) & 7.25 & 6.86 & 6.91 & 7.20 & 6.92 \\
 & Ryd. & 10.01 & 9.88 & 8.68 & (8.60) & 9.61 & (9.50) & \
8.65 & (8.58) & 9.24 & 8.72 & 8.77 & 9.20 & 8.91 \\
 & Ryd. & 10.10 & 9.87 & 9.33 & (9.09) & 9.81 & (9.67) & \
8.82 & (8.75) & 9.54 & 9.15 & 9.20 & 9.49 & 9.30 \\
\ce{N2} & Val. & 6.16 & 3.36 & 8.88 & (7.41) & 8.03 & (7.38) & \
6.17 & (5.91) & 8.20 & 8.15 & 8.19 & 7.66 & 7.70 \\
 & Val. & 7.95 & 7.57 & 9.04 & (8.10) & 8.66 & (8.10) & \
6.30 & (6.12) & 8.33 & 8.20 & 8.19 & 8.09 & 8.01 \\
 & Val. & 7.23 & 5.72 & 9.94 & (8.67) & 9.04 & (8.48) & \
7.11 & (6.90) & 9.30 & 9.25 & 9.30 & 8.91 & 8.87 \\
 & Val. & 8.43 & 7.86 & 10.91 & (9.88) & 10.11 & (9.66) & \
7.99 & (7.85) & 10.29 & 10.23 & 10.29 & 9.83 & 9.66 \\
\ce{CO} & Val. & 5.81 & 5.22 & 7.59 & (6.45) & 6.80 & (6.25) & \
4.99 & (4.76) & 6.51 & 6.45 & 6.42 & 6.36 & 6.28 \\
 & Val. & 7.68 & 6.21 & 8.80 & (7.71) & 8.57 & (8.07) & \
7.02 & (6.81) & 8.63 & 8.54 & 8.72 & 8.34 & 8.45 \\
 & Val. & 8.61 & 7.71 & 9.58 & (8.68) & 9.39 & (8.96) & \
7.78 & (7.62) & 9.44 & 9.33 & 9.56 & 9.23 & 9.27 \\
 & Val. & 9.61 & 9.25 & 10.24 & (9.56) & 10.25 & (9.91) & \
8.49 & (8.39) & 10.10 & 10.01 & 10.27 & 9.81 & 9.80 \\
 & Ryd. & 11.13 & 11.03 & 10.86 & (10.71) & 11.17 & (11.07) \
& 10.48 & (10.41) & 10.98 & 10.83 & 10.60 & 10.71 & 10.47 \\
\ce{C2H2} & Val. & 4.51 & 2.16 & 7.09 & (6.13) & 5.83 & (5.32) & \
4.18 & (3.99) & 5.79 & 5.75 & 5.76 & 5.45 & 5.53 \\
 & Val. & 5.41 & 4.44 & 7.60 & (6.81) & 6.64 & (6.24) & \
4.97 & (4.83) & 6.62 & 6.57 & 6.60 & 6.41 & 6.40 \\
 & Val. & 6.27 & 5.90 & 8.05 & (7.43) & 7.37 & (7.05) & \
5.66 & (5.57) & 7.31 & 7.27 & 7.29 & 7.12 & 7.08 \\
\ce{C2H4} & Val. & 3.61 & 0.76 & 6.15 & (5.20) & 4.96 & (4.50) & \
3.15 & (2.07) & 4.62 & 4.59 & 4.59 & 4.46 & 4.54 \\
 & Ryd. & 6.92 & 6.88 & 7.40 & (7.25) & 7.46 & (7.42) & \
6.83 & (6.07) & 7.26 & 7.23 & 7.19 & 7.29 & 7.23 \\
 & Ryd. & 7.65 & 7.62 & 8.04 & (7.96) & 8.23 & (8.19) & \
7.58 & (7.17) & 7.97 & 7.95 & 7.91 & 8.03 & 7.98 \\
\ce{CH2O} & Val. & 3.75 & 3.40 & 4.52 & (3.83) & 4.28 & (3.88) & \
2.17 & (2.02) & 3.58 & 3.46 & 3.59 & 3.56 & 3.58 \\
 & Val. & 4.88 & 1.95 & 5.96 & (4.31) & 6.32 & (5.76) & \
4.26 & (4.03) & 6.27 & 6.20 & 6.30 & 5.97 & 6.06 \\
 & Ryd. & 8.25 & 8.17 & 6.32 & (6.28) & 7.60 & (7.56) & \
6.79 & (6.75) & 6.66 & 6.39 & 6.44 & 7.08 & 7.06 \\
\hline 
MAE  &  & 0.82  & 1.65  & 0.72 & (0.39)  & 0.41 & (0.27) & 1.10  & (1.33)  & 0.27 & 0.21 & 0.24 & 0.10 &  \\
MSE  &  & -0.34 & -1.25 & 0.61 & (-0.11) & 0.41 & (0.06) & -1.10 & (-1.33) & 0.23 & 0.10 & 0.14 & 0.05 &  \\
RMSE &  & 0.92  & 2.10  & 0.88 & (0.54)  & 0.46 & (0.33) & 1.25  & (1.48)  & 0.31 & 0.27 & 0.30 & 0.13 &  \\
Max  &  & 1.64  & 4.34  & 1.61 & (1.75)  & 0.70 & (0.60) & 1.80  & (2.47)  & 0.63 & 0.67 & 0.63 & 0.29 &  \\
\hline
MAE  & Val. & 0.83 & 2.12 & 0.95 & (0.47)  & 0.36 & (0.19)  & 1.50  & (1.74)  & 0.27 & 0.21 & 0.27 & 0.06  \\
MSE  & Val. & -0.81 & -2.12 & 0.94 & (-0.08) & 0.36 & (-0.12) & -1.50 & (-1.74) & 0.27 & 0.20 & 0.27 & -0.00 \\
RMSE & Val. & 0.96 & 2.52 & 1.06 & (0.62)  & 0.40 & (0.23)  & 1.51  & (1.76)  & 0.31 & 0.26 & 0.32 & 0.08  \\
Max  & Val. & 1.64 & 4.34 & 1.61 & (1.75)  & 0.70 & (0.39)  & 1.80  & (2.47)  & 0.63 & 0.57 & 0.63 & 0.17  \\
\hline
MAE  & Ryd. & 0.78 & 0.70 & 0.25  & (0.24)  & 0.52 & (0.43) & 0.31  & (0.51)  & 0.26 & 0.21  & 0.16  & 0.16 \\
MSE  & Ryd. & 0.60 & 0.49 & -0.03 & (-0.17) & 0.52 & (0.43) & -0.30 & (-0.51) & 0.15 & -0.11 & -0.12 & 0.16 \\
RMSE & Ryd. & 0.85 & 0.75 & 0.34  & (0.34)  & 0.55 & (0.46) & 0.34  & (0.62)  & 0.32 & 0.30  & 0.25  & 0.19 \\
Max  & Ryd. & 1.19 & 1.11 & 0.74  & (0.78)  & 0.70 & (0.60) & 0.48  & (1.16)  & 0.51 & 0.67  & 0.62  & 0.29 \\
		\end{tabular}
	\end{ruledtabular}
\end{table*}
\end{squeezetable}

The results of our calculations for vertical transition energies using the aug-cc-pVTZ basis set are summarized in Tables \ref{Tab:singlets} and \ref{Tab:triplets} for the singlet and triplet excited states, respectively.
They also report separate statistical errors for different classes of singlet and triplet excitations: valence (Val.) and Rydberg (Ryd.) excited states. 

As expected, both CIS and TDHF exhibit large statistical errors compared to the TBEs. 
It is well known that TDHF provides a poor description of triplet excitations, often leading to triplet instabilities. \cite{Cizek_1967, Dreuw_2005} 
One notices that TDHF is particularly bad at valence excitations.
On the other hand, CIS provides a more balanced description of singlets and triplets, thanks to error cancellation.

In the {\SupMat}, we report additional TDHF calculations using the GF2, $GW$, and $GT$ quasiparticles (without their corresponding kernel) instead of HF orbital energies. 
These calculations, referred to as TDHF@GF2, TDHF@$GW$, and TDHF@$GT$, allow us to observe the effects of different kernels and quasiparticles on the excitation energies. 
We find that the sole introduction of quasiparticle energies does not improve the description of singlet excitations. 
It should be noted that these calculations for triplet excitations resulted in instabilities and are not shown.

The inclusion of the corresponding BSE kernel significantly improves the description of both singlet and triplet excitations. 
This highlights the key role of the excitonic effect (\ie, the attractive interaction of the excited electron and the hole left behind), which is captured by the BSE kernel and is crucial for an accurate description of neutral excitations. 
Importantly, BSE@GF2 (MAE and MSE of \SI{0.52}{\eV} and \SI{0.15}{\eV}, respectively) provides better excitation energies for singlet states compared to BSE@$GW$ (MAE and MSE of \SI{0.64}{\eV}), as indicated by their respective statistical descriptors. 
This observation suggests that the versatility of the GF2 kernel, which contains ph, hp, pp, and hh terms, is a key factor behind its superior performance in describing singlet excitations (see Sec.~\ref{sec:Xi_GF2}). 
However, these trends might be different for larger chemical systems where screening effects become predominant. 
Furthermore, while BSE@GF2 exhibits a similar accuracy to the second-order method CIS(D) for singlet excitations, BSE@$GW$ outperforms BSE@GF2 for triplet excitations.
Another notable observation is that the static GF2 kernel provides a better description of Rydberg excitations compared to valence states. 
Conversely, the static $GW$ kernel performs better for valence than Rydberg excitations. 
These hold for both singlet and triplet transitions.
A last point worth highlighting is the contrasted performance of BSE@$GT$ for the two classes of excitations: while the accuracy of BSE@$GT$ is poor for the valence states (MAEs of \SI{1.34}{\eV} and \SI{1.50}{\eV} for singlets and triplets, respectively), it can be considered accurate for Rydberg transitions (MAEs of \SI{0.23}{\eV} and \SI{0.31}{\eV} for singlets and triplets, respectively), where the excited-state density is much lower than the ground-state one, a situation where ladder diagrams are known to be relevant (see Sec.~\ref{sec:Sig_GT}). 

By taking into account the dynamical corrections, we observe an overall improvement in the description of both singlet and triplet excitations, except at the BSE@$GT$ level.
From a general point of view, as previously mentioned and analyzed in Refs.~\onlinecite{Loos_2020h}, Rydberg excitations are less affected by dynamical effects than valence excitations across all BSE kernels. 
For singlet excitations, dBSE@GF2 outperforms CIS(D), especially for singlet valence excitations where its performance surpasses that of all second-order methods, except for EOM-CCSD, which is known to be highly accurate for small molecular systems. \cite{Loos_2018a,Veril_2021}
Although dBSE@$GW$ shows an improvement compared to its static version, it does not reach the accuracy of dBSE@GF2 or second-order methods. 
However, for triplet excitations, dBSE@$GW$ is on par with CIS(D), ADC(2), CC2, and EOM-CCSD, while dBSE@GF2 falls short of the accuracy of CIS(D).
In particular, for triplet valence excitations, dBSE@$GW$ outperforms all second-order methods, except EOM-CCSD. 
For these small molecular systems, both at the static and dynamic levels, the second-order scheme BSE2@$GW$ does not bring any improvement upon its first-order version.

\subsection{Singlet-triplet gap of cycl[3,3,3]zine}
\label{sec:2T}

\begin{figure*}
\centering
	\includegraphics[width=\linewidth]{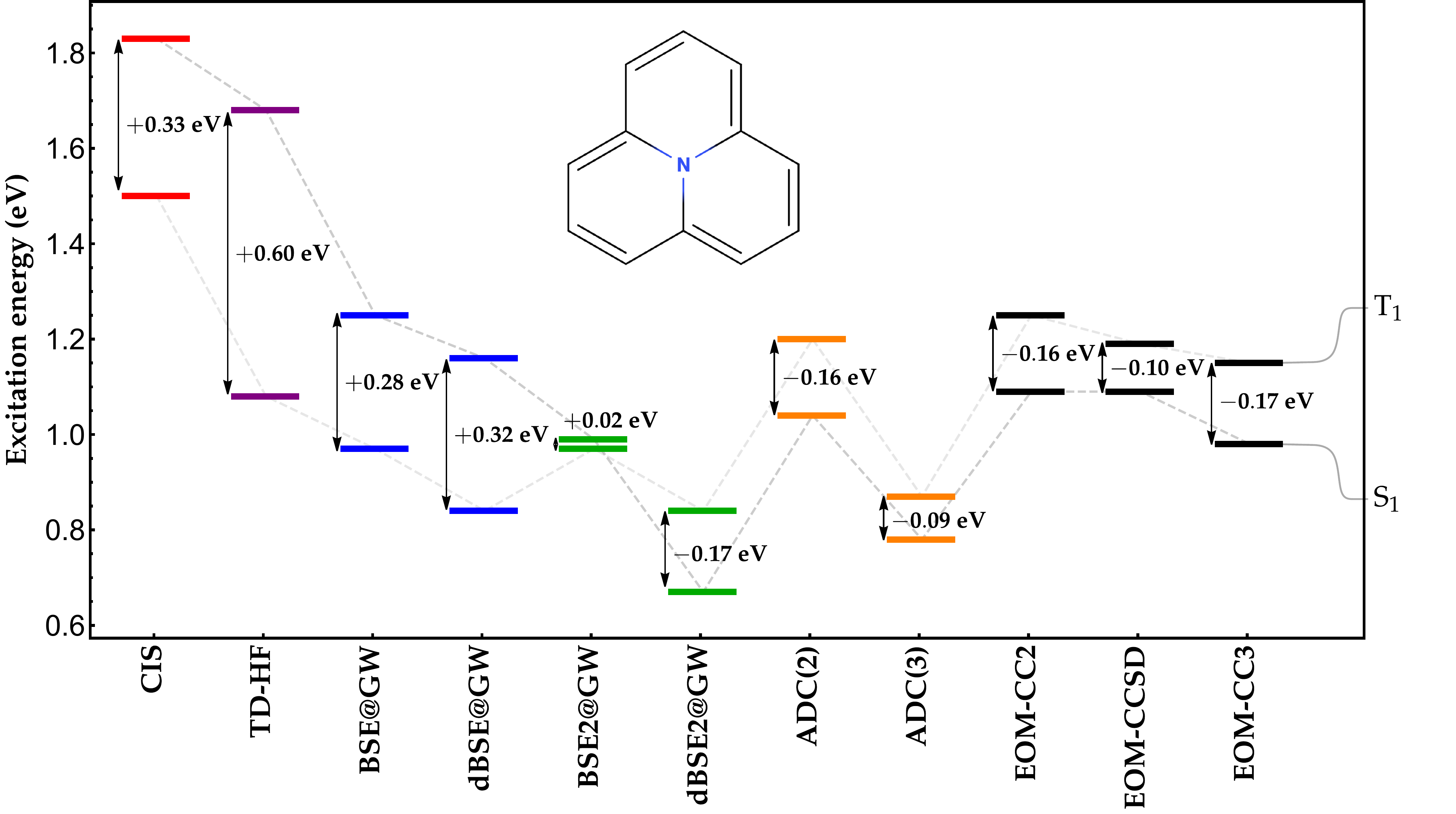}
	\caption{Evolution of the lowest singlet and triplet vertical excitation energies (in \si{\eV}) of cycl[3,3,3]zine evaluated with different computational methods using the cc-pVDZ basis.}
\label{fig:ST}
\end{figure*}

\begin{table}
\caption{Lowest singlet and triplet vertical excitation energies, $E_\text{S}$ and $E_\text{T}$, and resulting singlet-triplet gap $\Delta E_\text{ST}$ (in \si{\eV}) of cycl[3,3,3]zine computed at various levels of theory using the cc-pVDZ basis.
The percentage of single excitations involved in each transition, $\%T_1$, computed at the EOM-CC3 level is reported in parenthesis.}
\label{Tab:2T}
\begin{ruledtabular}
\begin{tabular}{lddd}
Method & \tabc{$E_\text{S}$} & \tabc{$E_\text{T}$} & \tabc{$\Delta E_\text{ST}$} \\
\hline
CIS & 1.83 & 1.50 & +0.33 \\
TDHF & 1.68 & 1.08 & +0.60  \\
BSE@$GW$ & 1.25 & 0.97 & +0.28  \\
dBSE@$GW$ & 1.16 & 0.84 & +0.32 \\
BSE2@$GW$ & 0.99 & 0.97 & +0.02  \\
dBSE2@$GW$ & 0.67 & 0.84 & -0.17  \\
CIS(D) & 1.07 & 1.37 & -0.30  \\
ADC(2) & 1.04 & 1.20 &  -0.16 \\
ADC(3) & 0.78 & 0.87 & -0.09  \\
EOM-CC2 & 1.09 & 1.25 & -0.16 \\
EOM-CCSD & 1.09 & 1.19 & -0.10  \\
EOM-CC3 & 0.98 (87\%) & 1.15 (96\%) & -0.17  \\
\end{tabular}
\end{ruledtabular}
\end{table}
 
Molecules with an inverted singlet-triplet gap (\ie, where the lowest singlet excited state is \alert{lower} in energy than the lowest triplet state) are of particular interest in TADF \cite{Endo_2009,Uoyama_2012} because they can harness both singlet and triplet excitons for emission, thereby enhancing the efficiency of OLEDs. \cite{Baldo_1998,Adachi_2001}
Thanks to this inverted gap, the system can undergo efficient reverse intersystem crossing, a process in which the population from the triplet state can be thermally activated and transferred back to the singlet state, resulting in delayed fluorescence.

Recently, such systems have been scrutinized at different computational levels, including TD-DFT and second-order wave function methods, such as CIS(D), ADC(2), and EOM-CCSD. \cite{deSilva_2019,Sanz-Rodrigo_2021,Ricci_2021,Olivier_2017,Olivier_2018,SanchoGarca_2022,Curtis_2023}
In particular, de Silva has shown that this inversion requires a substantial contribution from the double excitations. \cite{deSilva_2019}
This explains why adiabatic TD-DFT is not able to reproduce this particular feature, and second- or higher-order methods are required where double excitations are explicitly treated.

Following the computational protocol of Ref.~\onlinecite{deSilva_2019}, we compute the lowest singlet and triplet excitation energies, $E_\text{S}$ and $E_\text{T}$, as well as the corresponding singlet-triplet gap, $\Delta E_\text{ST}$, of cycl[3,3,3]zine (see Fig.~\ref{fig:ST}), a model molecular emitter for TADF, with the cc-pVDZ basis at various levels of theory.
The geometry of cycl[3,3,3]zine has been optimized at the B3LYP/cc-pVDZ level and is reported in {\SupMat} for the sake of completeness.
Additionally, we have been able to compute the singlet-triplet gap with third-order methods such as ADC(3) \cite{Trofimov_2002,Harbach_2014,Dreuw_2015} and EOM-CC3. \cite{Christiansen_1995b,Koch_1997}

Our results are gathered in Table \ref{Tab:2T} and shown in Fig.~\ref{fig:ST}.
As expected, the BSE@$GW$ and BSE2@$GW$ calculations do not produce an inverted singlet-triplet gap due to the static nature of the kernel.
Because the dynamical correction of the singlet and triplet excitation energies cancel each other pretty much exactly, dBSE@$GW$ yields the same state ordering. 
However, the second-order dynamical $GW$ kernel (which only corrects singlet states as explained in Sec.~\ref{sec:Theta_GW}) faithfully predicts this inversion although the corresponding excitation energies are underestimated compared to other approaches, except ADC(3), which is known to exhibit this trend. \cite{Loos_2020b}
Interestingly, ADC(2), EOM-CC2, EOM-CC3, and BSE2@$GW$ yield essentially the same value, while EOM-CCSD and ADC(3) slightly underestimate the gap. 
Because the percentage of single excitations involved in these two valence transitions ($\%T_1$, see Table \ref{Tab:2T}) is high (although not negligible for the singlet state), the EOM-CC3 value is likely to be accurate. \cite{Veril_2021}
Note that, because of the poor quality of the GF2 quasiparticle energies, we could not compute excitation energies at the (d)BSE@GF2 level as spurious poles appeared in the BSE kernel.

\section{Concluding remarks}
\label{sec:ccl}
In this study, our focus was on examining the relationships between different Green's function methods, specifically exploring various approximations for the self-energy (GF2, $GW$, and $GT$) and their corresponding BSE kernels at the static and dynamic levels.
Additionally, we extended the upfolding process, previously confined to the GF2 and $GW$ frameworks, to the $T$-matrix approximation.
The introduction of this upfolding framework allowed us to uncover connections between GF2 and the ADC(2) scheme concerning both charged and neutral excitations, and to propose new directions for the development of accurate kernels at the $GW$ level. 

Subsequently, we applied these three distinct approximations to calculate the principal IPs and vertical transition energies for both singlet and triplet states of small molecules. 
Our findings can be summarized as follows:
\begin{itemize}
	\item Confirming previous knowledge, the $GW$ approximation surpasses the GF2 method in accurately calculating IPs, emphasizing the significance of screening even in small molecular systems.
	\item The $T$-matrix approximation exhibits comparable accuracy to $GW$, although it falls slightly short.
	\item For the singlet excited states of small molecules, the GF2 kernel generally outperforms its $GW$ counterpart.
	Conversely, for triplet excitations, BSE@$GW$ provides more accurate vertical excitation energies.
	\item Importantly, our investigations highlight the sensitivity of BSE kernels to the nature of the excited states.
	For example, BSE@$GT$ is poor for valence states while it is accurate for Rydberg transitions. 
	\item Overall, except in the $T$-matrix approximation, dynamical corrections are almost systematically beneficial.
\end{itemize}
It is important to note that these conclusions are drawn specifically for small molecules, and it would be intriguing to explore if similar trends persist in larger systems. \alert{Moreover, it is worth mentioning that because we only rely on one-shot schemes, the quality of our results strongly depends on the starting point (HF orbitals). It is clear that more accurate results can be obtained using partially self-consistent schemes or by tuning the starting point using an adequate exchange-correlation functional.}

To initiate our pursuit of this objective, we examined the capability of our various schemes to replicate the inversion of the singlet-triplet gap in cycl[3,3,3]zine, a prototypical molecular emitter for TADF. With the exception of one case, we observed that all static and dynamic BSE-based schemes failed to reproduce this unique characteristic. The only exception was the dynamically-corrected BSE2@$GW$ scheme, which yielded a gap value consistent with that obtained from EOM-CC3 calculations.
This observation effectively highlights the significance of higher-order terms and dynamic effects within the BSE formalism, and we anticipate that these findings will stimulate further advancements in this area of research.

\section*{Supplementary Material}
\label{sec:supmat}
See the supplementary material for linearized \textit{vs} dynamical quasiparticle energies of the $GW20$ set in the cc-pVDZ basis, TDHF@GF2, TDHF@$GW$, and TDHF@$GT$ singlet excitation energies in the aug-cc-pVTZ basis, and optimize ground-state geometry of the cycl[3,3,3]zine molecule at the B3LYP/cc-pVDZ level.

\acknowledgements{
The authors thank Pina Romaniello, Fabien Bruneval, and Xavier Blase for insightful discussions.
They also thank Antoine Marie for useful comments on this manuscript.
This work used the HPC resources from CALMIP (Toulouse) under allocation 2023-18005.
This project has received funding from the European Research Council (ERC) under the European Union's Horizon 2020 research and innovation programme (Grant agreement No.~863481).}

\section*{Data availability statement}
The data that supports the findings of this study are available within the article and its supplementary material.


\bibliography{ufMBPT}

\begin{thebibliography}{279}%
\makeatletter
\providecommand \@ifxundefined [1]{%
 \@ifx{#1\undefined}
}%
\providecommand \@ifnum [1]{%
 \ifnum #1\expandafter \@firstoftwo
 \else \expandafter \@secondoftwo
 \fi
}%
\providecommand \@ifx [1]{%
 \ifx #1\expandafter \@firstoftwo
 \else \expandafter \@secondoftwo
 \fi
}%
\providecommand \natexlab [1]{#1}%
\providecommand \enquote  [1]{``#1''}%
\providecommand \bibnamefont  [1]{#1}%
\providecommand \bibfnamefont [1]{#1}%
\providecommand \citenamefont [1]{#1}%
\providecommand \href@noop [0]{\@secondoftwo}%
\providecommand \href [0]{\begingroup \@sanitize@url \@href}%
\providecommand \@href[1]{\@@startlink{#1}\@@href}%
\providecommand \@@href[1]{\endgroup#1\@@endlink}%
\providecommand \@sanitize@url [0]{\catcode `\\12\catcode `\$12\catcode
  `\&12\catcode `\#12\catcode `\^12\catcode `\_12\catcode `\%12\relax}%
\providecommand \@@startlink[1]{}%
\providecommand \@@endlink[0]{}%
\providecommand \url  [0]{\begingroup\@sanitize@url \@url }%
\providecommand \@url [1]{\endgroup\@href {#1}{\urlprefix }}%
\providecommand \urlprefix  [0]{URL }%
\providecommand \Eprint [0]{\href }%
\providecommand \doibase [0]{http://dx.doi.org/}%
\providecommand \selectlanguage [0]{\@gobble}%
\providecommand \bibinfo  [0]{\@secondoftwo}%
\providecommand \bibfield  [0]{\@secondoftwo}%
\providecommand \translation [1]{[#1]}%
\providecommand \BibitemOpen [0]{}%
\providecommand \bibitemStop [0]{}%
\providecommand \bibitemNoStop [0]{.\EOS\space}%
\providecommand \EOS [0]{\spacefactor3000\relax}%
\providecommand \BibitemShut  [1]{\csname bibitem#1\endcsname}%
\let\auto@bib@innerbib\@empty
\bibitem [{\citenamefont {Csanak}, \citenamefont {Taylor},\ and\ \citenamefont
  {Yaris}(1971)}]{CsanakBook}%
  \BibitemOpen
  \bibfield  {author} {\bibinfo {author} {\bibfnamefont {G.}~\bibnamefont
  {Csanak}}, \bibinfo {author} {\bibfnamefont {H.}~\bibnamefont {Taylor}}, \
  and\ \bibinfo {author} {\bibfnamefont {R.}~\bibnamefont {Yaris}},\ }in\
  \href@noop {} {\emph {\bibinfo {booktitle} {Advances in atomic and molecular
  physics}}},\ Vol.~\bibinfo {volume} {7}\ (\bibinfo  {publisher} {Elsevier},\
  \bibinfo {year} {1971})\ pp.\ \bibinfo {pages} {287--361}\BibitemShut
  {NoStop}%
\bibitem [{\citenamefont {Fetter}\ and\ \citenamefont
  {Waleck}(1971)}]{FetterBook}%
  \BibitemOpen
  \bibfield  {author} {\bibinfo {author} {\bibfnamefont {A.~L.}\ \bibnamefont
  {Fetter}}\ and\ \bibinfo {author} {\bibfnamefont {J.~D.}\ \bibnamefont
  {Waleck}},\ }\href@noop {} {\emph {\bibinfo {title} {Quantum Theory of Many
  Particle Systems}}}\ (\bibinfo  {publisher} {McGraw Hill, San Francisco},\
  \bibinfo {year} {1971})\BibitemShut {NoStop}%
\bibitem [{\citenamefont {Martin}, \citenamefont {Reining},\ and\ \citenamefont
  {Ceperley}(2016)}]{Martin_2016}%
  \BibitemOpen
  \bibfield  {author} {\bibinfo {author} {\bibfnamefont {R.~M.}\ \bibnamefont
  {Martin}}, \bibinfo {author} {\bibfnamefont {L.}~\bibnamefont {Reining}}, \
  and\ \bibinfo {author} {\bibfnamefont {D.~M.}\ \bibnamefont {Ceperley}},\
  }\href@noop {} {\emph {\bibinfo {title} {Interacting Electrons: Theory and
  Computational Approaches}}}\ (\bibinfo  {publisher} {Cambridge University
  Press},\ \bibinfo {year} {2016})\BibitemShut {NoStop}%
\bibitem [{\citenamefont {Bruneval}\ \emph {et~al.}(2016)\citenamefont
  {Bruneval}, \citenamefont {Rangel}, \citenamefont {Hamed}, \citenamefont
  {Shao}, \citenamefont {Yang},\ and\ \citenamefont {Neaton}}]{Bruneval_2016}%
  \BibitemOpen
  \bibfield  {author} {\bibinfo {author} {\bibfnamefont {F.}~\bibnamefont
  {Bruneval}}, \bibinfo {author} {\bibfnamefont {T.}~\bibnamefont {Rangel}},
  \bibinfo {author} {\bibfnamefont {S.~M.}\ \bibnamefont {Hamed}}, \bibinfo
  {author} {\bibfnamefont {M.}~\bibnamefont {Shao}}, \bibinfo {author}
  {\bibfnamefont {C.}~\bibnamefont {Yang}}, \ and\ \bibinfo {author}
  {\bibfnamefont {J.~B.}\ \bibnamefont {Neaton}},\ }\href {\doibase
  10.1016/j.cpc.2016.06.019} {\bibfield  {journal} {\bibinfo  {journal}
  {Comput. Phys. Commun.}\ }\textbf {\bibinfo {volume} {208}},\ \bibinfo
  {pages} {149} (\bibinfo {year} {2016})}\BibitemShut {NoStop}%
\bibitem [{\citenamefont {Golze}, \citenamefont {Dvorak},\ and\ \citenamefont
  {Rinke}(2019)}]{Golze_2019}%
  \BibitemOpen
  \bibfield  {author} {\bibinfo {author} {\bibfnamefont {D.}~\bibnamefont
  {Golze}}, \bibinfo {author} {\bibfnamefont {M.}~\bibnamefont {Dvorak}}, \
  and\ \bibinfo {author} {\bibfnamefont {P.}~\bibnamefont {Rinke}},\ }\href
  {\doibase 10.3389/fchem.2019.00377} {\bibfield  {journal} {\bibinfo
  {journal} {Front. Chem.}\ }\textbf {\bibinfo {volume} {7}},\ \bibinfo {pages}
  {377} (\bibinfo {year} {2019})}\BibitemShut {NoStop}%
\bibitem [{\citenamefont {Blase}, \citenamefont {Duchemin},\ and\ \citenamefont
  {Jacquemin}(2018)}]{Blase_2018}%
  \BibitemOpen
  \bibfield  {author} {\bibinfo {author} {\bibfnamefont {X.}~\bibnamefont
  {Blase}}, \bibinfo {author} {\bibfnamefont {I.}~\bibnamefont {Duchemin}}, \
  and\ \bibinfo {author} {\bibfnamefont {D.}~\bibnamefont {Jacquemin}},\ }\href
  {\doibase 10.1039/C7CS00049A} {\bibfield  {journal} {\bibinfo  {journal}
  {Chem. Soc. Rev.}\ }\textbf {\bibinfo {volume} {47}},\ \bibinfo {pages}
  {1022} (\bibinfo {year} {2018})}\BibitemShut {NoStop}%
\bibitem [{\citenamefont {Blase}\ \emph {et~al.}(2020)\citenamefont {Blase},
  \citenamefont {Duchemin}, \citenamefont {Jacquemin},\ and\ \citenamefont
  {Loos}}]{Blase_2020}%
  \BibitemOpen
  \bibfield  {author} {\bibinfo {author} {\bibfnamefont {X.}~\bibnamefont
  {Blase}}, \bibinfo {author} {\bibfnamefont {I.}~\bibnamefont {Duchemin}},
  \bibinfo {author} {\bibfnamefont {D.}~\bibnamefont {Jacquemin}}, \ and\
  \bibinfo {author} {\bibfnamefont {P.-F.}\ \bibnamefont {Loos}},\ }\href
  {\doibase 10.1021/acs.jpclett.0c01875} {\bibfield  {journal} {\bibinfo
  {journal} {J. Phys. Chem. Lett.}\ }\textbf {\bibinfo {volume} {11}},\
  \bibinfo {pages} {7371} (\bibinfo {year} {2020})}\BibitemShut {NoStop}%
\bibitem [{\citenamefont {Schirmer}(2018)}]{Schirmer_2018}%
  \BibitemOpen
  \bibfield  {author} {\bibinfo {author} {\bibfnamefont {J.}~\bibnamefont
  {Schirmer}},\ }\href@noop {} {\emph {\bibinfo {title} {Many-Body Methods for
  Atoms, Molecules and Clusters}}}\ (\bibinfo  {publisher} {Springer},\
  \bibinfo {year} {2018})\BibitemShut {NoStop}%
\bibitem [{\citenamefont {Das}(1973)}]{Das_1973}%
  \BibitemOpen
  \bibfield  {author} {\bibinfo {author} {\bibfnamefont {G.}~\bibnamefont
  {Das}},\ }\href {\doibase 10.1063/1.1679100} {\bibfield  {journal} {\bibinfo
  {journal} {J. Chem. Phys.}\ }\textbf {\bibinfo {volume} {58}},\ \bibinfo
  {pages} {5104} (\bibinfo {year} {1973})}\BibitemShut {NoStop}%
\bibitem [{\citenamefont {Dalgaard}\ and\ \citenamefont
  {J{\o}rgensen}(1978)}]{Dalgaard_1978}%
  \BibitemOpen
  \bibfield  {author} {\bibinfo {author} {\bibfnamefont {E.}~\bibnamefont
  {Dalgaard}}\ and\ \bibinfo {author} {\bibfnamefont {P.}~\bibnamefont
  {J{\o}rgensen}},\ }\href {\doibase 10.1063/1.437049} {\bibfield  {journal}
  {\bibinfo  {journal} {J. Chem. Phys.}\ }\textbf {\bibinfo {volume} {69}},\
  \bibinfo {pages} {3833} (\bibinfo {year} {1978})}\BibitemShut {NoStop}%
\bibitem [{\citenamefont {Lengsfield}(1980)}]{Lengsfield_1980}%
  \BibitemOpen
  \bibfield  {author} {\bibinfo {author} {\bibfnamefont {B.~H.}\ \bibnamefont
  {Lengsfield}},\ }\href {\doibase 10.1063/1.439885} {\bibfield  {journal}
  {\bibinfo  {journal} {J. Chem. Phys.}\ }\textbf {\bibinfo {volume} {73}},\
  \bibinfo {pages} {382} (\bibinfo {year} {1980})}\BibitemShut {NoStop}%
\bibitem [{\citenamefont {Bauschlicher}\ and\ \citenamefont
  {Yarkony}(1980)}]{Bauschlicher_1980a}%
  \BibitemOpen
  \bibfield  {author} {\bibinfo {author} {\bibfnamefont {C.~W.}\ \bibnamefont
  {Bauschlicher}}\ and\ \bibinfo {author} {\bibfnamefont {D.~R.}\ \bibnamefont
  {Yarkony}},\ }\href {\doibase 10.1063/1.439255} {\bibfield  {journal}
  {\bibinfo  {journal} {J. Chem. Phys.}\ }\textbf {\bibinfo {volume} {72}},\
  \bibinfo {pages} {1138} (\bibinfo {year} {1980})}\BibitemShut {NoStop}%
\bibitem [{\citenamefont {Bauschlicher}, \citenamefont {Silver},\ and\
  \citenamefont {Yarkony}(1980)}]{Bauschlicher_1980b}%
  \BibitemOpen
  \bibfield  {author} {\bibinfo {author} {\bibfnamefont {C.~W.}\ \bibnamefont
  {Bauschlicher}}, \bibinfo {author} {\bibfnamefont {D.~M.}\ \bibnamefont
  {Silver}}, \ and\ \bibinfo {author} {\bibfnamefont {D.~R.}\ \bibnamefont
  {Yarkony}},\ }\href {\doibase 10.1063/1.440456} {\bibfield  {journal}
  {\bibinfo  {journal} {J. Chem. Phys.}\ }\textbf {\bibinfo {volume} {73}},\
  \bibinfo {pages} {2867} (\bibinfo {year} {1980})}\BibitemShut {NoStop}%
\bibitem [{\citenamefont {Werner}\ and\ \citenamefont
  {Meyer}(1981)}]{Werner_1981}%
  \BibitemOpen
  \bibfield  {author} {\bibinfo {author} {\bibfnamefont {H.}~\bibnamefont
  {Werner}}\ and\ \bibinfo {author} {\bibfnamefont {W.}~\bibnamefont {Meyer}},\
  }\href {\doibase 10.1063/1.440892} {\bibfield  {journal} {\bibinfo  {journal}
  {J. Chem. Phys.}\ }\textbf {\bibinfo {volume} {74}},\ \bibinfo {pages} {5794}
  (\bibinfo {year} {1981})}\BibitemShut {NoStop}%
\bibitem [{\citenamefont {Golab}, \citenamefont {Yeager},\ and\ \citenamefont
  {J{\o}rgensen}(1983)}]{Golab_1983}%
  \BibitemOpen
  \bibfield  {author} {\bibinfo {author} {\bibfnamefont {J.~T.}\ \bibnamefont
  {Golab}}, \bibinfo {author} {\bibfnamefont {D.~L.}\ \bibnamefont {Yeager}}, \
  and\ \bibinfo {author} {\bibfnamefont {P.}~\bibnamefont {J{\o}rgensen}},\
  }\href {\doibase https://doi.org/10.1016/0301-0104(83)85106-4} {\bibfield
  {journal} {\bibinfo  {journal} {Chem. Phys.}\ }\textbf {\bibinfo {volume}
  {78}},\ \bibinfo {pages} {175} (\bibinfo {year} {1983})}\BibitemShut
  {NoStop}%
\bibitem [{\citenamefont {Ziegler}, \citenamefont {Rauk},\ and\ \citenamefont
  {Baerends}(1977)}]{Ziegler_1977}%
  \BibitemOpen
  \bibfield  {author} {\bibinfo {author} {\bibfnamefont {T.}~\bibnamefont
  {Ziegler}}, \bibinfo {author} {\bibfnamefont {A.}~\bibnamefont {Rauk}}, \
  and\ \bibinfo {author} {\bibfnamefont {E.}~\bibnamefont {Baerends}},\ }\href
  {\doibase https://doi.org/10.1007/BF00551551} {\bibfield  {journal} {\bibinfo
   {journal} {Theor. Chim. Acta}\ }\textbf {\bibinfo {volume} {43}},\ \bibinfo
  {pages} {261} (\bibinfo {year} {1977})}\BibitemShut {NoStop}%
\bibitem [{\citenamefont {Kowalczyk}, \citenamefont {Yost},\ and\ \citenamefont
  {Voorhis}(2011)}]{Kowalczyk_2011}%
  \BibitemOpen
  \bibfield  {author} {\bibinfo {author} {\bibfnamefont {T.}~\bibnamefont
  {Kowalczyk}}, \bibinfo {author} {\bibfnamefont {S.}~\bibnamefont {Yost}}, \
  and\ \bibinfo {author} {\bibfnamefont {T.}~\bibnamefont {Voorhis}},\ }\href
  {\doibase 10.1063/1.3530801} {\bibfield  {journal} {\bibinfo  {journal}
  {Chem. Phys.}\ }\textbf {\bibinfo {volume} {134}},\ \bibinfo {pages} {054128}
  (\bibinfo {year} {2011})}\BibitemShut {NoStop}%
\bibitem [{\citenamefont {Aryasetiawan}\ and\ \citenamefont
  {Gunnarsson}(1998)}]{Aryasetiawan_1998}%
  \BibitemOpen
  \bibfield  {author} {\bibinfo {author} {\bibfnamefont {F.}~\bibnamefont
  {Aryasetiawan}}\ and\ \bibinfo {author} {\bibfnamefont {O.}~\bibnamefont
  {Gunnarsson}},\ }\href {\doibase 10.1088/0034-4885/61/3/002} {\bibfield
  {journal} {\bibinfo  {journal} {Rep. Prog. Phys.}\ }\textbf {\bibinfo
  {volume} {61}},\ \bibinfo {pages} {237} (\bibinfo {year} {1998})}\BibitemShut
  {NoStop}%
\bibitem [{\citenamefont {Onida}, \citenamefont {Reining},\ and\ \citenamefont
  {Rubio}(2002)}]{Onida_2002}%
  \BibitemOpen
  \bibfield  {author} {\bibinfo {author} {\bibfnamefont {G.}~\bibnamefont
  {Onida}}, \bibinfo {author} {\bibfnamefont {L.}~\bibnamefont {Reining}}, \
  and\ \bibinfo {author} {\bibfnamefont {A.}~\bibnamefont {Rubio}},\ }\href
  {\doibase 10.1103/RevModPhys.74.601} {\bibfield  {journal} {\bibinfo
  {journal} {Rev. Mod. Phys.}\ }\textbf {\bibinfo {volume} {74}},\ \bibinfo
  {pages} {601} (\bibinfo {year} {2002})}\BibitemShut {NoStop}%
\bibitem [{\citenamefont {Reining}(2017)}]{Reining_2017}%
  \BibitemOpen
  \bibfield  {author} {\bibinfo {author} {\bibfnamefont {L.}~\bibnamefont
  {Reining}},\ }\href {\doibase 10.1002/wcms.1344} {\bibfield  {journal}
  {\bibinfo  {journal} {Wiley Interdiscip. Rev. Comput. Mol. Sci.}\ }\textbf
  {\bibinfo {volume} {8}},\ \bibinfo {pages} {e1344} (\bibinfo {year}
  {2017})}\BibitemShut {NoStop}%
\bibitem [{\citenamefont {Bruneval}, \citenamefont {Dattani},\ and\
  \citenamefont {van Setten}(2021)}]{Bruneval_2021}%
  \BibitemOpen
  \bibfield  {author} {\bibinfo {author} {\bibfnamefont {F.}~\bibnamefont
  {Bruneval}}, \bibinfo {author} {\bibfnamefont {N.}~\bibnamefont {Dattani}}, \
  and\ \bibinfo {author} {\bibfnamefont {M.~J.}\ \bibnamefont {van Setten}},\
  }\href {\doibase 10.3389/fchem.2021.749779} {\bibfield  {journal} {\bibinfo
  {journal} {Front. Chem.}\ }\textbf {\bibinfo {volume} {9}},\ \bibinfo {pages}
  {749779} (\bibinfo {year} {2021})}\BibitemShut {NoStop}%
\bibitem [{\citenamefont {Strinati}, \citenamefont {Mattausch},\ and\
  \citenamefont {Hanke}(1980)}]{Strinati_1980}%
  \BibitemOpen
  \bibfield  {author} {\bibinfo {author} {\bibfnamefont {G.}~\bibnamefont
  {Strinati}}, \bibinfo {author} {\bibfnamefont {H.~J.}\ \bibnamefont
  {Mattausch}}, \ and\ \bibinfo {author} {\bibfnamefont {W.}~\bibnamefont
  {Hanke}},\ }\href {\doibase 10.1103/PhysRevLett.45.290} {\bibfield  {journal}
  {\bibinfo  {journal} {Phys. Rev. Lett.}\ }\textbf {\bibinfo {volume} {45}},\
  \bibinfo {pages} {290} (\bibinfo {year} {1980})}\BibitemShut {NoStop}%
\bibitem [{\citenamefont {Strinati}, \citenamefont {Mattausch},\ and\
  \citenamefont {Hanke}(1982)}]{Strinati_1982a}%
  \BibitemOpen
  \bibfield  {author} {\bibinfo {author} {\bibfnamefont {G.}~\bibnamefont
  {Strinati}}, \bibinfo {author} {\bibfnamefont {H.~J.}\ \bibnamefont
  {Mattausch}}, \ and\ \bibinfo {author} {\bibfnamefont {W.}~\bibnamefont
  {Hanke}},\ }\href {\doibase 10.1103/PhysRevB.25.2867} {\bibfield  {journal}
  {\bibinfo  {journal} {Phys. Rev. B}\ }\textbf {\bibinfo {volume} {25}},\
  \bibinfo {pages} {2867} (\bibinfo {year} {1982})}\BibitemShut {NoStop}%
\bibitem [{\citenamefont {Strinati}(1982)}]{Strinati_1982b}%
  \BibitemOpen
  \bibfield  {author} {\bibinfo {author} {\bibfnamefont {G.}~\bibnamefont
  {Strinati}},\ }\href {\doibase 10.1103/PhysRevLett.49.1519} {\bibfield
  {journal} {\bibinfo  {journal} {Phys. Rev. Lett.}\ }\textbf {\bibinfo
  {volume} {49}},\ \bibinfo {pages} {1519} (\bibinfo {year}
  {1982})}\BibitemShut {NoStop}%
\bibitem [{\citenamefont {Hybertsen}\ and\ \citenamefont
  {Louie}(1985{\natexlab{a}})}]{Hybertsen_1985}%
  \BibitemOpen
  \bibfield  {author} {\bibinfo {author} {\bibfnamefont {M.~S.}\ \bibnamefont
  {Hybertsen}}\ and\ \bibinfo {author} {\bibfnamefont {S.~G.}\ \bibnamefont
  {Louie}},\ }\href@noop {} {\bibfield  {journal} {\bibinfo  {journal} {Phys.
  Rev. Lett.}\ }\textbf {\bibinfo {volume} {55}},\ \bibinfo {pages} {1418}
  (\bibinfo {year} {1985}{\natexlab{a}})}\BibitemShut {NoStop}%
\bibitem [{\citenamefont {Hybertsen}\ and\ \citenamefont
  {Louie}(1986)}]{Hybertsen_1986}%
  \BibitemOpen
  \bibfield  {author} {\bibinfo {author} {\bibfnamefont {M.~S.}\ \bibnamefont
  {Hybertsen}}\ and\ \bibinfo {author} {\bibfnamefont {S.~G.}\ \bibnamefont
  {Louie}},\ }\href {\doibase 10.1103/PhysRevB.34.5390} {\bibfield  {journal}
  {\bibinfo  {journal} {Phys. Rev. B}\ }\textbf {\bibinfo {volume} {34}},\
  \bibinfo {pages} {5390} (\bibinfo {year} {1986})}\BibitemShut {NoStop}%
\bibitem [{\citenamefont {Godby}, \citenamefont {Schl{\"u}ter},\ and\
  \citenamefont {Sham}(1986)}]{Godby_1986}%
  \BibitemOpen
  \bibfield  {author} {\bibinfo {author} {\bibfnamefont {R.~W.}\ \bibnamefont
  {Godby}}, \bibinfo {author} {\bibfnamefont {M.}~\bibnamefont {Schl{\"u}ter}},
  \ and\ \bibinfo {author} {\bibfnamefont {L.~J.}\ \bibnamefont {Sham}},\
  }\href {\doibase 10.1103/PhysRevLett.56.2415} {\bibfield  {journal} {\bibinfo
   {journal} {Phys. Rev. Lett.}\ }\textbf {\bibinfo {volume} {56}},\ \bibinfo
  {pages} {2415} (\bibinfo {year} {1986})}\BibitemShut {NoStop}%
\bibitem [{\citenamefont {Godby}, \citenamefont {Schl{\"u}ter},\ and\
  \citenamefont {Sham}(1987{\natexlab{a}})}]{Godby_1987}%
  \BibitemOpen
  \bibfield  {author} {\bibinfo {author} {\bibfnamefont {R.~W.}\ \bibnamefont
  {Godby}}, \bibinfo {author} {\bibfnamefont {M.}~\bibnamefont {Schl{\"u}ter}},
  \ and\ \bibinfo {author} {\bibfnamefont {L.~J.}\ \bibnamefont {Sham}},\
  }\href {\doibase 10.1103/PhysRevB.36.6497} {\bibfield  {journal} {\bibinfo
  {journal} {Phys. Rev. B}\ }\textbf {\bibinfo {volume} {36}},\ \bibinfo
  {pages} {6497} (\bibinfo {year} {1987}{\natexlab{a}})}\BibitemShut {NoStop}%
\bibitem [{\citenamefont {Godby}, \citenamefont {Schl{\"u}ter},\ and\
  \citenamefont {Sham}(1987{\natexlab{b}})}]{Godby_1987a}%
  \BibitemOpen
  \bibfield  {author} {\bibinfo {author} {\bibfnamefont {R.~W.}\ \bibnamefont
  {Godby}}, \bibinfo {author} {\bibfnamefont {M.}~\bibnamefont {Schl{\"u}ter}},
  \ and\ \bibinfo {author} {\bibfnamefont {L.~J.}\ \bibnamefont {Sham}},\
  }\href {\doibase 10.1103/PhysRevB.35.4170} {\bibfield  {journal} {\bibinfo
  {journal} {Phys. Rev. B}\ }\textbf {\bibinfo {volume} {35}},\ \bibinfo
  {pages} {4170} (\bibinfo {year} {1987}{\natexlab{b}})}\BibitemShut {NoStop}%
\bibitem [{\citenamefont {Godby}, \citenamefont {Schl\"uter},\ and\
  \citenamefont {Sham}(1988)}]{Godby_1988}%
  \BibitemOpen
  \bibfield  {author} {\bibinfo {author} {\bibfnamefont {R.~W.}\ \bibnamefont
  {Godby}}, \bibinfo {author} {\bibfnamefont {M.}~\bibnamefont {Schl\"uter}}, \
  and\ \bibinfo {author} {\bibfnamefont {L.~J.}\ \bibnamefont {Sham}},\ }\href
  {\doibase 10.1103/PhysRevB.37.10159} {\bibfield  {journal} {\bibinfo
  {journal} {Phys. Rev. B}\ }\textbf {\bibinfo {volume} {37}},\ \bibinfo
  {pages} {10159} (\bibinfo {year} {1988})}\BibitemShut {NoStop}%
\bibitem [{\citenamefont {Blase}\ \emph {et~al.}(1995)\citenamefont {Blase},
  \citenamefont {Rubio}, \citenamefont {Louie},\ and\ \citenamefont
  {Cohen}}]{Blase_1995}%
  \BibitemOpen
  \bibfield  {author} {\bibinfo {author} {\bibfnamefont {X.}~\bibnamefont
  {Blase}}, \bibinfo {author} {\bibfnamefont {A.}~\bibnamefont {Rubio}},
  \bibinfo {author} {\bibfnamefont {S.~G.}\ \bibnamefont {Louie}}, \ and\
  \bibinfo {author} {\bibfnamefont {M.~L.}\ \bibnamefont {Cohen}},\ }\href
  {\doibase 10.1103/PhysRevB.51.6868} {\bibfield  {journal} {\bibinfo
  {journal} {Phys. Rev. B}\ }\textbf {\bibinfo {volume} {51}},\ \bibinfo
  {pages} {6868} (\bibinfo {year} {1995})}\BibitemShut {NoStop}%
\bibitem [{\citenamefont {Rohlfing}\ and\ \citenamefont
  {Louie}(1999)}]{Rohlfing_1999a}%
  \BibitemOpen
  \bibfield  {author} {\bibinfo {author} {\bibfnamefont {M.}~\bibnamefont
  {Rohlfing}}\ and\ \bibinfo {author} {\bibfnamefont {S.~G.}\ \bibnamefont
  {Louie}},\ }\href {\doibase 10.1103/PhysRevLett.82.1959} {\bibfield
  {journal} {\bibinfo  {journal} {Phys. Rev. Lett.}\ }\textbf {\bibinfo
  {volume} {82}},\ \bibinfo {pages} {1959} (\bibinfo {year}
  {1999})}\BibitemShut {NoStop}%
\bibitem [{\citenamefont {van~der Horst}\ \emph {et~al.}(1999)\citenamefont
  {van~der Horst}, \citenamefont {Bobbert}, \citenamefont {Michels},
  \citenamefont {Brocks},\ and\ \citenamefont {Kelly}}]{Horst_1999}%
  \BibitemOpen
  \bibfield  {author} {\bibinfo {author} {\bibfnamefont {J.-W.}\ \bibnamefont
  {van~der Horst}}, \bibinfo {author} {\bibfnamefont {P.~A.}\ \bibnamefont
  {Bobbert}}, \bibinfo {author} {\bibfnamefont {M.~A.~J.}\ \bibnamefont
  {Michels}}, \bibinfo {author} {\bibfnamefont {G.}~\bibnamefont {Brocks}}, \
  and\ \bibinfo {author} {\bibfnamefont {P.~J.}\ \bibnamefont {Kelly}},\ }\href
  {\doibase 10.1103/PhysRevLett.83.4413} {\bibfield  {journal} {\bibinfo
  {journal} {Phys. Rev. Lett.}\ }\textbf {\bibinfo {volume} {83}},\ \bibinfo
  {pages} {4413} (\bibinfo {year} {1999})}\BibitemShut {NoStop}%
\bibitem [{\citenamefont {Puschnig}\ and\ \citenamefont
  {Ambrosch-Draxl}(2002)}]{Puschnig_2002}%
  \BibitemOpen
  \bibfield  {author} {\bibinfo {author} {\bibfnamefont {P.}~\bibnamefont
  {Puschnig}}\ and\ \bibinfo {author} {\bibfnamefont {C.}~\bibnamefont
  {Ambrosch-Draxl}},\ }\href {\doibase 10.1103/PhysRevLett.89.056405}
  {\bibfield  {journal} {\bibinfo  {journal} {Phys. Rev. Lett.}\ }\textbf
  {\bibinfo {volume} {89}},\ \bibinfo {pages} {056405} (\bibinfo {year}
  {2002})}\BibitemShut {NoStop}%
\bibitem [{\citenamefont {Tiago}, \citenamefont {Northrup},\ and\ \citenamefont
  {Louie}(2003)}]{Tiago_2003}%
  \BibitemOpen
  \bibfield  {author} {\bibinfo {author} {\bibfnamefont {M.~L.}\ \bibnamefont
  {Tiago}}, \bibinfo {author} {\bibfnamefont {J.~E.}\ \bibnamefont {Northrup}},
  \ and\ \bibinfo {author} {\bibfnamefont {S.~G.}\ \bibnamefont {Louie}},\
  }\href {\doibase 10.1103/PhysRevB.67.115212} {\bibfield  {journal} {\bibinfo
  {journal} {Phys. Rev. B}\ }\textbf {\bibinfo {volume} {67}},\ \bibinfo
  {pages} {115212} (\bibinfo {year} {2003})}\BibitemShut {NoStop}%
\bibitem [{\citenamefont {Rocca}, \citenamefont {Lu},\ and\ \citenamefont
  {Galli}(2010)}]{Rocca_2010}%
  \BibitemOpen
  \bibfield  {author} {\bibinfo {author} {\bibfnamefont {D.}~\bibnamefont
  {Rocca}}, \bibinfo {author} {\bibfnamefont {D.}~\bibnamefont {Lu}}, \ and\
  \bibinfo {author} {\bibfnamefont {G.}~\bibnamefont {Galli}},\ }\href
  {\doibase 10.1063/1.3494540} {\bibfield  {journal} {\bibinfo  {journal} {J.
  Chem. Phys.}\ }\textbf {\bibinfo {volume} {133}},\ \bibinfo {pages} {164109}
  (\bibinfo {year} {2010})}\BibitemShut {NoStop}%
\bibitem [{\citenamefont {Boulanger}\ \emph {et~al.}(2014)\citenamefont
  {Boulanger}, \citenamefont {Jacquemin}, \citenamefont {Duchemin},\ and\
  \citenamefont {Blase}}]{Boulanger_2014}%
  \BibitemOpen
  \bibfield  {author} {\bibinfo {author} {\bibfnamefont {P.}~\bibnamefont
  {Boulanger}}, \bibinfo {author} {\bibfnamefont {D.}~\bibnamefont
  {Jacquemin}}, \bibinfo {author} {\bibfnamefont {I.}~\bibnamefont {Duchemin}},
  \ and\ \bibinfo {author} {\bibfnamefont {X.}~\bibnamefont {Blase}},\ }\href
  {\doibase 10.1021/ct401101u} {\bibfield  {journal} {\bibinfo  {journal} {J.
  Chem. Theory Comput.}\ }\textbf {\bibinfo {volume} {10}},\ \bibinfo {pages}
  {1212} (\bibinfo {year} {2014})}\BibitemShut {NoStop}%
\bibitem [{\citenamefont {Jacquemin}, \citenamefont {Duchemin},\ and\
  \citenamefont {Blase}(2015{\natexlab{a}})}]{Jacquemin_2015a}%
  \BibitemOpen
  \bibfield  {author} {\bibinfo {author} {\bibfnamefont {D.}~\bibnamefont
  {Jacquemin}}, \bibinfo {author} {\bibfnamefont {I.}~\bibnamefont {Duchemin}},
  \ and\ \bibinfo {author} {\bibfnamefont {X.}~\bibnamefont {Blase}},\ }\href
  {\doibase 10.1021/acs.jctc.5b00304} {\bibfield  {journal} {\bibinfo
  {journal} {J. Chem. Theory Comput.}\ }\textbf {\bibinfo {volume} {11}},\
  \bibinfo {pages} {3290} (\bibinfo {year} {2015}{\natexlab{a}})}\BibitemShut
  {NoStop}%
\bibitem [{\citenamefont {Bruneval}, \citenamefont {Hamed},\ and\ \citenamefont
  {Neaton}(2015)}]{Bruneval_2015}%
  \BibitemOpen
  \bibfield  {author} {\bibinfo {author} {\bibfnamefont {F.}~\bibnamefont
  {Bruneval}}, \bibinfo {author} {\bibfnamefont {S.~M.}\ \bibnamefont {Hamed}},
  \ and\ \bibinfo {author} {\bibfnamefont {J.~B.}\ \bibnamefont {Neaton}},\
  }\href {\doibase 10.1063/1.4922489} {\bibfield  {journal} {\bibinfo
  {journal} {J. Chem. Phys.}\ }\textbf {\bibinfo {volume} {142}},\ \bibinfo
  {pages} {244101} (\bibinfo {year} {2015})}\BibitemShut {NoStop}%
\bibitem [{\citenamefont {Jacquemin}, \citenamefont {Duchemin},\ and\
  \citenamefont {Blase}(2015{\natexlab{b}})}]{Jacquemin_2015b}%
  \BibitemOpen
  \bibfield  {author} {\bibinfo {author} {\bibfnamefont {D.}~\bibnamefont
  {Jacquemin}}, \bibinfo {author} {\bibfnamefont {I.}~\bibnamefont {Duchemin}},
  \ and\ \bibinfo {author} {\bibfnamefont {X.}~\bibnamefont {Blase}},\ }\href
  {\doibase 10.1021/acs.jctc.5b00619} {\bibfield  {journal} {\bibinfo
  {journal} {J. Chem. Theory Comput.}\ }\textbf {\bibinfo {volume} {11}},\
  \bibinfo {pages} {5340} (\bibinfo {year} {2015}{\natexlab{b}})}\BibitemShut
  {NoStop}%
\bibitem [{\citenamefont {Hirose}, \citenamefont {Noguchi},\ and\ \citenamefont
  {Sugino}(2015)}]{Hirose_2015}%
  \BibitemOpen
  \bibfield  {author} {\bibinfo {author} {\bibfnamefont {D.}~\bibnamefont
  {Hirose}}, \bibinfo {author} {\bibfnamefont {Y.}~\bibnamefont {Noguchi}}, \
  and\ \bibinfo {author} {\bibfnamefont {O.}~\bibnamefont {Sugino}},\ }\href
  {\doibase 10.1103/PhysRevB.91.205111} {\bibfield  {journal} {\bibinfo
  {journal} {Phys. Rev. B}\ }\textbf {\bibinfo {volume} {91}},\ \bibinfo
  {pages} {205111} (\bibinfo {year} {2015})}\BibitemShut {NoStop}%
\bibitem [{\citenamefont {Jacquemin}, \citenamefont {Duchemin},\ and\
  \citenamefont {Blase}(2017)}]{Jacquemin_2017a}%
  \BibitemOpen
  \bibfield  {author} {\bibinfo {author} {\bibfnamefont {D.}~\bibnamefont
  {Jacquemin}}, \bibinfo {author} {\bibfnamefont {I.}~\bibnamefont {Duchemin}},
  \ and\ \bibinfo {author} {\bibfnamefont {X.}~\bibnamefont {Blase}},\ }\href
  {\doibase 10.1021/acs.jpclett.7b00381} {\bibfield  {journal} {\bibinfo
  {journal} {J. Phys. Chem. Lett.}\ }\textbf {\bibinfo {volume} {8}},\ \bibinfo
  {pages} {1524} (\bibinfo {year} {2017})}\BibitemShut {NoStop}%
\bibitem [{\citenamefont {Jacquemin}\ \emph {et~al.}(2017)\citenamefont
  {Jacquemin}, \citenamefont {Duchemin}, \citenamefont {Blondel},\ and\
  \citenamefont {Blase}}]{Jacquemin_2017b}%
  \BibitemOpen
  \bibfield  {author} {\bibinfo {author} {\bibfnamefont {D.}~\bibnamefont
  {Jacquemin}}, \bibinfo {author} {\bibfnamefont {I.}~\bibnamefont {Duchemin}},
  \bibinfo {author} {\bibfnamefont {A.}~\bibnamefont {Blondel}}, \ and\
  \bibinfo {author} {\bibfnamefont {X.}~\bibnamefont {Blase}},\ }\href
  {\doibase 10.1021/acs.jctc.6b01169} {\bibfield  {journal} {\bibinfo
  {journal} {J. Chem. Theory Comput.}\ }\textbf {\bibinfo {volume} {13}},\
  \bibinfo {pages} {767} (\bibinfo {year} {2017})}\BibitemShut {NoStop}%
\bibitem [{\citenamefont {Rangel}\ \emph {et~al.}(2017)\citenamefont {Rangel},
  \citenamefont {Hamed}, \citenamefont {Bruneval},\ and\ \citenamefont
  {Neaton}}]{Rangel_2017}%
  \BibitemOpen
  \bibfield  {author} {\bibinfo {author} {\bibfnamefont {T.}~\bibnamefont
  {Rangel}}, \bibinfo {author} {\bibfnamefont {S.~M.}\ \bibnamefont {Hamed}},
  \bibinfo {author} {\bibfnamefont {F.}~\bibnamefont {Bruneval}}, \ and\
  \bibinfo {author} {\bibfnamefont {J.~B.}\ \bibnamefont {Neaton}},\ }\href
  {\doibase 10.1063/1.4983126} {\bibfield  {journal} {\bibinfo  {journal} {J.
  Chem. Phys.}\ }\textbf {\bibinfo {volume} {146}},\ \bibinfo {pages} {194108}
  (\bibinfo {year} {2017})}\BibitemShut {NoStop}%
\bibitem [{\citenamefont {Krause}\ and\ \citenamefont
  {Klopper}(2017)}]{Krause_2017}%
  \BibitemOpen
  \bibfield  {author} {\bibinfo {author} {\bibfnamefont {K.}~\bibnamefont
  {Krause}}\ and\ \bibinfo {author} {\bibfnamefont {W.}~\bibnamefont
  {Klopper}},\ }\href {\doibase 10.1002/jcc.24688} {\bibfield  {journal}
  {\bibinfo  {journal} {J. Comput. Chem.}\ }\textbf {\bibinfo {volume} {38}},\
  \bibinfo {pages} {383} (\bibinfo {year} {2017})}\BibitemShut {NoStop}%
\bibitem [{\citenamefont {Gui}, \citenamefont {Holzer},\ and\ \citenamefont
  {Klopper}(2018)}]{Gui_2018}%
  \BibitemOpen
  \bibfield  {author} {\bibinfo {author} {\bibfnamefont {X.}~\bibnamefont
  {Gui}}, \bibinfo {author} {\bibfnamefont {C.}~\bibnamefont {Holzer}}, \ and\
  \bibinfo {author} {\bibfnamefont {W.}~\bibnamefont {Klopper}},\ }\href
  {\doibase 10.1021/acs.jctc.8b00014} {\bibfield  {journal} {\bibinfo
  {journal} {J. Chem. Theory Comput.}\ }\textbf {\bibinfo {volume} {14}},\
  \bibinfo {pages} {2127} (\bibinfo {year} {2018})}\BibitemShut {NoStop}%
\bibitem [{\citenamefont {Liu}\ \emph {et~al.}(2020)\citenamefont {Liu},
  \citenamefont {Kloppenburg}, \citenamefont {Yao}, \citenamefont {Ren},
  \citenamefont {Appel}, \citenamefont {Kanai},\ and\ \citenamefont
  {Blum}}]{Liu_2020}%
  \BibitemOpen
  \bibfield  {author} {\bibinfo {author} {\bibfnamefont {C.}~\bibnamefont
  {Liu}}, \bibinfo {author} {\bibfnamefont {J.}~\bibnamefont {Kloppenburg}},
  \bibinfo {author} {\bibfnamefont {Y.}~\bibnamefont {Yao}}, \bibinfo {author}
  {\bibfnamefont {X.}~\bibnamefont {Ren}}, \bibinfo {author} {\bibfnamefont
  {H.}~\bibnamefont {Appel}}, \bibinfo {author} {\bibfnamefont
  {Y.}~\bibnamefont {Kanai}}, \ and\ \bibinfo {author} {\bibfnamefont
  {V.}~\bibnamefont {Blum}},\ }\href {\doibase 10.1063/1.5123290} {\bibfield
  {journal} {\bibinfo  {journal} {J. Chem. Phys.}\ }\textbf {\bibinfo {volume}
  {152}},\ \bibinfo {pages} {044105} (\bibinfo {year} {2020})}\BibitemShut
  {NoStop}%
\bibitem [{\citenamefont {Li}\ \emph {et~al.}(2017)\citenamefont {Li},
  \citenamefont {Holzmann}, \citenamefont {Duchemin}, \citenamefont {Blase},\
  and\ \citenamefont {Olevano}}]{Li_2017}%
  \BibitemOpen
  \bibfield  {author} {\bibinfo {author} {\bibfnamefont {J.}~\bibnamefont
  {Li}}, \bibinfo {author} {\bibfnamefont {M.}~\bibnamefont {Holzmann}},
  \bibinfo {author} {\bibfnamefont {I.}~\bibnamefont {Duchemin}}, \bibinfo
  {author} {\bibfnamefont {X.}~\bibnamefont {Blase}}, \ and\ \bibinfo {author}
  {\bibfnamefont {V.}~\bibnamefont {Olevano}},\ }\href {\doibase
  10.1103/PhysRevLett.118.163001} {\bibfield  {journal} {\bibinfo  {journal}
  {Phys. Rev. Lett.}\ }\textbf {\bibinfo {volume} {118}},\ \bibinfo {pages}
  {163001} (\bibinfo {year} {2017})}\BibitemShut {NoStop}%
\bibitem [{\citenamefont {Li}\ \emph {et~al.}(2019)\citenamefont {Li},
  \citenamefont {Drummond}, \citenamefont {Schuck},\ and\ \citenamefont
  {Olevano}}]{Li_2019}%
  \BibitemOpen
  \bibfield  {author} {\bibinfo {author} {\bibfnamefont {J.}~\bibnamefont
  {Li}}, \bibinfo {author} {\bibfnamefont {N.~D.}\ \bibnamefont {Drummond}},
  \bibinfo {author} {\bibfnamefont {P.}~\bibnamefont {Schuck}}, \ and\ \bibinfo
  {author} {\bibfnamefont {V.}~\bibnamefont {Olevano}},\ }\href {\doibase
  10.21468/SciPostPhys.6.4.040} {\bibfield  {journal} {\bibinfo  {journal}
  {SciPost Phys.}\ }\textbf {\bibinfo {volume} {6}},\ \bibinfo {pages} {040}
  (\bibinfo {year} {2019})}\BibitemShut {NoStop}%
\bibitem [{\citenamefont {Li}\ \emph {et~al.}(2020)\citenamefont {Li},
  \citenamefont {Duchemin}, \citenamefont {Blase},\ and\ \citenamefont
  {Olevano}}]{Li_2020}%
  \BibitemOpen
  \bibfield  {author} {\bibinfo {author} {\bibfnamefont {J.}~\bibnamefont
  {Li}}, \bibinfo {author} {\bibfnamefont {I.}~\bibnamefont {Duchemin}},
  \bibinfo {author} {\bibfnamefont {X.}~\bibnamefont {Blase}}, \ and\ \bibinfo
  {author} {\bibfnamefont {V.}~\bibnamefont {Olevano}},\ }\href {\doibase
  10.21468/SciPostPhys.8.2.020} {\bibfield  {journal} {\bibinfo  {journal}
  {SciPost Phys.}\ }\textbf {\bibinfo {volume} {8}},\ \bibinfo {pages} {20}
  (\bibinfo {year} {2020})}\BibitemShut {NoStop}%
\bibitem [{\citenamefont {Li}\ and\ \citenamefont {Olevano}(2021)}]{Li_2021}%
  \BibitemOpen
  \bibfield  {author} {\bibinfo {author} {\bibfnamefont {J.}~\bibnamefont
  {Li}}\ and\ \bibinfo {author} {\bibfnamefont {V.}~\bibnamefont {Olevano}},\
  }\href {\doibase 10.1103/PhysRevA.103.012809} {\bibfield  {journal} {\bibinfo
   {journal} {Phys. Rev. A}\ }\textbf {\bibinfo {volume} {103}},\ \bibinfo
  {pages} {012809} (\bibinfo {year} {2021})}\BibitemShut {NoStop}%
\bibitem [{\citenamefont {Holzer}\ and\ \citenamefont
  {Klopper}(2018)}]{Holzer_2018a}%
  \BibitemOpen
  \bibfield  {author} {\bibinfo {author} {\bibfnamefont {C.}~\bibnamefont
  {Holzer}}\ and\ \bibinfo {author} {\bibfnamefont {W.}~\bibnamefont
  {Klopper}},\ }\href {\doibase 10.1063/1.5051028} {\bibfield  {journal}
  {\bibinfo  {journal} {J. Chem. Phys.}\ }\textbf {\bibinfo {volume} {149}},\
  \bibinfo {pages} {101101} (\bibinfo {year} {2018})}\BibitemShut {NoStop}%
\bibitem [{\citenamefont {Holzer}\ \emph {et~al.}(2018)\citenamefont {Holzer},
  \citenamefont {Gui}, \citenamefont {Harding}, \citenamefont {Kresse},
  \citenamefont {Helgaker},\ and\ \citenamefont {Klopper}}]{Holzer_2018b}%
  \BibitemOpen
  \bibfield  {author} {\bibinfo {author} {\bibfnamefont {C.}~\bibnamefont
  {Holzer}}, \bibinfo {author} {\bibfnamefont {X.}~\bibnamefont {Gui}},
  \bibinfo {author} {\bibfnamefont {M.~E.}\ \bibnamefont {Harding}}, \bibinfo
  {author} {\bibfnamefont {G.}~\bibnamefont {Kresse}}, \bibinfo {author}
  {\bibfnamefont {T.}~\bibnamefont {Helgaker}}, \ and\ \bibinfo {author}
  {\bibfnamefont {W.}~\bibnamefont {Klopper}},\ }\href {\doibase
  10.1063/1.5047030} {\bibfield  {journal} {\bibinfo  {journal} {J. Chem.
  Phys.}\ }\textbf {\bibinfo {volume} {149}},\ \bibinfo {pages} {144106}
  (\bibinfo {year} {2018})}\BibitemShut {NoStop}%
\bibitem [{\citenamefont {Loos}\ \emph
  {et~al.}(2020{\natexlab{a}})\citenamefont {Loos}, \citenamefont {Scemama},
  \citenamefont {Duchemin}, \citenamefont {Jacquemin},\ and\ \citenamefont
  {Blase}}]{Loos_2020e}%
  \BibitemOpen
  \bibfield  {author} {\bibinfo {author} {\bibfnamefont {P.-F.}\ \bibnamefont
  {Loos}}, \bibinfo {author} {\bibfnamefont {A.}~\bibnamefont {Scemama}},
  \bibinfo {author} {\bibfnamefont {I.}~\bibnamefont {Duchemin}}, \bibinfo
  {author} {\bibfnamefont {D.}~\bibnamefont {Jacquemin}}, \ and\ \bibinfo
  {author} {\bibfnamefont {X.}~\bibnamefont {Blase}},\ }\href {\doibase
  10.1021/acs.jpclett.0c00460} {\bibfield  {journal} {\bibinfo  {journal} {J.
  Phys. Chem. Lett.}\ }\textbf {\bibinfo {volume} {11}},\ \bibinfo {pages}
  {3536} (\bibinfo {year} {2020}{\natexlab{a}})}\BibitemShut {NoStop}%
\bibitem [{\citenamefont {Loos}\ \emph {et~al.}(2021)\citenamefont {Loos},
  \citenamefont {Comin}, \citenamefont {Blase},\ and\ \citenamefont
  {Jacquemin}}]{Loos_2021}%
  \BibitemOpen
  \bibfield  {author} {\bibinfo {author} {\bibfnamefont {P.-F.}\ \bibnamefont
  {Loos}}, \bibinfo {author} {\bibfnamefont {M.}~\bibnamefont {Comin}},
  \bibinfo {author} {\bibfnamefont {X.}~\bibnamefont {Blase}}, \ and\ \bibinfo
  {author} {\bibfnamefont {D.}~\bibnamefont {Jacquemin}},\ }\href {\doibase
  10.1021/acs.jctc.1c00226} {\bibfield  {journal} {\bibinfo  {journal} {J.
  Chem. Theory Comput.}\ }\textbf {\bibinfo {volume} {17}},\ \bibinfo {pages}
  {3666} (\bibinfo {year} {2021})}\BibitemShut {NoStop}%
\bibitem [{\citenamefont {McKeon}\ \emph {et~al.}(2022)\citenamefont {McKeon},
  \citenamefont {Hamed}, \citenamefont {Bruneval},\ and\ \citenamefont
  {Neaton}}]{McKeon_2022}%
  \BibitemOpen
  \bibfield  {author} {\bibinfo {author} {\bibfnamefont {C.~A.}\ \bibnamefont
  {McKeon}}, \bibinfo {author} {\bibfnamefont {S.~M.}\ \bibnamefont {Hamed}},
  \bibinfo {author} {\bibfnamefont {F.}~\bibnamefont {Bruneval}}, \ and\
  \bibinfo {author} {\bibfnamefont {J.~B.}\ \bibnamefont {Neaton}},\ }\href
  {\doibase 10.1063/5.0097582} {\bibfield  {journal} {\bibinfo  {journal} {J.
  Chem. Phys.}\ }\textbf {\bibinfo {volume} {157}},\ \bibinfo {pages} {074103}
  (\bibinfo {year} {2022})}\BibitemShut {NoStop}%
\bibitem [{\citenamefont {{van Setten}}\ \emph {et~al.}(2018)\citenamefont
  {{van Setten}}, \citenamefont {Costa}, \citenamefont {Vi{\~n}es},\ and\
  \citenamefont {Illas}}]{vanSetten_2018}%
  \BibitemOpen
  \bibfield  {author} {\bibinfo {author} {\bibfnamefont {M.~J.}\ \bibnamefont
  {{van Setten}}}, \bibinfo {author} {\bibfnamefont {R.}~\bibnamefont {Costa}},
  \bibinfo {author} {\bibfnamefont {F.}~\bibnamefont {Vi{\~n}es}}, \ and\
  \bibinfo {author} {\bibfnamefont {F.}~\bibnamefont {Illas}},\ }\href
  {\doibase 10.1021/acs.jctc.7b01192} {\bibfield  {journal} {\bibinfo
  {journal} {J. Chem. Theory Comput.}\ }\textbf {\bibinfo {volume} {14}},\
  \bibinfo {pages} {877} (\bibinfo {year} {2018})}\BibitemShut {NoStop}%
\bibitem [{\citenamefont {Jin}, \citenamefont {Su},\ and\ \citenamefont
  {Yang}(2019)}]{Jin_2019a}%
  \BibitemOpen
  \bibfield  {author} {\bibinfo {author} {\bibfnamefont {Y.}~\bibnamefont
  {Jin}}, \bibinfo {author} {\bibfnamefont {N.~Q.}\ \bibnamefont {Su}}, \ and\
  \bibinfo {author} {\bibfnamefont {W.}~\bibnamefont {Yang}},\ }\href {\doibase
  10.1021/acs.jpclett.8b03337} {\bibfield  {journal} {\bibinfo  {journal} {J.
  Phys. Chem. Lett.}\ }\textbf {\bibinfo {volume} {10}},\ \bibinfo {pages}
  {447} (\bibinfo {year} {2019})}\BibitemShut {NoStop}%
\bibitem [{\citenamefont {Jin}\ and\ \citenamefont {Yang}(2019)}]{Jin_2019b}%
  \BibitemOpen
  \bibfield  {author} {\bibinfo {author} {\bibfnamefont {Y.}~\bibnamefont
  {Jin}}\ and\ \bibinfo {author} {\bibfnamefont {W.}~\bibnamefont {Yang}},\
  }\href {\doibase 10.1021/acs.jpca.9b02379} {\bibfield  {journal} {\bibinfo
  {journal} {J. Phys. Chem. A}\ }\textbf {\bibinfo {volume} {123}},\ \bibinfo
  {pages} {3199} (\bibinfo {year} {2019})}\BibitemShut {NoStop}%
\bibitem [{\citenamefont {Golze}\ \emph {et~al.}(2018)\citenamefont {Golze},
  \citenamefont {Wilhelm}, \citenamefont {van Setten},\ and\ \citenamefont
  {Rinke}}]{Golze_2018}%
  \BibitemOpen
  \bibfield  {author} {\bibinfo {author} {\bibfnamefont {D.}~\bibnamefont
  {Golze}}, \bibinfo {author} {\bibfnamefont {J.}~\bibnamefont {Wilhelm}},
  \bibinfo {author} {\bibfnamefont {M.~J.}\ \bibnamefont {van Setten}}, \ and\
  \bibinfo {author} {\bibfnamefont {P.}~\bibnamefont {Rinke}},\ }\href
  {\doibase 10.1021/acs.jctc.8b00458} {\bibfield  {journal} {\bibinfo
  {journal} {J. Chem. Theory Comput.}\ }\textbf {\bibinfo {volume} {14}},\
  \bibinfo {pages} {4856} (\bibinfo {year} {2018})}\BibitemShut {NoStop}%
\bibitem [{\citenamefont {Golze}, \citenamefont {Keller},\ and\ \citenamefont
  {Rinke}(2020)}]{Golze_2020}%
  \BibitemOpen
  \bibfield  {author} {\bibinfo {author} {\bibfnamefont {D.}~\bibnamefont
  {Golze}}, \bibinfo {author} {\bibfnamefont {L.}~\bibnamefont {Keller}}, \
  and\ \bibinfo {author} {\bibfnamefont {P.}~\bibnamefont {Rinke}},\ }\href
  {\doibase 10.1021/acs.jpclett.9b03423} {\bibfield  {journal} {\bibinfo
  {journal} {J. Phys. Chem. Lett.}\ }\textbf {\bibinfo {volume} {11}},\
  \bibinfo {pages} {1840} (\bibinfo {year} {2020})}\BibitemShut {NoStop}%
\bibitem [{\citenamefont {Li}\ \emph {et~al.}(2022)\citenamefont {Li},
  \citenamefont {Jin}, \citenamefont {Rinke}, \citenamefont {Yang},\ and\
  \citenamefont {Golze}}]{Li_2022a}%
  \BibitemOpen
  \bibfield  {author} {\bibinfo {author} {\bibfnamefont {J.}~\bibnamefont
  {Li}}, \bibinfo {author} {\bibfnamefont {Y.}~\bibnamefont {Jin}}, \bibinfo
  {author} {\bibfnamefont {P.}~\bibnamefont {Rinke}}, \bibinfo {author}
  {\bibfnamefont {W.}~\bibnamefont {Yang}}, \ and\ \bibinfo {author}
  {\bibfnamefont {D.}~\bibnamefont {Golze}},\ }\href {\doibase
  10.1021/acs.jctc.2c00617} {\bibfield  {journal} {\bibinfo  {journal} {J.
  Chem. Theory Comput.}\ }\textbf {\bibinfo {volume} {18}},\ \bibinfo {pages}
  {7570} (\bibinfo {year} {2022})}\BibitemShut {NoStop}%
\bibitem [{\citenamefont {Li}, \citenamefont {Golze},\ and\ \citenamefont
  {Yang}(2022)}]{Li_2022b}%
  \BibitemOpen
  \bibfield  {author} {\bibinfo {author} {\bibfnamefont {J.}~\bibnamefont
  {Li}}, \bibinfo {author} {\bibfnamefont {D.}~\bibnamefont {Golze}}, \ and\
  \bibinfo {author} {\bibfnamefont {W.}~\bibnamefont {Yang}},\ }\href {\doibase
  10.1021/acs.jctc.2c00686} {\bibfield  {journal} {\bibinfo  {journal} {J.
  Chem. Theory Comput.}\ }\textbf {\bibinfo {volume} {18}},\ \bibinfo {pages}
  {6637} (\bibinfo {year} {2022})}\BibitemShut {NoStop}%
\bibitem [{\citenamefont {Li}\ and\ \citenamefont {Yang}(2022)}]{Li_2022c}%
  \BibitemOpen
  \bibfield  {author} {\bibinfo {author} {\bibfnamefont {J.}~\bibnamefont
  {Li}}\ and\ \bibinfo {author} {\bibfnamefont {W.}~\bibnamefont {Yang}},\
  }\href {\doibase 10.1021/acs.jpclett.2c02051} {\bibfield  {journal} {\bibinfo
   {journal} {J. Phys. Chem. Lett.}\ }\textbf {\bibinfo {volume} {13}},\
  \bibinfo {pages} {9372} (\bibinfo {year} {2022})}\BibitemShut {NoStop}%
\bibitem [{\citenamefont {F{\"o}rster}\ and\ \citenamefont
  {Visscher}(2022{\natexlab{a}})}]{Forster_2022a}%
  \BibitemOpen
  \bibfield  {author} {\bibinfo {author} {\bibfnamefont {A.}~\bibnamefont
  {F{\"o}rster}}\ and\ \bibinfo {author} {\bibfnamefont {L.}~\bibnamefont
  {Visscher}},\ }\href {\doibase 10.1021/acs.jctc.2c00531} {\bibfield
  {journal} {\bibinfo  {journal} {J. Chem. Theory Comput.}\ }\textbf {\bibinfo
  {volume} {18}},\ \bibinfo {pages} {6779} (\bibinfo {year}
  {2022}{\natexlab{a}})}\BibitemShut {NoStop}%
\bibitem [{\citenamefont {Casida}\ and\ \citenamefont
  {Chong}(1989)}]{Casida_1989}%
  \BibitemOpen
  \bibfield  {author} {\bibinfo {author} {\bibfnamefont {M.~E.}\ \bibnamefont
  {Casida}}\ and\ \bibinfo {author} {\bibfnamefont {D.~P.}\ \bibnamefont
  {Chong}},\ }\href {\doibase 10.1103/PhysRevA.40.4837} {\bibfield  {journal}
  {\bibinfo  {journal} {Phys. Rev. A}\ }\textbf {\bibinfo {volume} {40}},\
  \bibinfo {pages} {4837} (\bibinfo {year} {1989})}\BibitemShut {NoStop}%
\bibitem [{\citenamefont {Casida}\ and\ \citenamefont
  {Chong}(1991)}]{Casida_1991}%
  \BibitemOpen
  \bibfield  {author} {\bibinfo {author} {\bibfnamefont {M.~E.}\ \bibnamefont
  {Casida}}\ and\ \bibinfo {author} {\bibfnamefont {D.~P.}\ \bibnamefont
  {Chong}},\ }\href {\doibase 10.1103/PhysRevA.44.5773} {\bibfield  {journal}
  {\bibinfo  {journal} {Phys. Rev. A}\ }\textbf {\bibinfo {volume} {44}},\
  \bibinfo {pages} {5773} (\bibinfo {year} {1991})}\BibitemShut {NoStop}%
\bibitem [{\citenamefont {Ortiz}(2013)}]{Ortiz_2013}%
  \BibitemOpen
  \bibfield  {author} {\bibinfo {author} {\bibfnamefont {J.~V.}\ \bibnamefont
  {Ortiz}},\ }\href {\doibase 10.1002/wcms.1116} {\bibfield  {journal}
  {\bibinfo  {journal} {Wiley Interdiscip. Rev. Comput. Mol. Sci.}\ }\textbf
  {\bibinfo {volume} {3}},\ \bibinfo {pages} {123} (\bibinfo {year}
  {2013})}\BibitemShut {NoStop}%
\bibitem [{\citenamefont {Phillips}\ and\ \citenamefont
  {Zgid}(2014)}]{Phillips_2014}%
  \BibitemOpen
  \bibfield  {author} {\bibinfo {author} {\bibfnamefont {J.~J.}\ \bibnamefont
  {Phillips}}\ and\ \bibinfo {author} {\bibfnamefont {D.}~\bibnamefont
  {Zgid}},\ }\href {\doibase 10.1063/1.4884951} {\bibfield  {journal} {\bibinfo
   {journal} {J. Chem. Phys.}\ }\textbf {\bibinfo {volume} {140}},\ \bibinfo
  {pages} {241101} (\bibinfo {year} {2014})}\BibitemShut {NoStop}%
\bibitem [{\citenamefont {Phillips}, \citenamefont {Kananenka},\ and\
  \citenamefont {Zgid}(2015)}]{Phillips_2015}%
  \BibitemOpen
  \bibfield  {author} {\bibinfo {author} {\bibfnamefont {J.~J.}\ \bibnamefont
  {Phillips}}, \bibinfo {author} {\bibfnamefont {A.~A.}\ \bibnamefont
  {Kananenka}}, \ and\ \bibinfo {author} {\bibfnamefont {D.}~\bibnamefont
  {Zgid}},\ }\href {\doibase 10.1063/1.4921259} {\bibfield  {journal} {\bibinfo
   {journal} {J. Chem. Phys.}\ }\textbf {\bibinfo {volume} {142}},\ \bibinfo
  {pages} {194108} (\bibinfo {year} {2015})}\BibitemShut {NoStop}%
\bibitem [{\citenamefont {Rusakov}, \citenamefont {Phillips},\ and\
  \citenamefont {Zgid}(2014)}]{Rusakov_2014}%
  \BibitemOpen
  \bibfield  {author} {\bibinfo {author} {\bibfnamefont {A.~A.}\ \bibnamefont
  {Rusakov}}, \bibinfo {author} {\bibfnamefont {J.~J.}\ \bibnamefont
  {Phillips}}, \ and\ \bibinfo {author} {\bibfnamefont {D.}~\bibnamefont
  {Zgid}},\ }\href {\doibase 10.1063/1.4901432} {\bibfield  {journal} {\bibinfo
   {journal} {J. Chem. Phys.}\ }\textbf {\bibinfo {volume} {141}},\ \bibinfo
  {pages} {194105} (\bibinfo {year} {2014})}\BibitemShut {NoStop}%
\bibitem [{\citenamefont {Rusakov}\ and\ \citenamefont
  {Zgid}(2016)}]{Rusakov_2016}%
  \BibitemOpen
  \bibfield  {author} {\bibinfo {author} {\bibfnamefont {A.~A.}\ \bibnamefont
  {Rusakov}}\ and\ \bibinfo {author} {\bibfnamefont {D.}~\bibnamefont {Zgid}},\
  }\href {\doibase 10.1063/1.4940900} {\bibfield  {journal} {\bibinfo
  {journal} {J. Chem. Phys.}\ }\textbf {\bibinfo {volume} {144}},\ \bibinfo
  {pages} {054106} (\bibinfo {year} {2016})}\BibitemShut {NoStop}%
\bibitem [{\citenamefont {Hirata}\ \emph {et~al.}(2015)\citenamefont {Hirata},
  \citenamefont {Hermes}, \citenamefont {Simons},\ and\ \citenamefont
  {Ortiz}}]{Hirata_2015}%
  \BibitemOpen
  \bibfield  {author} {\bibinfo {author} {\bibfnamefont {S.}~\bibnamefont
  {Hirata}}, \bibinfo {author} {\bibfnamefont {M.~R.}\ \bibnamefont {Hermes}},
  \bibinfo {author} {\bibfnamefont {J.}~\bibnamefont {Simons}}, \ and\ \bibinfo
  {author} {\bibfnamefont {J.~V.}\ \bibnamefont {Ortiz}},\ }\href {\doibase
  10.1021/acs.jctc.5b00005} {\bibfield  {journal} {\bibinfo  {journal} {J.
  Chem. Theory Comput.}\ }\textbf {\bibinfo {volume} {11}},\ \bibinfo {pages}
  {1595} (\bibinfo {year} {2015})}\BibitemShut {NoStop}%
\bibitem [{\citenamefont {Hirata}\ \emph {et~al.}(2017)\citenamefont {Hirata},
  \citenamefont {Doran}, \citenamefont {Knowles},\ and\ \citenamefont
  {Ortiz}}]{Hirata_2017}%
  \BibitemOpen
  \bibfield  {author} {\bibinfo {author} {\bibfnamefont {S.}~\bibnamefont
  {Hirata}}, \bibinfo {author} {\bibfnamefont {A.~E.}\ \bibnamefont {Doran}},
  \bibinfo {author} {\bibfnamefont {P.~J.}\ \bibnamefont {Knowles}}, \ and\
  \bibinfo {author} {\bibfnamefont {J.~V.}\ \bibnamefont {Ortiz}},\ }\href
  {\doibase 10.1063/1.4994837} {\bibfield  {journal} {\bibinfo  {journal} {J.
  Chem. Phys.}\ }\textbf {\bibinfo {volume} {147}},\ \bibinfo {pages} {044108}
  (\bibinfo {year} {2017})}\BibitemShut {NoStop}%
\bibitem [{\citenamefont {Backhouse}, \citenamefont {Santana-Bonilla},\ and\
  \citenamefont {Booth}(2021)}]{Backhouse_2021}%
  \BibitemOpen
  \bibfield  {author} {\bibinfo {author} {\bibfnamefont {O.~J.}\ \bibnamefont
  {Backhouse}}, \bibinfo {author} {\bibfnamefont {A.}~\bibnamefont
  {Santana-Bonilla}}, \ and\ \bibinfo {author} {\bibfnamefont {G.~H.}\
  \bibnamefont {Booth}},\ }\href {\doibase 10.1021/acs.jpclett.1c02383}
  {\bibfield  {journal} {\bibinfo  {journal} {J. Phys. Chem. Lett.}\ }\textbf
  {\bibinfo {volume} {12}},\ \bibinfo {pages} {7650} (\bibinfo {year}
  {2021})}\BibitemShut {NoStop}%
\bibitem [{\citenamefont {Backhouse}\ and\ \citenamefont
  {Booth}(2020)}]{Backhouse_2020b}%
  \BibitemOpen
  \bibfield  {author} {\bibinfo {author} {\bibfnamefont {O.~J.}\ \bibnamefont
  {Backhouse}}\ and\ \bibinfo {author} {\bibfnamefont {G.~H.}\ \bibnamefont
  {Booth}},\ }\href {\doibase 10.1021/acs.jctc.0c00701} {\bibfield  {journal}
  {\bibinfo  {journal} {J. Chem. Theory Comput.}\ }\textbf {\bibinfo {volume}
  {16}},\ \bibinfo {pages} {6294} (\bibinfo {year} {2020})}\BibitemShut
  {NoStop}%
\bibitem [{\citenamefont {Backhouse}, \citenamefont {Nusspickel},\ and\
  \citenamefont {Booth}(2020)}]{Backhouse_2020a}%
  \BibitemOpen
  \bibfield  {author} {\bibinfo {author} {\bibfnamefont {O.~J.}\ \bibnamefont
  {Backhouse}}, \bibinfo {author} {\bibfnamefont {M.}~\bibnamefont
  {Nusspickel}}, \ and\ \bibinfo {author} {\bibfnamefont {G.~H.}\ \bibnamefont
  {Booth}},\ }\href {\doibase 10.1021/acs.jctc.9b01182} {\bibfield  {journal}
  {\bibinfo  {journal} {J. Chem. Theory Comput.}\ }\textbf {\bibinfo {volume}
  {16}},\ \bibinfo {pages} {1090} (\bibinfo {year} {2020})}\BibitemShut
  {NoStop}%
\bibitem [{\citenamefont {Pokhilko}\ and\ \citenamefont
  {Zgid}(2021)}]{Pokhilko_2021a}%
  \BibitemOpen
  \bibfield  {author} {\bibinfo {author} {\bibfnamefont {P.}~\bibnamefont
  {Pokhilko}}\ and\ \bibinfo {author} {\bibfnamefont {D.}~\bibnamefont
  {Zgid}},\ }\href {\doibase 10.1063/5.0055191} {\bibfield  {journal} {\bibinfo
   {journal} {J. Chem. Phys.}\ }\textbf {\bibinfo {volume} {155}},\ \bibinfo
  {pages} {024101} (\bibinfo {year} {2021})}\BibitemShut {NoStop}%
\bibitem [{\citenamefont {Pokhilko}\ \emph {et~al.}(2021)\citenamefont
  {Pokhilko}, \citenamefont {Iskakov}, \citenamefont {Yeh},\ and\ \citenamefont
  {Zgid}}]{Pokhilko_2021b}%
  \BibitemOpen
  \bibfield  {author} {\bibinfo {author} {\bibfnamefont {P.}~\bibnamefont
  {Pokhilko}}, \bibinfo {author} {\bibfnamefont {S.}~\bibnamefont {Iskakov}},
  \bibinfo {author} {\bibfnamefont {C.-N.}\ \bibnamefont {Yeh}}, \ and\
  \bibinfo {author} {\bibfnamefont {D.}~\bibnamefont {Zgid}},\ }\href {\doibase
  10.1063/5.0054661} {\bibfield  {journal} {\bibinfo  {journal} {J. Chem.
  Phys.}\ }\textbf {\bibinfo {volume} {155}},\ \bibinfo {pages} {024119}
  (\bibinfo {year} {2021})}\BibitemShut {NoStop}%
\bibitem [{\citenamefont {Pokhilko}, \citenamefont {Yeh},\ and\ \citenamefont
  {Zgid}(2022)}]{Pokhilko_2022}%
  \BibitemOpen
  \bibfield  {author} {\bibinfo {author} {\bibfnamefont {P.}~\bibnamefont
  {Pokhilko}}, \bibinfo {author} {\bibfnamefont {C.-N.}\ \bibnamefont {Yeh}}, \
  and\ \bibinfo {author} {\bibfnamefont {D.}~\bibnamefont {Zgid}},\ }\href
  {\doibase 10.1063/5.0082586} {\bibfield  {journal} {\bibinfo  {journal} {J.
  Chem. Phys.}\ }\textbf {\bibinfo {volume} {156}},\ \bibinfo {pages} {094101}
  (\bibinfo {year} {2022})}\BibitemShut {NoStop}%
\bibitem [{\citenamefont {Stefanucci}\ and\ \citenamefont {van
  Leeuwen}(2013)}]{Stefanucci_2013}%
  \BibitemOpen
  \bibfield  {author} {\bibinfo {author} {\bibfnamefont {G.}~\bibnamefont
  {Stefanucci}}\ and\ \bibinfo {author} {\bibfnamefont {R.}~\bibnamefont {van
  Leeuwen}},\ }\href@noop {} {\emph {\bibinfo {title} {Nonequilibrium Many-Body
  Theory of Quantum Systems: A Modern Introduction}}}\ (\bibinfo  {publisher}
  {{Cambridge University Press}},\ \bibinfo {address} {Cambridge},\ \bibinfo
  {year} {2013})\BibitemShut {NoStop}%
\bibitem [{\citenamefont {Liebsch}(1981)}]{Liebsch_1981}%
  \BibitemOpen
  \bibfield  {author} {\bibinfo {author} {\bibfnamefont {A.}~\bibnamefont
  {Liebsch}},\ }\href {\doibase 10.1103/PhysRevB.23.5203} {\bibfield  {journal}
  {\bibinfo  {journal} {Phys. Rev. B}\ }\textbf {\bibinfo {volume} {23}},\
  \bibinfo {pages} {5203} (\bibinfo {year} {1981})}\BibitemShut {NoStop}%
\bibitem [{\citenamefont {Bickers}, \citenamefont {Scalapino},\ and\
  \citenamefont {White}(1989)}]{Bickers_1989a}%
  \BibitemOpen
  \bibfield  {author} {\bibinfo {author} {\bibfnamefont {N.~E.}\ \bibnamefont
  {Bickers}}, \bibinfo {author} {\bibfnamefont {D.~J.}\ \bibnamefont
  {Scalapino}}, \ and\ \bibinfo {author} {\bibfnamefont {S.~R.}\ \bibnamefont
  {White}},\ }\href {\doibase 10.1103/PhysRevLett.62.961} {\bibfield  {journal}
  {\bibinfo  {journal} {Phys. Rev. Lett.}\ }\textbf {\bibinfo {volume} {62}},\
  \bibinfo {pages} {961} (\bibinfo {year} {1989})}\BibitemShut {NoStop}%
\bibitem [{\citenamefont {Bickers}\ and\ \citenamefont
  {White}(1991)}]{Bickers_1991}%
  \BibitemOpen
  \bibfield  {author} {\bibinfo {author} {\bibfnamefont {N.~E.}\ \bibnamefont
  {Bickers}}\ and\ \bibinfo {author} {\bibfnamefont {S.~R.}\ \bibnamefont
  {White}},\ }\href {\doibase 10.1103/PhysRevB.43.8044} {\bibfield  {journal}
  {\bibinfo  {journal} {Phys. Rev. B}\ }\textbf {\bibinfo {volume} {43}},\
  \bibinfo {pages} {8044} (\bibinfo {year} {1991})}\BibitemShut {NoStop}%
\bibitem [{\citenamefont {Katsnelson}\ and\ \citenamefont
  {Lichtenstein}(1999)}]{Katsnelson_1999}%
  \BibitemOpen
  \bibfield  {author} {\bibinfo {author} {\bibfnamefont {M.~I.}\ \bibnamefont
  {Katsnelson}}\ and\ \bibinfo {author} {\bibfnamefont {A.~I.}\ \bibnamefont
  {Lichtenstein}},\ }\href {\doibase 10.1088/0953-8984/11/4/011} {\bibfield
  {journal} {\bibinfo  {journal} {J. Phys. Condens. Matter}\ }\textbf {\bibinfo
  {volume} {11}},\ \bibinfo {pages} {1037} (\bibinfo {year}
  {1999})}\BibitemShut {NoStop}%
\bibitem [{\citenamefont {Katsnelson}\ and\ \citenamefont
  {Lichtenstein}(2002)}]{Katsnelson_2002}%
  \BibitemOpen
  \bibfield  {author} {\bibinfo {author} {\bibfnamefont {M.}~\bibnamefont
  {Katsnelson}}\ and\ \bibinfo {author} {\bibfnamefont {A.}~\bibnamefont
  {Lichtenstein}},\ }\href {\doibase 10.1140/epjb/e2002-00352-1} {\bibfield
  {journal} {\bibinfo  {journal} {Eur. Phys. J. B}\ }\textbf {\bibinfo {volume}
  {30}},\ \bibinfo {pages} {9} (\bibinfo {year} {2002})}\BibitemShut {NoStop}%
\bibitem [{\citenamefont {Zhukov}, \citenamefont {Chulkov},\ and\ \citenamefont
  {Echenique}(2005)}]{Zhukov_2005}%
  \BibitemOpen
  \bibfield  {author} {\bibinfo {author} {\bibfnamefont {V.~P.}\ \bibnamefont
  {Zhukov}}, \bibinfo {author} {\bibfnamefont {E.~V.}\ \bibnamefont {Chulkov}},
  \ and\ \bibinfo {author} {\bibfnamefont {P.~M.}\ \bibnamefont {Echenique}},\
  }\href {\doibase 10.1103/PhysRevB.72.155109} {\bibfield  {journal} {\bibinfo
  {journal} {Phys. Rev. B}\ }\textbf {\bibinfo {volume} {72}},\ \bibinfo
  {pages} {72.155109} (\bibinfo {year} {2005})}\BibitemShut {NoStop}%
\bibitem [{\citenamefont {Puig~von Friesen}, \citenamefont {Verdozzi},\ and\
  \citenamefont {Almbladh}(2010)}]{vonFriesen_2010}%
  \BibitemOpen
  \bibfield  {author} {\bibinfo {author} {\bibfnamefont {M.}~\bibnamefont
  {Puig~von Friesen}}, \bibinfo {author} {\bibfnamefont {C.}~\bibnamefont
  {Verdozzi}}, \ and\ \bibinfo {author} {\bibfnamefont {C.-O.}\ \bibnamefont
  {Almbladh}},\ }\href {\doibase 10.1103/PhysRevB.82.155108} {\bibfield
  {journal} {\bibinfo  {journal} {Phys. Rev. B}\ }\textbf {\bibinfo {volume}
  {82}},\ \bibinfo {pages} {155108} (\bibinfo {year} {2010})}\BibitemShut
  {NoStop}%
\bibitem [{\citenamefont {Romaniello}, \citenamefont {Bechstedt},\ and\
  \citenamefont {Reining}(2012)}]{Romaniello_2012}%
  \BibitemOpen
  \bibfield  {author} {\bibinfo {author} {\bibfnamefont {P.}~\bibnamefont
  {Romaniello}}, \bibinfo {author} {\bibfnamefont {F.}~\bibnamefont
  {Bechstedt}}, \ and\ \bibinfo {author} {\bibfnamefont {L.}~\bibnamefont
  {Reining}},\ }\href {\doibase 10.1103/PhysRevB.85.155131} {\bibfield
  {journal} {\bibinfo  {journal} {Phys. Rev. B}\ }\textbf {\bibinfo {volume}
  {85}},\ \bibinfo {pages} {155131} (\bibinfo {year} {2012})}\BibitemShut
  {NoStop}%
\bibitem [{\citenamefont {Gukelberger}, \citenamefont {Huang},\ and\
  \citenamefont {Werner}(2015)}]{Gukelberger_2015}%
  \BibitemOpen
  \bibfield  {author} {\bibinfo {author} {\bibfnamefont {J.}~\bibnamefont
  {Gukelberger}}, \bibinfo {author} {\bibfnamefont {L.}~\bibnamefont {Huang}},
  \ and\ \bibinfo {author} {\bibfnamefont {P.}~\bibnamefont {Werner}},\ }\href
  {\doibase 10.1103/PhysRevB.91.235114} {\bibfield  {journal} {\bibinfo
  {journal} {Phys. Rev. B}\ }\textbf {\bibinfo {volume} {91}},\ \bibinfo
  {pages} {235114} (\bibinfo {year} {2015})}\BibitemShut {NoStop}%
\bibitem [{\citenamefont {M\"uller}, \citenamefont {Bl\"ugel},\ and\
  \citenamefont {Friedrich}(2019)}]{Muller_2019}%
  \BibitemOpen
  \bibfield  {author} {\bibinfo {author} {\bibfnamefont {M.~C. T.~D.}\
  \bibnamefont {M\"uller}}, \bibinfo {author} {\bibfnamefont {S.}~\bibnamefont
  {Bl\"ugel}}, \ and\ \bibinfo {author} {\bibfnamefont {C.}~\bibnamefont
  {Friedrich}},\ }\href {\doibase 10.1103/PhysRevB.100.045130} {\bibfield
  {journal} {\bibinfo  {journal} {Phys. Rev. B}\ }\textbf {\bibinfo {volume}
  {100}},\ \bibinfo {pages} {045130} (\bibinfo {year} {2019})}\BibitemShut
  {NoStop}%
\bibitem [{\citenamefont {Friedrich}(2019)}]{Friedrich_2019}%
  \BibitemOpen
  \bibfield  {author} {\bibinfo {author} {\bibfnamefont {C.}~\bibnamefont
  {Friedrich}},\ }\href {\doibase 10.1103/PhysRevB.100.075142} {\bibfield
  {journal} {\bibinfo  {journal} {Phys. Rev. B}\ }\textbf {\bibinfo {volume}
  {100}},\ \bibinfo {pages} {075142} (\bibinfo {year} {2019})}\BibitemShut
  {NoStop}%
\bibitem [{\citenamefont {Biswas}\ and\ \citenamefont
  {Singh}(2021)}]{Biswas_2021}%
  \BibitemOpen
  \bibfield  {author} {\bibinfo {author} {\bibfnamefont {T.}~\bibnamefont
  {Biswas}}\ and\ \bibinfo {author} {\bibfnamefont {A.}~\bibnamefont {Singh}},\
  }\href {\doibase 10.1038/s41524-021-00640-3} {\bibfield  {journal} {\bibinfo
  {journal} {npj Comput. Mater.}\ }\textbf {\bibinfo {volume} {7}},\ \bibinfo
  {pages} {189} (\bibinfo {year} {2021})}\BibitemShut {NoStop}%
\bibitem [{\citenamefont {Zhang}, \citenamefont {Su},\ and\ \citenamefont
  {Yang}(2017)}]{Zhang_2017}%
  \BibitemOpen
  \bibfield  {author} {\bibinfo {author} {\bibfnamefont {D.}~\bibnamefont
  {Zhang}}, \bibinfo {author} {\bibfnamefont {N.~Q.}\ \bibnamefont {Su}}, \
  and\ \bibinfo {author} {\bibfnamefont {W.}~\bibnamefont {Yang}},\ }\href
  {\doibase 10.1021/acs.jpclett.7b01275} {\bibfield  {journal} {\bibinfo
  {journal} {J. Phys. Chem. Lett.}\ }\textbf {\bibinfo {volume} {8}},\ \bibinfo
  {pages} {3223} (\bibinfo {year} {2017})}\BibitemShut {NoStop}%
\bibitem [{\citenamefont {Li}, \citenamefont {Chen},\ and\ \citenamefont
  {Yang}(2021)}]{Li_2021b}%
  \BibitemOpen
  \bibfield  {author} {\bibinfo {author} {\bibfnamefont {J.}~\bibnamefont
  {Li}}, \bibinfo {author} {\bibfnamefont {Z.}~\bibnamefont {Chen}}, \ and\
  \bibinfo {author} {\bibfnamefont {W.}~\bibnamefont {Yang}},\ }\href {\doibase
  10.1021/acs.jpclett.1c01723} {\bibfield  {journal} {\bibinfo  {journal} {J.
  Phys. Chem. Lett.}\ }\textbf {\bibinfo {volume} {12}},\ \bibinfo {pages}
  {6203} (\bibinfo {year} {2021})}\BibitemShut {NoStop}%
\bibitem [{\citenamefont {Loos}\ and\ \citenamefont
  {Romaniello}(2022)}]{Loos_2022}%
  \BibitemOpen
  \bibfield  {author} {\bibinfo {author} {\bibfnamefont {P.-F.}\ \bibnamefont
  {Loos}}\ and\ \bibinfo {author} {\bibfnamefont {P.}~\bibnamefont
  {Romaniello}},\ }\href {\doibase 10.1063/5.0088364} {\bibfield  {journal}
  {\bibinfo  {journal} {J. Chem. Phys.}\ }\textbf {\bibinfo {volume} {156}},\
  \bibinfo {pages} {164101} (\bibinfo {year} {2022})}\BibitemShut {NoStop}%
\bibitem [{\citenamefont {Bethe}\ and\ \citenamefont
  {Goldstone}(1957)}]{Bethe_1957}%
  \BibitemOpen
  \bibfield  {author} {\bibinfo {author} {\bibfnamefont {H.~A.}\ \bibnamefont
  {Bethe}}\ and\ \bibinfo {author} {\bibfnamefont {J.}~\bibnamefont
  {Goldstone}},\ }\href {http://www.jstor.org/stable/100108} {\bibfield
  {journal} {\bibinfo  {journal} {Proc. Math. Phys. Eng. Sci.}\ }\textbf
  {\bibinfo {volume} {238}},\ \bibinfo {pages} {551} (\bibinfo {year}
  {1957})}\BibitemShut {NoStop}%
\bibitem [{\citenamefont {Baym}\ and\ \citenamefont
  {Kadanoff}(1961)}]{Baym_1961}%
  \BibitemOpen
  \bibfield  {author} {\bibinfo {author} {\bibfnamefont {G.}~\bibnamefont
  {Baym}}\ and\ \bibinfo {author} {\bibfnamefont {L.~P.}\ \bibnamefont
  {Kadanoff}},\ }\href {\doibase 10.1103/PhysRev.124.287} {\bibfield  {journal}
  {\bibinfo  {journal} {Phys. Rev.}\ }\textbf {\bibinfo {volume} {124}},\
  \bibinfo {pages} {287} (\bibinfo {year} {1961})}\BibitemShut {NoStop}%
\bibitem [{\citenamefont {Baym}(1962)}]{Baym_1962}%
  \BibitemOpen
  \bibfield  {author} {\bibinfo {author} {\bibfnamefont {G.}~\bibnamefont
  {Baym}},\ }\href {\doibase 10.1103/PhysRev.127.1391} {\bibfield  {journal}
  {\bibinfo  {journal} {Phys. Rev.}\ }\textbf {\bibinfo {volume} {127}},\
  \bibinfo {pages} {1391} (\bibinfo {year} {1962})}\BibitemShut {NoStop}%
\bibitem [{\citenamefont
  {Danielewicz}(1984{\natexlab{a}})}]{Danielewicz_1984a}%
  \BibitemOpen
  \bibfield  {author} {\bibinfo {author} {\bibfnamefont {P.}~\bibnamefont
  {Danielewicz}},\ }\href {\doibase
  https://doi.org/10.1016/0003-4916(84)90092-7} {\bibfield  {journal} {\bibinfo
   {journal} {Ann. Phys.}\ }\textbf {\bibinfo {volume} {152}},\ \bibinfo
  {pages} {239} (\bibinfo {year} {1984}{\natexlab{a}})}\BibitemShut {NoStop}%
\bibitem [{\citenamefont
  {Danielewicz}(1984{\natexlab{b}})}]{Danielewicz_1984b}%
  \BibitemOpen
  \bibfield  {author} {\bibinfo {author} {\bibfnamefont {P.}~\bibnamefont
  {Danielewicz}},\ }\href {\doibase
  https://doi.org/10.1016/0003-4916(84)90093-9} {\bibfield  {journal} {\bibinfo
   {journal} {Ann. Phys.}\ }\textbf {\bibinfo {volume} {152}},\ \bibinfo
  {pages} {305} (\bibinfo {year} {1984}{\natexlab{b}})}\BibitemShut {NoStop}%
\bibitem [{\citenamefont {De~Dominicis}\ and\ \citenamefont
  {Martin}(1964{\natexlab{a}})}]{DeDominicis_1964a}%
  \BibitemOpen
  \bibfield  {author} {\bibinfo {author} {\bibfnamefont {C.}~\bibnamefont
  {De~Dominicis}}\ and\ \bibinfo {author} {\bibfnamefont {P.~C.}\ \bibnamefont
  {Martin}},\ }\href {\doibase 10.1063/1.1704062} {\bibfield  {journal}
  {\bibinfo  {journal} {J. Math. Phys.}\ }\textbf {\bibinfo {volume} {5}},\
  \bibinfo {pages} {14} (\bibinfo {year} {1964}{\natexlab{a}})}\BibitemShut
  {NoStop}%
\bibitem [{\citenamefont {De~Dominicis}\ and\ \citenamefont
  {Martin}(1964{\natexlab{b}})}]{DeDominicis_1964b}%
  \BibitemOpen
  \bibfield  {author} {\bibinfo {author} {\bibfnamefont {C.}~\bibnamefont
  {De~Dominicis}}\ and\ \bibinfo {author} {\bibfnamefont {P.~C.}\ \bibnamefont
  {Martin}},\ }\href {\doibase 10.1063/1.1704064} {\bibfield  {journal}
  {\bibinfo  {journal} {J. Math. Phys.}\ }\textbf {\bibinfo {volume} {5}},\
  \bibinfo {pages} {31} (\bibinfo {year} {1964}{\natexlab{b}})}\BibitemShut
  {NoStop}%
\bibitem [{\citenamefont {Bickers}\ and\ \citenamefont
  {Scalapino}(1989)}]{Bickers_1989b}%
  \BibitemOpen
  \bibfield  {author} {\bibinfo {author} {\bibfnamefont {N.}~\bibnamefont
  {Bickers}}\ and\ \bibinfo {author} {\bibfnamefont {D.}~\bibnamefont
  {Scalapino}},\ }\href {\doibase https://doi.org/10.1016/0003-4916(89)90359-X}
  {\bibfield  {journal} {\bibinfo  {journal} {Ann. Phys.}\ }\textbf {\bibinfo
  {volume} {193}},\ \bibinfo {pages} {206} (\bibinfo {year}
  {1989})}\BibitemShut {NoStop}%
\bibitem [{\citenamefont {Hedin}(1999)}]{Hedin_1999}%
  \BibitemOpen
  \bibfield  {author} {\bibinfo {author} {\bibfnamefont {L.}~\bibnamefont
  {Hedin}},\ }\href {\doibase 10.1088/0953-8984/11/42/201} {\bibfield
  {journal} {\bibinfo  {journal} {J. Phys. Condens. Matter}\ }\textbf {\bibinfo
  {volume} {11}},\ \bibinfo {pages} {R489} (\bibinfo {year}
  {1999})}\BibitemShut {NoStop}%
\bibitem [{\citenamefont {Bickers}(2004)}]{Bickers_2004}%
  \BibitemOpen
  \bibfield  {author} {\bibinfo {author} {\bibfnamefont {N.~E.}\ \bibnamefont
  {Bickers}},\ }\enquote {\bibinfo {title} {Self-consistent many-body theory
  for condensed matter systems},}\ in\ \href {\doibase 10.1007/0-387-21717-7_6}
  {\emph {\bibinfo {booktitle} {Theoretical Methods for Strongly Correlated
  Electrons}}},\ \bibinfo {editor} {edited by\ \bibinfo {editor} {\bibfnamefont
  {D.}~\bibnamefont {S{\'e}n{\'e}chal}}, \bibinfo {editor} {\bibfnamefont
  {A.-M.}\ \bibnamefont {Tremblay}}, \ and\ \bibinfo {editor} {\bibfnamefont
  {C.}~\bibnamefont {Bourbonnais}}}\ (\bibinfo  {publisher} {Springer New
  York},\ \bibinfo {year} {2004})\ pp.\ \bibinfo {pages} {237--296}\BibitemShut
  {NoStop}%
\bibitem [{\citenamefont {Shirley}(1996)}]{Shirley_1996}%
  \BibitemOpen
  \bibfield  {author} {\bibinfo {author} {\bibfnamefont {E.~L.}\ \bibnamefont
  {Shirley}},\ }\href {\doibase 10.1103/PhysRevB.54.7758} {\bibfield  {journal}
  {\bibinfo  {journal} {Phys. Rev. B}\ }\textbf {\bibinfo {volume} {54}},\
  \bibinfo {pages} {7758} (\bibinfo {year} {1996})}\BibitemShut {NoStop}%
\bibitem [{\citenamefont {Del~Sole}, \citenamefont {Reining},\ and\
  \citenamefont {Godby}(1994)}]{DelSol_1994}%
  \BibitemOpen
  \bibfield  {author} {\bibinfo {author} {\bibfnamefont {R.}~\bibnamefont
  {Del~Sole}}, \bibinfo {author} {\bibfnamefont {L.}~\bibnamefont {Reining}}, \
  and\ \bibinfo {author} {\bibfnamefont {R.~W.}\ \bibnamefont {Godby}},\ }\href
  {\doibase 10.1103/PhysRevB.49.8024} {\bibfield  {journal} {\bibinfo
  {journal} {Phys. Rev. B}\ }\textbf {\bibinfo {volume} {49}},\ \bibinfo
  {pages} {8024} (\bibinfo {year} {1994})}\BibitemShut {NoStop}%
\bibitem [{\citenamefont {Schindlmayr}\ and\ \citenamefont
  {Godby}(1998)}]{Schindlmayr_1998}%
  \BibitemOpen
  \bibfield  {author} {\bibinfo {author} {\bibfnamefont {A.}~\bibnamefont
  {Schindlmayr}}\ and\ \bibinfo {author} {\bibfnamefont {R.~W.}\ \bibnamefont
  {Godby}},\ }\href {\doibase 10.1103/PhysRevLett.80.1702} {\bibfield
  {journal} {\bibinfo  {journal} {Phys. Rev. Lett.}\ }\textbf {\bibinfo
  {volume} {80}},\ \bibinfo {pages} {1702} (\bibinfo {year}
  {1998})}\BibitemShut {NoStop}%
\bibitem [{\citenamefont {Morris}\ \emph {et~al.}(2007)\citenamefont {Morris},
  \citenamefont {Stankovski}, \citenamefont {Delaney}, \citenamefont {Rinke},
  \citenamefont {Garc\'{\i}a-Gonz\'alez},\ and\ \citenamefont
  {Godby}}]{Morris_2007}%
  \BibitemOpen
  \bibfield  {author} {\bibinfo {author} {\bibfnamefont {A.~J.}\ \bibnamefont
  {Morris}}, \bibinfo {author} {\bibfnamefont {M.}~\bibnamefont {Stankovski}},
  \bibinfo {author} {\bibfnamefont {K.~T.}\ \bibnamefont {Delaney}}, \bibinfo
  {author} {\bibfnamefont {P.}~\bibnamefont {Rinke}}, \bibinfo {author}
  {\bibfnamefont {P.}~\bibnamefont {Garc\'{\i}a-Gonz\'alez}}, \ and\ \bibinfo
  {author} {\bibfnamefont {R.~W.}\ \bibnamefont {Godby}},\ }\href {\doibase
  10.1103/PhysRevB.76.155106} {\bibfield  {journal} {\bibinfo  {journal} {Phys.
  Rev. B}\ }\textbf {\bibinfo {volume} {76}},\ \bibinfo {pages} {155106}
  (\bibinfo {year} {2007})}\BibitemShut {NoStop}%
\bibitem [{\citenamefont {Shishkin}, \citenamefont {Marsman},\ and\
  \citenamefont {Kresse}(2007)}]{Shishkin_2007b}%
  \BibitemOpen
  \bibfield  {author} {\bibinfo {author} {\bibfnamefont {M.}~\bibnamefont
  {Shishkin}}, \bibinfo {author} {\bibfnamefont {M.}~\bibnamefont {Marsman}}, \
  and\ \bibinfo {author} {\bibfnamefont {G.}~\bibnamefont {Kresse}},\ }\href
  {\doibase 10.1103/PhysRevLett.99.246403} {\bibfield  {journal} {\bibinfo
  {journal} {Phys. Rev. Lett.}\ }\textbf {\bibinfo {volume} {99}},\ \bibinfo
  {pages} {246403} (\bibinfo {year} {2007})}\BibitemShut {NoStop}%
\bibitem [{\citenamefont {Romaniello}, \citenamefont {Guyot},\ and\
  \citenamefont {Reining}(2009)}]{Romaniello_2009a}%
  \BibitemOpen
  \bibfield  {author} {\bibinfo {author} {\bibfnamefont {P.}~\bibnamefont
  {Romaniello}}, \bibinfo {author} {\bibfnamefont {S.}~\bibnamefont {Guyot}}, \
  and\ \bibinfo {author} {\bibfnamefont {L.}~\bibnamefont {Reining}},\ }\href
  {\doibase 10.1063/1.3249965} {\bibfield  {journal} {\bibinfo  {journal} {J.
  Chem. Phys.}\ }\textbf {\bibinfo {volume} {131}},\ \bibinfo {pages} {154111}
  (\bibinfo {year} {2009})}\BibitemShut {NoStop}%
\bibitem [{\citenamefont {Gr\"uneis}\ \emph {et~al.}(2014)\citenamefont
  {Gr\"uneis}, \citenamefont {Kresse}, \citenamefont {Hinuma},\ and\
  \citenamefont {Oba}}]{Gruneis_2014}%
  \BibitemOpen
  \bibfield  {author} {\bibinfo {author} {\bibfnamefont {A.}~\bibnamefont
  {Gr\"uneis}}, \bibinfo {author} {\bibfnamefont {G.}~\bibnamefont {Kresse}},
  \bibinfo {author} {\bibfnamefont {Y.}~\bibnamefont {Hinuma}}, \ and\ \bibinfo
  {author} {\bibfnamefont {F.}~\bibnamefont {Oba}},\ }\href {\doibase
  10.1103/PhysRevLett.112.096401} {\bibfield  {journal} {\bibinfo  {journal}
  {Phys. Rev. Lett.}\ }\textbf {\bibinfo {volume} {112}},\ \bibinfo {pages}
  {096401} (\bibinfo {year} {2014})}\BibitemShut {NoStop}%
\bibitem [{\citenamefont {Hung}\ \emph {et~al.}(2017)\citenamefont {Hung},
  \citenamefont {Bruneval}, \citenamefont {Baishya},\ and\ \citenamefont
  {{\"O}{\u g}{\"u}t}}]{Hung_2017}%
  \BibitemOpen
  \bibfield  {author} {\bibinfo {author} {\bibfnamefont {L.}~\bibnamefont
  {Hung}}, \bibinfo {author} {\bibfnamefont {F.}~\bibnamefont {Bruneval}},
  \bibinfo {author} {\bibfnamefont {K.}~\bibnamefont {Baishya}}, \ and\
  \bibinfo {author} {\bibfnamefont {S.}~\bibnamefont {{\"O}{\u g}{\"u}t}},\
  }\href {\doibase 10.1021/acs.jctc.7b00123} {\bibfield  {journal} {\bibinfo
  {journal} {J. Chem. Theory Comput.}\ }\textbf {\bibinfo {volume} {13}},\
  \bibinfo {pages} {2135} (\bibinfo {year} {2017})}\BibitemShut {NoStop}%
\bibitem [{\citenamefont {Maggio}\ and\ \citenamefont
  {Kresse}(2017)}]{Maggio_2017b}%
  \BibitemOpen
  \bibfield  {author} {\bibinfo {author} {\bibfnamefont {E.}~\bibnamefont
  {Maggio}}\ and\ \bibinfo {author} {\bibfnamefont {G.}~\bibnamefont
  {Kresse}},\ }\href {\doibase 10.1021/acs.jctc.7b00586} {\bibfield  {journal}
  {\bibinfo  {journal} {J. Chem. Theory Comput.}\ }\textbf {\bibinfo {volume}
  {13}},\ \bibinfo {pages} {4765} (\bibinfo {year} {2017})}\BibitemShut
  {NoStop}%
\bibitem [{\citenamefont {Mejuto-Zaera}\ and\ \citenamefont
  {Vl\ifmmode~\check{c}\else \v{c}\fi{}ek}(2022)}]{Mejuto-Zaera_2022}%
  \BibitemOpen
  \bibfield  {author} {\bibinfo {author} {\bibfnamefont {C.}~\bibnamefont
  {Mejuto-Zaera}}\ and\ \bibinfo {author} {\bibfnamefont {V.~c.~v.}\
  \bibnamefont {Vl\ifmmode~\check{c}\else \v{c}\fi{}ek}},\ }\href {\doibase
  10.1103/PhysRevB.106.165129} {\bibfield  {journal} {\bibinfo  {journal}
  {Phys. Rev. B}\ }\textbf {\bibinfo {volume} {106}},\ \bibinfo {pages}
  {165129} (\bibinfo {year} {2022})}\BibitemShut {NoStop}%
\bibitem [{\citenamefont {Ren}\ \emph {et~al.}(2015)\citenamefont {Ren},
  \citenamefont {Marom}, \citenamefont {Caruso}, \citenamefont {Scheffler},\
  and\ \citenamefont {Rinke}}]{Ren_2015}%
  \BibitemOpen
  \bibfield  {author} {\bibinfo {author} {\bibfnamefont {X.}~\bibnamefont
  {Ren}}, \bibinfo {author} {\bibfnamefont {N.}~\bibnamefont {Marom}}, \bibinfo
  {author} {\bibfnamefont {F.}~\bibnamefont {Caruso}}, \bibinfo {author}
  {\bibfnamefont {M.}~\bibnamefont {Scheffler}}, \ and\ \bibinfo {author}
  {\bibfnamefont {P.}~\bibnamefont {Rinke}},\ }\href {\doibase
  10.1103/PhysRevB.92.081104} {\bibfield  {journal} {\bibinfo  {journal} {Phys.
  Rev. B}\ }\textbf {\bibinfo {volume} {92}},\ \bibinfo {pages} {081104}
  (\bibinfo {year} {2015})}\BibitemShut {NoStop}%
\bibitem [{\citenamefont {Wang}, \citenamefont {Rinke},\ and\ \citenamefont
  {Ren}(2021)}]{Wang_2021}%
  \BibitemOpen
  \bibfield  {author} {\bibinfo {author} {\bibfnamefont {Y.}~\bibnamefont
  {Wang}}, \bibinfo {author} {\bibfnamefont {P.}~\bibnamefont {Rinke}}, \ and\
  \bibinfo {author} {\bibfnamefont {X.}~\bibnamefont {Ren}},\ }\href {\doibase
  10.1021/acs.jctc.1c00488} {\bibfield  {journal} {\bibinfo  {journal} {J.
  Chem. Theory Comput.}\ }\textbf {\bibinfo {volume} {17}},\ \bibinfo {pages}
  {5140} (\bibinfo {year} {2021})}\BibitemShut {NoStop}%
\bibitem [{\citenamefont {Wang}\ and\ \citenamefont {Ren}(2022)}]{Wang_2022}%
  \BibitemOpen
  \bibfield  {author} {\bibinfo {author} {\bibfnamefont {Y.}~\bibnamefont
  {Wang}}\ and\ \bibinfo {author} {\bibfnamefont {X.}~\bibnamefont {Ren}},\
  }\href {\doibase 10.1063/5.0122425} {\bibfield  {journal} {\bibinfo
  {journal} {J. Chem. Theory Comput.}\ }\textbf {\bibinfo {volume} {157}},\
  \bibinfo {pages} {214115} (\bibinfo {year} {2022})}\BibitemShut {NoStop}%
\bibitem [{\citenamefont {F{\"o}rster}\ and\ \citenamefont
  {Visscher}(2022{\natexlab{b}})}]{Forster_2022b}%
  \BibitemOpen
  \bibfield  {author} {\bibinfo {author} {\bibfnamefont {A.}~\bibnamefont
  {F{\"o}rster}}\ and\ \bibinfo {author} {\bibfnamefont {L.}~\bibnamefont
  {Visscher}},\ }\href {\doibase 10.1103/PhysRevB.105.125121} {\bibfield
  {journal} {\bibinfo  {journal} {Phys. Rev. B}\ }\textbf {\bibinfo {volume}
  {105}},\ \bibinfo {pages} {125121} (\bibinfo {year}
  {2022}{\natexlab{b}})}\BibitemShut {NoStop}%
\bibitem [{\citenamefont {Vl{\v c}ek}(2019)}]{Vlcek_2019}%
  \BibitemOpen
  \bibfield  {author} {\bibinfo {author} {\bibfnamefont {V.}~\bibnamefont
  {Vl{\v c}ek}},\ }\href {\doibase https://doi.org/10.1021/acs.jctc.9b00317}
  {\bibfield  {journal} {\bibinfo  {journal} {J. Chem. Theory Comput.}\
  }\textbf {\bibinfo {volume} {15}},\ \bibinfo {pages} {6254–6266} (\bibinfo
  {year} {2019})}\BibitemShut {NoStop}%
\bibitem [{\citenamefont {Pavlyukh}, \citenamefont {Stefanucci},\ and\
  \citenamefont {van Leeuwen}(2020)}]{Pavlyukh_2020}%
  \BibitemOpen
  \bibfield  {author} {\bibinfo {author} {\bibfnamefont {Y.}~\bibnamefont
  {Pavlyukh}}, \bibinfo {author} {\bibfnamefont {G.}~\bibnamefont
  {Stefanucci}}, \ and\ \bibinfo {author} {\bibfnamefont {R.}~\bibnamefont {van
  Leeuwen}},\ }\href {\doibase 10.1103/PhysRevB.102.045121} {\bibfield
  {journal} {\bibinfo  {journal} {Phys. Rev. B}\ }\textbf {\bibinfo {volume}
  {102}},\ \bibinfo {pages} {045121} (\bibinfo {year} {2020})}\BibitemShut
  {NoStop}%
\bibitem [{\citenamefont {Lewis}\ and\ \citenamefont
  {Berkelbach}(2019)}]{Lewis_2019a}%
  \BibitemOpen
  \bibfield  {author} {\bibinfo {author} {\bibfnamefont {A.~M.}\ \bibnamefont
  {Lewis}}\ and\ \bibinfo {author} {\bibfnamefont {T.~C.}\ \bibnamefont
  {Berkelbach}},\ }\href {\doibase 10.1021/acs.jctc.8b00995} {\bibfield
  {journal} {\bibinfo  {journal} {J. Chem. Theory Comput.}\ }\textbf {\bibinfo
  {volume} {15}},\ \bibinfo {pages} {2925} (\bibinfo {year}
  {2019})}\BibitemShut {NoStop}%
\bibitem [{\citenamefont {Hybertsen}\ and\ \citenamefont
  {Louie}(1985{\natexlab{b}})}]{Hybertsen_1985a}%
  \BibitemOpen
  \bibfield  {author} {\bibinfo {author} {\bibfnamefont {M.~S.}\ \bibnamefont
  {Hybertsen}}\ and\ \bibinfo {author} {\bibfnamefont {S.~G.}\ \bibnamefont
  {Louie}},\ }\href {\doibase 10.1103/PhysRevLett.55.1418} {\bibfield
  {journal} {\bibinfo  {journal} {Phys. Rev. Lett.}\ }\textbf {\bibinfo
  {volume} {55}},\ \bibinfo {pages} {1418} (\bibinfo {year}
  {1985}{\natexlab{b}})}\BibitemShut {NoStop}%
\bibitem [{\citenamefont {von~der Linden}\ and\ \citenamefont
  {Horsch}(1988)}]{Linden_1988}%
  \BibitemOpen
  \bibfield  {author} {\bibinfo {author} {\bibfnamefont {W.}~\bibnamefont
  {von~der Linden}}\ and\ \bibinfo {author} {\bibfnamefont {P.}~\bibnamefont
  {Horsch}},\ }\href {\doibase 10.1103/PhysRevB.37.8351} {\bibfield  {journal}
  {\bibinfo  {journal} {Phys. Rev. B}\ }\textbf {\bibinfo {volume} {37}},\
  \bibinfo {pages} {8351} (\bibinfo {year} {1988})}\BibitemShut {NoStop}%
\bibitem [{\citenamefont {Northrup}, \citenamefont {Hybertsen},\ and\
  \citenamefont {Louie}(1991)}]{Northrup_1991}%
  \BibitemOpen
  \bibfield  {author} {\bibinfo {author} {\bibfnamefont {J.~E.}\ \bibnamefont
  {Northrup}}, \bibinfo {author} {\bibfnamefont {M.~S.}\ \bibnamefont
  {Hybertsen}}, \ and\ \bibinfo {author} {\bibfnamefont {S.~G.}\ \bibnamefont
  {Louie}},\ }\href {\doibase 10.1103/PhysRevLett.66.500} {\bibfield  {journal}
  {\bibinfo  {journal} {Phys. Rev. Lett.}\ }\textbf {\bibinfo {volume} {66}},\
  \bibinfo {pages} {500} (\bibinfo {year} {1991})}\BibitemShut {NoStop}%
\bibitem [{\citenamefont {Blase}, \citenamefont {Zhu},\ and\ \citenamefont
  {Louie}(1994)}]{Blase_1994}%
  \BibitemOpen
  \bibfield  {author} {\bibinfo {author} {\bibfnamefont {X.}~\bibnamefont
  {Blase}}, \bibinfo {author} {\bibfnamefont {X.}~\bibnamefont {Zhu}}, \ and\
  \bibinfo {author} {\bibfnamefont {S.~G.}\ \bibnamefont {Louie}},\ }\href
  {\doibase 10.1103/PhysRevB.49.4973} {\bibfield  {journal} {\bibinfo
  {journal} {Phys. Rev. B}\ }\textbf {\bibinfo {volume} {49}},\ \bibinfo
  {pages} {4973} (\bibinfo {year} {1994})}\BibitemShut {NoStop}%
\bibitem [{\citenamefont {Rohlfing}, \citenamefont {Kr{\"u}ger},\ and\
  \citenamefont {Pollmann}(1995)}]{Rohlfing_1995}%
  \BibitemOpen
  \bibfield  {author} {\bibinfo {author} {\bibfnamefont {M.}~\bibnamefont
  {Rohlfing}}, \bibinfo {author} {\bibfnamefont {P.}~\bibnamefont
  {Kr{\"u}ger}}, \ and\ \bibinfo {author} {\bibfnamefont {J.}~\bibnamefont
  {Pollmann}},\ }\href {\doibase 10.1103/PhysRevB.52.1905} {\bibfield
  {journal} {\bibinfo  {journal} {Phys. Rev. B}\ }\textbf {\bibinfo {volume}
  {52}},\ \bibinfo {pages} {1905} (\bibinfo {year} {1995})}\BibitemShut
  {NoStop}%
\bibitem [{\citenamefont {Shishkin}\ and\ \citenamefont
  {Kresse}(2007)}]{Shishkin_2007a}%
  \BibitemOpen
  \bibfield  {author} {\bibinfo {author} {\bibfnamefont {M.}~\bibnamefont
  {Shishkin}}\ and\ \bibinfo {author} {\bibfnamefont {G.}~\bibnamefont
  {Kresse}},\ }\href {\doibase 10.1103/PhysRevB.75.235102} {\bibfield
  {journal} {\bibinfo  {journal} {Phys. Rev. B}\ }\textbf {\bibinfo {volume}
  {75}},\ \bibinfo {pages} {235102} (\bibinfo {year} {2007})}\BibitemShut
  {NoStop}%
\bibitem [{\citenamefont {Blase}\ and\ \citenamefont
  {Attaccalite}(2011)}]{Blase_2011}%
  \BibitemOpen
  \bibfield  {author} {\bibinfo {author} {\bibfnamefont {X.}~\bibnamefont
  {Blase}}\ and\ \bibinfo {author} {\bibfnamefont {C.}~\bibnamefont
  {Attaccalite}},\ }\href {\doibase 10.1063/1.3655352} {\bibfield  {journal}
  {\bibinfo  {journal} {Appl. Phys. Lett.}\ }\textbf {\bibinfo {volume} {99}},\
  \bibinfo {pages} {171909} (\bibinfo {year} {2011})}\BibitemShut {NoStop}%
\bibitem [{\citenamefont {Faber}\ \emph {et~al.}(2011)\citenamefont {Faber},
  \citenamefont {Attaccalite}, \citenamefont {Olevano}, \citenamefont {Runge},\
  and\ \citenamefont {Blase}}]{Faber_2011}%
  \BibitemOpen
  \bibfield  {author} {\bibinfo {author} {\bibfnamefont {C.}~\bibnamefont
  {Faber}}, \bibinfo {author} {\bibfnamefont {C.}~\bibnamefont {Attaccalite}},
  \bibinfo {author} {\bibfnamefont {V.}~\bibnamefont {Olevano}}, \bibinfo
  {author} {\bibfnamefont {E.}~\bibnamefont {Runge}}, \ and\ \bibinfo {author}
  {\bibfnamefont {X.}~\bibnamefont {Blase}},\ }\href {\doibase
  10.1103/PhysRevB.83.115123} {\bibfield  {journal} {\bibinfo  {journal} {Phys.
  Rev. B}\ }\textbf {\bibinfo {volume} {83}},\ \bibinfo {pages} {115123}
  (\bibinfo {year} {2011})}\BibitemShut {NoStop}%
\bibitem [{\citenamefont {Rangel}\ \emph {et~al.}(2016)\citenamefont {Rangel},
  \citenamefont {Hamed}, \citenamefont {Bruneval},\ and\ \citenamefont
  {Neaton}}]{Rangel_2016}%
  \BibitemOpen
  \bibfield  {author} {\bibinfo {author} {\bibfnamefont {T.}~\bibnamefont
  {Rangel}}, \bibinfo {author} {\bibfnamefont {S.~M.}\ \bibnamefont {Hamed}},
  \bibinfo {author} {\bibfnamefont {F.}~\bibnamefont {Bruneval}}, \ and\
  \bibinfo {author} {\bibfnamefont {J.~B.}\ \bibnamefont {Neaton}},\ }\href
  {\doibase 10.1021/acs.jctc.6b00163} {\bibfield  {journal} {\bibinfo
  {journal} {J. Chem. Theory Comput.}\ }\textbf {\bibinfo {volume} {12}},\
  \bibinfo {pages} {2834} (\bibinfo {year} {2016})}\BibitemShut {NoStop}%
\bibitem [{\citenamefont {Faleev}, \citenamefont {{van Schilfgaarde}},\ and\
  \citenamefont {Kotani}(2004)}]{Faleev_2004}%
  \BibitemOpen
  \bibfield  {author} {\bibinfo {author} {\bibfnamefont {S.~V.}\ \bibnamefont
  {Faleev}}, \bibinfo {author} {\bibfnamefont {M.}~\bibnamefont {{van
  Schilfgaarde}}}, \ and\ \bibinfo {author} {\bibfnamefont {T.}~\bibnamefont
  {Kotani}},\ }\href {\doibase 10.1103/PhysRevLett.93.126406} {\bibfield
  {journal} {\bibinfo  {journal} {Phys. Rev. Lett.}\ }\textbf {\bibinfo
  {volume} {93}},\ \bibinfo {pages} {126406} (\bibinfo {year}
  {2004})}\BibitemShut {NoStop}%
\bibitem [{\citenamefont {{van Schilfgaarde}}, \citenamefont {Kotani},\ and\
  \citenamefont {Faleev}(2006)}]{vanSchilfgaarde_2006}%
  \BibitemOpen
  \bibfield  {author} {\bibinfo {author} {\bibfnamefont {M.}~\bibnamefont {{van
  Schilfgaarde}}}, \bibinfo {author} {\bibfnamefont {T.}~\bibnamefont
  {Kotani}}, \ and\ \bibinfo {author} {\bibfnamefont {S.}~\bibnamefont
  {Faleev}},\ }\href {\doibase 10.1103/PhysRevLett.96.226402} {\bibfield
  {journal} {\bibinfo  {journal} {Phys. Rev. Lett.}\ }\textbf {\bibinfo
  {volume} {96}},\ \bibinfo {pages} {226402} (\bibinfo {year}
  {2006})}\BibitemShut {NoStop}%
\bibitem [{\citenamefont {Kotani}, \citenamefont {{van Schilfgaarde}},\ and\
  \citenamefont {Faleev}(2007)}]{Kotani_2007}%
  \BibitemOpen
  \bibfield  {author} {\bibinfo {author} {\bibfnamefont {T.}~\bibnamefont
  {Kotani}}, \bibinfo {author} {\bibfnamefont {M.}~\bibnamefont {{van
  Schilfgaarde}}}, \ and\ \bibinfo {author} {\bibfnamefont {S.~V.}\
  \bibnamefont {Faleev}},\ }\href {\doibase 10.1103/PhysRevB.76.165106}
  {\bibfield  {journal} {\bibinfo  {journal} {Phys. Rev. B}\ }\textbf {\bibinfo
  {volume} {76}},\ \bibinfo {pages} {165106} (\bibinfo {year}
  {2007})}\BibitemShut {NoStop}%
\bibitem [{\citenamefont {Ke}(2011)}]{Ke_2011}%
  \BibitemOpen
  \bibfield  {author} {\bibinfo {author} {\bibfnamefont {S.-H.}\ \bibnamefont
  {Ke}},\ }\href {\doibase 10.1103/PhysRevB.84.205415} {\bibfield  {journal}
  {\bibinfo  {journal} {Phys. Rev. B}\ }\textbf {\bibinfo {volume} {84}},\
  \bibinfo {pages} {205415} (\bibinfo {year} {2011})}\BibitemShut {NoStop}%
\bibitem [{\citenamefont {Kaplan}\ \emph {et~al.}(2016)\citenamefont {Kaplan},
  \citenamefont {Harding}, \citenamefont {Seiler}, \citenamefont {Weigend},
  \citenamefont {Evers},\ and\ \citenamefont {{van Setten}}}]{Kaplan_2016}%
  \BibitemOpen
  \bibfield  {author} {\bibinfo {author} {\bibfnamefont {F.}~\bibnamefont
  {Kaplan}}, \bibinfo {author} {\bibfnamefont {M.~E.}\ \bibnamefont {Harding}},
  \bibinfo {author} {\bibfnamefont {C.}~\bibnamefont {Seiler}}, \bibinfo
  {author} {\bibfnamefont {F.}~\bibnamefont {Weigend}}, \bibinfo {author}
  {\bibfnamefont {F.}~\bibnamefont {Evers}}, \ and\ \bibinfo {author}
  {\bibfnamefont {M.~J.}\ \bibnamefont {{van Setten}}},\ }\href {\doibase
  10.1021/acs.jctc.5b01238} {\bibfield  {journal} {\bibinfo  {journal} {J.
  Chem. Theory Comput.}\ }\textbf {\bibinfo {volume} {12}},\ \bibinfo {pages}
  {2528} (\bibinfo {year} {2016})}\BibitemShut {NoStop}%
\bibitem [{\citenamefont {Salpeter}\ and\ \citenamefont
  {Bethe}(1951)}]{Salpeter_1951}%
  \BibitemOpen
  \bibfield  {author} {\bibinfo {author} {\bibfnamefont {E.~E.}\ \bibnamefont
  {Salpeter}}\ and\ \bibinfo {author} {\bibfnamefont {H.~A.}\ \bibnamefont
  {Bethe}},\ }\href {\doibase 10.1103/PhysRev.84.1232} {\bibfield  {journal}
  {\bibinfo  {journal} {Phys. Rev.}\ }\textbf {\bibinfo {volume} {84}},\
  \bibinfo {pages} {1232} (\bibinfo {year} {1951})}\BibitemShut {NoStop}%
\bibitem [{\citenamefont {Strinati}(1988)}]{Strinati_1988}%
  \BibitemOpen
  \bibfield  {author} {\bibinfo {author} {\bibfnamefont {G.}~\bibnamefont
  {Strinati}},\ }\href {\doibase 10.1007/BF02725962} {\bibfield  {journal}
  {\bibinfo  {journal} {Riv. Nuovo Cimento}\ }\textbf {\bibinfo {volume}
  {11}},\ \bibinfo {pages} {1} (\bibinfo {year} {1988})}\BibitemShut {NoStop}%
\bibitem [{\citenamefont {Runge}\ and\ \citenamefont
  {Gross}(1984)}]{Runge_1984}%
  \BibitemOpen
  \bibfield  {author} {\bibinfo {author} {\bibfnamefont {E.}~\bibnamefont
  {Runge}}\ and\ \bibinfo {author} {\bibfnamefont {E.~K.~U.}\ \bibnamefont
  {Gross}},\ }\href {\doibase 10.1103/PhysRevLett.52.997} {\bibfield  {journal}
  {\bibinfo  {journal} {Phys. Rev. Lett.}\ }\textbf {\bibinfo {volume} {52}},\
  \bibinfo {pages} {997} (\bibinfo {year} {1984})}\BibitemShut {NoStop}%
\bibitem [{\citenamefont {Casida}(1995)}]{Casida_1995}%
  \BibitemOpen
  \bibfield  {author} {\bibinfo {author} {\bibfnamefont {M.~E.}\ \bibnamefont
  {Casida}},\ }\enquote {\bibinfo {title} {Time-dependent density functional
  response theory for molecules},}\ \ (\bibinfo  {publisher} {World Scientific,
  Singapore},\ \bibinfo {year} {1995})\ pp.\ \bibinfo {pages}
  {155--192}\BibitemShut {NoStop}%
\bibitem [{\citenamefont {Petersilka}, \citenamefont {Gossmann},\ and\
  \citenamefont {Gross}(1996)}]{Petersilka_1996}%
  \BibitemOpen
  \bibfield  {author} {\bibinfo {author} {\bibfnamefont {M.}~\bibnamefont
  {Petersilka}}, \bibinfo {author} {\bibfnamefont {U.~J.}\ \bibnamefont
  {Gossmann}}, \ and\ \bibinfo {author} {\bibfnamefont {E.~K.~U.}\ \bibnamefont
  {Gross}},\ }\href {\doibase 10.1103/PhysRevLett.76.1212} {\bibfield
  {journal} {\bibinfo  {journal} {Phys. Rev. Lett.}\ }\textbf {\bibinfo
  {volume} {76}},\ \bibinfo {pages} {1212} (\bibinfo {year}
  {1996})}\BibitemShut {NoStop}%
\bibitem [{\citenamefont {Ullrich}(2012)}]{UlrichBook}%
  \BibitemOpen
  \bibfield  {author} {\bibinfo {author} {\bibfnamefont {C.}~\bibnamefont
  {Ullrich}},\ }\href@noop {} {\emph {\bibinfo {title} {Time-Dependent
  Density-Functional Theory: Concepts and Applications}}},\ Oxford Graduate
  Texts\ (\bibinfo  {publisher} {Oxford University Press},\ \bibinfo {address}
  {New York},\ \bibinfo {year} {2012})\BibitemShut {NoStop}%
\bibitem [{\citenamefont {Zhang}, \citenamefont {Steinmann},\ and\
  \citenamefont {Yang}(2013)}]{Zhang_2013}%
  \BibitemOpen
  \bibfield  {author} {\bibinfo {author} {\bibfnamefont {D.}~\bibnamefont
  {Zhang}}, \bibinfo {author} {\bibfnamefont {S.~N.}\ \bibnamefont
  {Steinmann}}, \ and\ \bibinfo {author} {\bibfnamefont {W.}~\bibnamefont
  {Yang}},\ }\href {\doibase 10.1063/1.4824907} {\bibfield  {journal} {\bibinfo
   {journal} {J. Chem. Phys.}\ }\textbf {\bibinfo {volume} {139}},\ \bibinfo
  {pages} {154109} (\bibinfo {year} {2013})}\BibitemShut {NoStop}%
\bibitem [{\citenamefont {Rebolini}\ and\ \citenamefont
  {Toulouse}(2016)}]{Rebolini_2016}%
  \BibitemOpen
  \bibfield  {author} {\bibinfo {author} {\bibfnamefont {E.}~\bibnamefont
  {Rebolini}}\ and\ \bibinfo {author} {\bibfnamefont {J.}~\bibnamefont
  {Toulouse}},\ }\href {\doibase 10.1063/1.4943003} {\bibfield  {journal}
  {\bibinfo  {journal} {J. Chem. Phys.}\ }\textbf {\bibinfo {volume} {144}},\
  \bibinfo {pages} {094107} (\bibinfo {year} {2016})}\BibitemShut {NoStop}%
\bibitem [{\citenamefont {Dou}\ \emph {et~al.}(2022)\citenamefont {Dou},
  \citenamefont {Lee}, \citenamefont {Zhu}, \citenamefont {Mej{\'\i}a},
  \citenamefont {Reichman}, \citenamefont {Baer},\ and\ \citenamefont
  {Rabani}}]{Dou_2022}%
  \BibitemOpen
  \bibfield  {author} {\bibinfo {author} {\bibfnamefont {W.}~\bibnamefont
  {Dou}}, \bibinfo {author} {\bibfnamefont {J.}~\bibnamefont {Lee}}, \bibinfo
  {author} {\bibfnamefont {J.}~\bibnamefont {Zhu}}, \bibinfo {author}
  {\bibfnamefont {L.}~\bibnamefont {Mej{\'\i}a}}, \bibinfo {author}
  {\bibfnamefont {D.~R.}\ \bibnamefont {Reichman}}, \bibinfo {author}
  {\bibfnamefont {R.}~\bibnamefont {Baer}}, \ and\ \bibinfo {author}
  {\bibfnamefont {E.}~\bibnamefont {Rabani}},\ }\href {\doibase
  10.1021/acs.jctc.2c00057} {\bibfield  {journal} {\bibinfo  {journal} {J.
  Chem. Theory Comput.}\ }\textbf {\bibinfo {volume} {18}},\ \bibinfo {pages}
  {5221} (\bibinfo {year} {2022})}\BibitemShut {NoStop}%
\bibitem [{\citenamefont {Loos}, \citenamefont {Romaniello},\ and\
  \citenamefont {Berger}(2018)}]{Loos_2018b}%
  \BibitemOpen
  \bibfield  {author} {\bibinfo {author} {\bibfnamefont {P.~F.}\ \bibnamefont
  {Loos}}, \bibinfo {author} {\bibfnamefont {P.}~\bibnamefont {Romaniello}}, \
  and\ \bibinfo {author} {\bibfnamefont {J.~A.}\ \bibnamefont {Berger}},\
  }\href {\doibase 10.1021/acs.jctc.8b00260} {\bibfield  {journal} {\bibinfo
  {journal} {J. Chem. Theory Comput.}\ }\textbf {\bibinfo {volume} {14}},\
  \bibinfo {pages} {3071} (\bibinfo {year} {2018})}\BibitemShut {NoStop}%
\bibitem [{\citenamefont {V{\'e}ril}\ \emph {et~al.}(2018)\citenamefont
  {V{\'e}ril}, \citenamefont {Romaniello}, \citenamefont {Berger},\ and\
  \citenamefont {Loos}}]{Veril_2018}%
  \BibitemOpen
  \bibfield  {author} {\bibinfo {author} {\bibfnamefont {M.}~\bibnamefont
  {V{\'e}ril}}, \bibinfo {author} {\bibfnamefont {P.}~\bibnamefont
  {Romaniello}}, \bibinfo {author} {\bibfnamefont {J.~A.}\ \bibnamefont
  {Berger}}, \ and\ \bibinfo {author} {\bibfnamefont {P.~F.}\ \bibnamefont
  {Loos}},\ }\href {\doibase 10.1021/acs.jctc.8b00745} {\bibfield  {journal}
  {\bibinfo  {journal} {J. Chem. Theory Comput.}\ }\textbf {\bibinfo {volume}
  {14}},\ \bibinfo {pages} {5220} (\bibinfo {year} {2018})}\BibitemShut
  {NoStop}%
\bibitem [{\citenamefont {Berger}, \citenamefont {Loos},\ and\ \citenamefont
  {Romaniello}(2020)}]{Berger_2021}%
  \BibitemOpen
  \bibfield  {author} {\bibinfo {author} {\bibfnamefont {J.~A.}\ \bibnamefont
  {Berger}}, \bibinfo {author} {\bibfnamefont {P.-F.}\ \bibnamefont {Loos}}, \
  and\ \bibinfo {author} {\bibfnamefont {P.}~\bibnamefont {Romaniello}},\
  }\href {\doibase 10.1021/acs.jctc.0c00896} {\bibfield  {journal} {\bibinfo
  {journal} {J. Chem. Theory Comput.}\ }\textbf {\bibinfo {volume} {17}},\
  \bibinfo {pages} {191} (\bibinfo {year} {2020})}\BibitemShut {NoStop}%
\bibitem [{\citenamefont {Di~Sabatino}, \citenamefont {Loos},\ and\
  \citenamefont {Romaniello}(2021)}]{DiSabatino_2021}%
  \BibitemOpen
  \bibfield  {author} {\bibinfo {author} {\bibfnamefont {S.}~\bibnamefont
  {Di~Sabatino}}, \bibinfo {author} {\bibfnamefont {P.-F.}\ \bibnamefont
  {Loos}}, \ and\ \bibinfo {author} {\bibfnamefont {P.}~\bibnamefont
  {Romaniello}},\ }\href {\doibase 10.3389/fchem.2021.751054} {\bibfield
  {journal} {\bibinfo  {journal} {Front. Chem.}\ }\textbf {\bibinfo {volume}
  {9}},\ \bibinfo {pages} {751054} (\bibinfo {year} {2021})}\BibitemShut
  {NoStop}%
\bibitem [{\citenamefont {Hedin}(1965)}]{Hedin_1965}%
  \BibitemOpen
  \bibfield  {author} {\bibinfo {author} {\bibfnamefont {L.}~\bibnamefont
  {Hedin}},\ }\href {\doibase 10.1103/PhysRev.139.A796} {\bibfield  {journal}
  {\bibinfo  {journal} {Phys. Rev.}\ }\textbf {\bibinfo {volume} {139}},\
  \bibinfo {pages} {A796} (\bibinfo {year} {1965})}\BibitemShut {NoStop}%
\bibitem [{\citenamefont {Bruneval}, \citenamefont {Vast},\ and\ \citenamefont
  {Reining}(2006)}]{Bruneval_2006}%
  \BibitemOpen
  \bibfield  {author} {\bibinfo {author} {\bibfnamefont {F.}~\bibnamefont
  {Bruneval}}, \bibinfo {author} {\bibfnamefont {N.}~\bibnamefont {Vast}}, \
  and\ \bibinfo {author} {\bibfnamefont {L.}~\bibnamefont {Reining}},\ }\href
  {\doibase 10.1103/PhysRevB.74.045102} {\bibfield  {journal} {\bibinfo
  {journal} {Phys. Rev. B}\ }\textbf {\bibinfo {volume} {74}},\ \bibinfo
  {pages} {045102} (\bibinfo {year} {2006})}\BibitemShut {NoStop}%
\bibitem [{\citenamefont {Koval}, \citenamefont {Foerster},\ and\ \citenamefont
  {S{\'a}nchez-Portal}(2014)}]{Koval_2014}%
  \BibitemOpen
  \bibfield  {author} {\bibinfo {author} {\bibfnamefont {P.}~\bibnamefont
  {Koval}}, \bibinfo {author} {\bibfnamefont {D.}~\bibnamefont {Foerster}}, \
  and\ \bibinfo {author} {\bibfnamefont {D.}~\bibnamefont
  {S{\'a}nchez-Portal}},\ }\href {\doibase 10.1103/PhysRevB.89.155417}
  {\bibfield  {journal} {\bibinfo  {journal} {Phys. Rev. B}\ }\textbf {\bibinfo
  {volume} {89}},\ \bibinfo {pages} {155417} (\bibinfo {year}
  {2014})}\BibitemShut {NoStop}%
\bibitem [{\citenamefont {Wilhelm}\ \emph {et~al.}(2018)\citenamefont
  {Wilhelm}, \citenamefont {Golze}, \citenamefont {Talirz}, \citenamefont
  {Hutter},\ and\ \citenamefont {Pignedoli}}]{Wilhelm_2018}%
  \BibitemOpen
  \bibfield  {author} {\bibinfo {author} {\bibfnamefont {J.}~\bibnamefont
  {Wilhelm}}, \bibinfo {author} {\bibfnamefont {D.}~\bibnamefont {Golze}},
  \bibinfo {author} {\bibfnamefont {L.}~\bibnamefont {Talirz}}, \bibinfo
  {author} {\bibfnamefont {J.}~\bibnamefont {Hutter}}, \ and\ \bibinfo {author}
  {\bibfnamefont {C.~A.}\ \bibnamefont {Pignedoli}},\ }\href {\doibase
  10.1021/acs.jpclett.7b02740} {\bibfield  {journal} {\bibinfo  {journal} {J.
  Phys. Chem. Lett.}\ }\textbf {\bibinfo {volume} {9}},\ \bibinfo {pages} {306}
  (\bibinfo {year} {2018})}\BibitemShut {NoStop}%
\bibitem [{\citenamefont {Stan}, \citenamefont {Dahlen},\ and\ \citenamefont
  {van Leeuwen}(2006)}]{Stan_2006}%
  \BibitemOpen
  \bibfield  {author} {\bibinfo {author} {\bibfnamefont {A.}~\bibnamefont
  {Stan}}, \bibinfo {author} {\bibfnamefont {N.~E.}\ \bibnamefont {Dahlen}}, \
  and\ \bibinfo {author} {\bibfnamefont {R.}~\bibnamefont {van Leeuwen}},\
  }\href {\doibase 10.1209/epl/i2006-10266-6} {\bibfield  {journal} {\bibinfo
  {journal} {Europhys. Lett. EPL}\ }\textbf {\bibinfo {volume} {76}},\ \bibinfo
  {pages} {298} (\bibinfo {year} {2006})}\BibitemShut {NoStop}%
\bibitem [{\citenamefont {Stan}, \citenamefont {Dahlen},\ and\ \citenamefont
  {{van Leeuwen}}(2009)}]{Stan_2009}%
  \BibitemOpen
  \bibfield  {author} {\bibinfo {author} {\bibfnamefont {A.}~\bibnamefont
  {Stan}}, \bibinfo {author} {\bibfnamefont {N.~E.}\ \bibnamefont {Dahlen}}, \
  and\ \bibinfo {author} {\bibfnamefont {R.}~\bibnamefont {{van Leeuwen}}},\
  }\href {\doibase 10.1063/1.3089567} {\bibfield  {journal} {\bibinfo
  {journal} {J. Chem. Phys.}\ }\textbf {\bibinfo {volume} {130}},\ \bibinfo
  {pages} {114105} (\bibinfo {year} {2009})}\BibitemShut {NoStop}%
\bibitem [{\citenamefont {Rostgaard}, \citenamefont {Jacobsen},\ and\
  \citenamefont {Thygesen}(2010)}]{Rostgaard_2010}%
  \BibitemOpen
  \bibfield  {author} {\bibinfo {author} {\bibfnamefont {C.}~\bibnamefont
  {Rostgaard}}, \bibinfo {author} {\bibfnamefont {K.~W.}\ \bibnamefont
  {Jacobsen}}, \ and\ \bibinfo {author} {\bibfnamefont {K.~S.}\ \bibnamefont
  {Thygesen}},\ }\href {\doibase 10.1103/PhysRevB.81.085103} {\bibfield
  {journal} {\bibinfo  {journal} {Phys. Rev. B}\ }\textbf {\bibinfo {volume}
  {81}},\ \bibinfo {pages} {085103} (\bibinfo {year} {2010})}\BibitemShut
  {NoStop}%
\bibitem [{\citenamefont {Caruso}\ \emph {et~al.}(2012)\citenamefont {Caruso},
  \citenamefont {Rinke}, \citenamefont {Ren}, \citenamefont {Scheffler},\ and\
  \citenamefont {Rubio}}]{Caruso_2012}%
  \BibitemOpen
  \bibfield  {author} {\bibinfo {author} {\bibfnamefont {F.}~\bibnamefont
  {Caruso}}, \bibinfo {author} {\bibfnamefont {P.}~\bibnamefont {Rinke}},
  \bibinfo {author} {\bibfnamefont {X.}~\bibnamefont {Ren}}, \bibinfo {author}
  {\bibfnamefont {M.}~\bibnamefont {Scheffler}}, \ and\ \bibinfo {author}
  {\bibfnamefont {A.}~\bibnamefont {Rubio}},\ }\href {\doibase
  10.1103/PhysRevB.86.081102} {\bibfield  {journal} {\bibinfo  {journal} {Phys.
  Rev. B}\ }\textbf {\bibinfo {volume} {86}},\ \bibinfo {pages} {081102(R)}
  (\bibinfo {year} {2012})}\BibitemShut {NoStop}%
\bibitem [{\citenamefont {Caruso}\ \emph
  {et~al.}(2013{\natexlab{a}})\citenamefont {Caruso}, \citenamefont {Rinke},
  \citenamefont {Ren}, \citenamefont {Rubio},\ and\ \citenamefont
  {Scheffler}}]{Caruso_2013a}%
  \BibitemOpen
  \bibfield  {author} {\bibinfo {author} {\bibfnamefont {F.}~\bibnamefont
  {Caruso}}, \bibinfo {author} {\bibfnamefont {P.}~\bibnamefont {Rinke}},
  \bibinfo {author} {\bibfnamefont {X.}~\bibnamefont {Ren}}, \bibinfo {author}
  {\bibfnamefont {A.}~\bibnamefont {Rubio}}, \ and\ \bibinfo {author}
  {\bibfnamefont {M.}~\bibnamefont {Scheffler}},\ }\href {\doibase
  10.1103/PhysRevB.88.075105} {\bibfield  {journal} {\bibinfo  {journal} {Phys.
  Rev. B}\ }\textbf {\bibinfo {volume} {88}},\ \bibinfo {pages} {075105}
  (\bibinfo {year} {2013}{\natexlab{a}})}\BibitemShut {NoStop}%
\bibitem [{\citenamefont {Caruso}(2013)}]{Caruso_2013b}%
  \BibitemOpen
  \bibfield  {author} {\bibinfo {author} {\bibfnamefont {F.}~\bibnamefont
  {Caruso}},\ }\emph {\bibinfo {title} {Self-Consistent {{GW}} Approach for the
  Unified Description of Ground and Excited States of Finite Systems}},\
  \href@noop {} {\bibinfo {type} {{{PhD Thesis}}}},\ \bibinfo  {school} {Freie
  Universit{\"a}t Berlin} (\bibinfo {year} {2013})\BibitemShut {NoStop}%
\bibitem [{\citenamefont {Caruso}\ \emph
  {et~al.}(2013{\natexlab{b}})\citenamefont {Caruso}, \citenamefont {Rohr},
  \citenamefont {Hellgren}, \citenamefont {Ren}, \citenamefont {Rinke},
  \citenamefont {Rubio},\ and\ \citenamefont {Scheffler}}]{Caruso_2013}%
  \BibitemOpen
  \bibfield  {author} {\bibinfo {author} {\bibfnamefont {F.}~\bibnamefont
  {Caruso}}, \bibinfo {author} {\bibfnamefont {D.~R.}\ \bibnamefont {Rohr}},
  \bibinfo {author} {\bibfnamefont {M.}~\bibnamefont {Hellgren}}, \bibinfo
  {author} {\bibfnamefont {X.}~\bibnamefont {Ren}}, \bibinfo {author}
  {\bibfnamefont {P.}~\bibnamefont {Rinke}}, \bibinfo {author} {\bibfnamefont
  {A.}~\bibnamefont {Rubio}}, \ and\ \bibinfo {author} {\bibfnamefont
  {M.}~\bibnamefont {Scheffler}},\ }\href {\doibase
  10.1103/PhysRevLett.110.146403} {\bibfield  {journal} {\bibinfo  {journal}
  {Phys. Rev. Lett.}\ }\textbf {\bibinfo {volume} {110}},\ \bibinfo {pages}
  {146403} (\bibinfo {year} {2013}{\natexlab{b}})}\BibitemShut {NoStop}%
\bibitem [{\citenamefont {Scott}, \citenamefont {Backhouse},\ and\
  \citenamefont {Booth}(2023)}]{Scott_2023}%
  \BibitemOpen
  \bibfield  {author} {\bibinfo {author} {\bibfnamefont {C.~J.~C.}\
  \bibnamefont {Scott}}, \bibinfo {author} {\bibfnamefont {O.~J.}\ \bibnamefont
  {Backhouse}}, \ and\ \bibinfo {author} {\bibfnamefont {G.~H.}\ \bibnamefont
  {Booth}},\ }\href {\doibase 10.1063/5.0143291} {\bibfield  {journal}
  {\bibinfo  {journal} {J. Chem. Phys.}\ }\textbf {\bibinfo {volume} {158}},\
  \bibinfo {pages} {124102} (\bibinfo {year} {2023})}\BibitemShut {NoStop}%
\bibitem [{\citenamefont {Monino}\ and\ \citenamefont
  {Loos}(2022)}]{Monino_2022}%
  \BibitemOpen
  \bibfield  {author} {\bibinfo {author} {\bibfnamefont {E.}~\bibnamefont
  {Monino}}\ and\ \bibinfo {author} {\bibfnamefont {P.-F.}\ \bibnamefont
  {Loos}},\ }\href {\doibase 10.1063/5.0089317} {\bibfield  {journal} {\bibinfo
   {journal} {J. Chem. Phys.}\ }\textbf {\bibinfo {volume} {156}},\ \bibinfo
  {pages} {231101} (\bibinfo {year} {2022})}\BibitemShut {NoStop}%
\bibitem [{\citenamefont {Marie}\ and\ \citenamefont
  {Loos}(2023)}]{Marie_2023}%
  \BibitemOpen
  \bibfield  {author} {\bibinfo {author} {\bibfnamefont {A.}~\bibnamefont
  {Marie}}\ and\ \bibinfo {author} {\bibfnamefont {P.-F.}\ \bibnamefont
  {Loos}},\ }\href@noop {} {\enquote {\bibinfo {title} {A similarity
  renormalization group approach to green's function methods},}\ } (\bibinfo
  {year} {2023}),\ \Eprint {http://arxiv.org/abs/2303.05984} {arXiv:2303.05984
  [physics.chem-ph]} \BibitemShut {NoStop}%
\bibitem [{\citenamefont {Ankudinov}, \citenamefont {Nesvizhskii},\ and\
  \citenamefont {Rehr}(2003)}]{Ankudinov_2003}%
  \BibitemOpen
  \bibfield  {author} {\bibinfo {author} {\bibfnamefont {A.~L.}\ \bibnamefont
  {Ankudinov}}, \bibinfo {author} {\bibfnamefont {A.~I.}\ \bibnamefont
  {Nesvizhskii}}, \ and\ \bibinfo {author} {\bibfnamefont {J.~J.}\ \bibnamefont
  {Rehr}},\ }\href {\doibase 10.1103/PhysRevB.67.115120} {\bibfield  {journal}
  {\bibinfo  {journal} {Phys. Rev. B}\ }\textbf {\bibinfo {volume} {67}},\
  \bibinfo {pages} {115120} (\bibinfo {year} {2003})}\BibitemShut {NoStop}%
\bibitem [{\citenamefont {Romaniello}\ \emph {et~al.}(2009)\citenamefont
  {Romaniello}, \citenamefont {Sangalli}, \citenamefont {Berger}, \citenamefont
  {Sottile}, \citenamefont {Molinari}, \citenamefont {Reining},\ and\
  \citenamefont {Onida}}]{Romaniello_2009b}%
  \BibitemOpen
  \bibfield  {author} {\bibinfo {author} {\bibfnamefont {P.}~\bibnamefont
  {Romaniello}}, \bibinfo {author} {\bibfnamefont {D.}~\bibnamefont
  {Sangalli}}, \bibinfo {author} {\bibfnamefont {J.~A.}\ \bibnamefont
  {Berger}}, \bibinfo {author} {\bibfnamefont {F.}~\bibnamefont {Sottile}},
  \bibinfo {author} {\bibfnamefont {L.~G.}\ \bibnamefont {Molinari}}, \bibinfo
  {author} {\bibfnamefont {L.}~\bibnamefont {Reining}}, \ and\ \bibinfo
  {author} {\bibfnamefont {G.}~\bibnamefont {Onida}},\ }\href {\doibase
  10.1063/1.3065669} {\bibfield  {journal} {\bibinfo  {journal} {J. Chem.
  Phys.}\ }\textbf {\bibinfo {volume} {130}},\ \bibinfo {pages} {044108}
  (\bibinfo {year} {2009})}\BibitemShut {NoStop}%
\bibitem [{\citenamefont {Sangalli}\ \emph {et~al.}(2011)\citenamefont
  {Sangalli}, \citenamefont {Romaniello}, \citenamefont {Onida},\ and\
  \citenamefont {Marini}}]{Sangalli_2011}%
  \BibitemOpen
  \bibfield  {author} {\bibinfo {author} {\bibfnamefont {D.}~\bibnamefont
  {Sangalli}}, \bibinfo {author} {\bibfnamefont {P.}~\bibnamefont
  {Romaniello}}, \bibinfo {author} {\bibfnamefont {G.}~\bibnamefont {Onida}}, \
  and\ \bibinfo {author} {\bibfnamefont {A.}~\bibnamefont {Marini}},\ }\href
  {\doibase 10.1063/1.3518705} {\bibfield  {journal} {\bibinfo  {journal} {J.
  Chem. Phys.}\ }\textbf {\bibinfo {volume} {134}},\ \bibinfo {pages} {034115}
  (\bibinfo {year} {2011})}\BibitemShut {NoStop}%
\bibitem [{\citenamefont {Loos}\ and\ \citenamefont
  {Blase}(2020)}]{Loos_2020h}%
  \BibitemOpen
  \bibfield  {author} {\bibinfo {author} {\bibfnamefont {P.-F.}\ \bibnamefont
  {Loos}}\ and\ \bibinfo {author} {\bibfnamefont {X.}~\bibnamefont {Blase}},\
  }\href {\doibase 10.1063/5.0023168} {\bibfield  {journal} {\bibinfo
  {journal} {J. Chem. Phys.}\ }\textbf {\bibinfo {volume} {153}},\ \bibinfo
  {pages} {114120} (\bibinfo {year} {2020})}\BibitemShut {NoStop}%
\bibitem [{\citenamefont {Authier}\ and\ \citenamefont
  {Loos}(2020)}]{Authier_2020}%
  \BibitemOpen
  \bibfield  {author} {\bibinfo {author} {\bibfnamefont {J.}~\bibnamefont
  {Authier}}\ and\ \bibinfo {author} {\bibfnamefont {P.-F.}\ \bibnamefont
  {Loos}},\ }\href {\doibase 10.1063/5.0028040} {\bibfield  {journal} {\bibinfo
   {journal} {J. Chem. Phys.}\ }\textbf {\bibinfo {volume} {153}},\ \bibinfo
  {pages} {184105} (\bibinfo {year} {2020})}\BibitemShut {NoStop}%
\bibitem [{\citenamefont {Levine}\ \emph {et~al.}(2006)\citenamefont {Levine},
  \citenamefont {Ko}, \citenamefont {Quenneville},\ and\ \citenamefont
  {Martinez}}]{Levine_2006}%
  \BibitemOpen
  \bibfield  {author} {\bibinfo {author} {\bibfnamefont {B.~G.}\ \bibnamefont
  {Levine}}, \bibinfo {author} {\bibfnamefont {C.}~\bibnamefont {Ko}}, \bibinfo
  {author} {\bibfnamefont {J.}~\bibnamefont {Quenneville}}, \ and\ \bibinfo
  {author} {\bibfnamefont {T.~J.}\ \bibnamefont {Martinez}},\ }\href {\doibase
  10.1080/00268970500417762} {\bibfield  {journal} {\bibinfo  {journal} {Mol.
  Phys.}\ }\textbf {\bibinfo {volume} {104}},\ \bibinfo {pages} {1039}
  (\bibinfo {year} {2006})}\BibitemShut {NoStop}%
\bibitem [{\citenamefont {Tozer}\ and\ \citenamefont
  {Handy}(2000)}]{Tozer_2000}%
  \BibitemOpen
  \bibfield  {author} {\bibinfo {author} {\bibfnamefont {D.~J.}\ \bibnamefont
  {Tozer}}\ and\ \bibinfo {author} {\bibfnamefont {N.~C.}\ \bibnamefont
  {Handy}},\ }\href {\doibase 10.1039/a910321j} {\bibfield  {journal} {\bibinfo
   {journal} {Phys. Chem. Chem. Phys.}\ }\textbf {\bibinfo {volume} {2}},\
  \bibinfo {pages} {2117} (\bibinfo {year} {2000})}\BibitemShut {NoStop}%
\bibitem [{\citenamefont {Elliott}\ \emph {et~al.}(2011)\citenamefont
  {Elliott}, \citenamefont {Goldson}, \citenamefont {Canahui},\ and\
  \citenamefont {Maitra}}]{Elliott_2011}%
  \BibitemOpen
  \bibfield  {author} {\bibinfo {author} {\bibfnamefont {P.}~\bibnamefont
  {Elliott}}, \bibinfo {author} {\bibfnamefont {S.}~\bibnamefont {Goldson}},
  \bibinfo {author} {\bibfnamefont {C.}~\bibnamefont {Canahui}}, \ and\
  \bibinfo {author} {\bibfnamefont {N.~T.}\ \bibnamefont {Maitra}},\ }\href
  {\doibase 10.1016/j.chemphys.2011.03.020} {\bibfield  {journal} {\bibinfo
  {journal} {Chem. Phys.}\ }\textbf {\bibinfo {volume} {391}},\ \bibinfo
  {pages} {110} (\bibinfo {year} {2011})}\BibitemShut {NoStop}%
\bibitem [{\citenamefont {Maitra}(2012)}]{Maitra_2012}%
  \BibitemOpen
  \bibfield  {author} {\bibinfo {author} {\bibfnamefont {N.~T.}\ \bibnamefont
  {Maitra}},\ }\enquote {\bibinfo {title} {Memory: History , initial-state
  dependence , and double-excitations},}\ in\ \href {\doibase
  10.1007/978-3-642-23518-4_8} {\emph {\bibinfo {booktitle} {Fundamentals of
  Time-Dependent Density Functional Theory}}},\ Vol.\ \bibinfo {volume} {837},\
  \bibinfo {editor} {edited by\ \bibinfo {editor} {\bibfnamefont {M.~A.}\
  \bibnamefont {Marques}}, \bibinfo {editor} {\bibfnamefont {N.~T.}\
  \bibnamefont {Maitra}}, \bibinfo {editor} {\bibfnamefont {F.~M.}\
  \bibnamefont {Nogueira}}, \bibinfo {editor} {\bibfnamefont {E.}~\bibnamefont
  {Gross}}, \ and\ \bibinfo {editor} {\bibfnamefont {A.}~\bibnamefont
  {Rubio}}}\ (\bibinfo  {publisher} {Springer Berlin Heidelberg},\ \bibinfo
  {address} {Berlin, Heidelberg},\ \bibinfo {year} {2012})\ pp.\ \bibinfo
  {pages} {167--184}\BibitemShut {NoStop}%
\bibitem [{\citenamefont {Maitra}(2016)}]{Maitra_2016}%
  \BibitemOpen
  \bibfield  {author} {\bibinfo {author} {\bibfnamefont {N.~T.}\ \bibnamefont
  {Maitra}},\ }\href {\doibase 10.1063/1.4953039} {\bibfield  {journal}
  {\bibinfo  {journal} {J. Chem. Phys.}\ }\textbf {\bibinfo {volume} {144}},\
  \bibinfo {pages} {220901} (\bibinfo {year} {2016})}\BibitemShut {NoStop}%
\bibitem [{\citenamefont {Strinati}(1984)}]{Strinati_1984}%
  \BibitemOpen
  \bibfield  {author} {\bibinfo {author} {\bibfnamefont {G.}~\bibnamefont
  {Strinati}},\ }\href {\doibase 10.1103/PhysRevB.29.5718} {\bibfield
  {journal} {\bibinfo  {journal} {Phys. Rev. B}\ }\textbf {\bibinfo {volume}
  {29}},\ \bibinfo {pages} {5718} (\bibinfo {year} {1984})}\BibitemShut
  {NoStop}%
\bibitem [{\citenamefont {Rohlfing}\ and\ \citenamefont
  {Louie}(2000)}]{Rohlfing_2000}%
  \BibitemOpen
  \bibfield  {author} {\bibinfo {author} {\bibfnamefont {M.}~\bibnamefont
  {Rohlfing}}\ and\ \bibinfo {author} {\bibfnamefont {S.~G.}\ \bibnamefont
  {Louie}},\ }\href {\doibase 10.1103/PhysRevB.62.4927} {\bibfield  {journal}
  {\bibinfo  {journal} {Phys. Rev. B}\ }\textbf {\bibinfo {volume} {62}},\
  \bibinfo {pages} {4927} (\bibinfo {year} {2000})}\BibitemShut {NoStop}%
\bibitem [{\citenamefont {Ma}, \citenamefont {Rohlfing},\ and\ \citenamefont
  {Molteni}(2009{\natexlab{a}})}]{Ma_2009a}%
  \BibitemOpen
  \bibfield  {author} {\bibinfo {author} {\bibfnamefont {Y.}~\bibnamefont
  {Ma}}, \bibinfo {author} {\bibfnamefont {M.}~\bibnamefont {Rohlfing}}, \ and\
  \bibinfo {author} {\bibfnamefont {C.}~\bibnamefont {Molteni}},\ }\href
  {\doibase 10.1103/PhysRevB.80.241405} {\bibfield  {journal} {\bibinfo
  {journal} {Phys. Rev. B}\ }\textbf {\bibinfo {volume} {80}},\ \bibinfo
  {pages} {241405} (\bibinfo {year} {2009}{\natexlab{a}})}\BibitemShut
  {NoStop}%
\bibitem [{\citenamefont {Ma}, \citenamefont {Rohlfing},\ and\ \citenamefont
  {Molteni}(2009{\natexlab{b}})}]{Ma_2009b}%
  \BibitemOpen
  \bibfield  {author} {\bibinfo {author} {\bibfnamefont {Y.}~\bibnamefont
  {Ma}}, \bibinfo {author} {\bibfnamefont {M.}~\bibnamefont {Rohlfing}}, \ and\
  \bibinfo {author} {\bibfnamefont {C.}~\bibnamefont {Molteni}},\ }\href
  {\doibase 10.1021/ct900528h} {\bibfield  {journal} {\bibinfo  {journal} {J.
  Chem. Theory. Comput.}\ }\textbf {\bibinfo {volume} {6}},\ \bibinfo {pages}
  {257} (\bibinfo {year} {2009}{\natexlab{b}})}\BibitemShut {NoStop}%
\bibitem [{\citenamefont {Kaczmarski}, \citenamefont {Ma},\ and\ \citenamefont
  {Rohlfing}(2010)}]{Kaczmarski_2010}%
  \BibitemOpen
  \bibfield  {author} {\bibinfo {author} {\bibfnamefont {M.~S.}\ \bibnamefont
  {Kaczmarski}}, \bibinfo {author} {\bibfnamefont {Y.}~\bibnamefont {Ma}}, \
  and\ \bibinfo {author} {\bibfnamefont {M.}~\bibnamefont {Rohlfing}},\ }\href
  {\doibase 10.1103/PhysRevB.81.115433} {\bibfield  {journal} {\bibinfo
  {journal} {Phys. Rev. B}\ }\textbf {\bibinfo {volume} {81}},\ \bibinfo
  {pages} {115433} (\bibinfo {year} {2010})}\BibitemShut {NoStop}%
\bibitem [{\citenamefont {Baumeier}, \citenamefont {Andrienko},\ and\
  \citenamefont {Rohlfing}(2012)}]{Baumeier_2012a}%
  \BibitemOpen
  \bibfield  {author} {\bibinfo {author} {\bibfnamefont {B.}~\bibnamefont
  {Baumeier}}, \bibinfo {author} {\bibfnamefont {D.}~\bibnamefont {Andrienko}},
  \ and\ \bibinfo {author} {\bibfnamefont {M.}~\bibnamefont {Rohlfing}},\
  }\href {\doibase 10.1021/ct300311x} {\bibfield  {journal} {\bibinfo
  {journal} {J. Chem. Theory Comput.}\ }\textbf {\bibinfo {volume} {8}},\
  \bibinfo {pages} {2790} (\bibinfo {year} {2012})}\BibitemShut {NoStop}%
\bibitem [{\citenamefont {Baumeier}\ \emph {et~al.}(2012)\citenamefont
  {Baumeier}, \citenamefont {Andrienko}, \citenamefont {Ma},\ and\
  \citenamefont {Rohlfing}}]{Baumeier_2012b}%
  \BibitemOpen
  \bibfield  {author} {\bibinfo {author} {\bibfnamefont {B.}~\bibnamefont
  {Baumeier}}, \bibinfo {author} {\bibfnamefont {D.}~\bibnamefont {Andrienko}},
  \bibinfo {author} {\bibfnamefont {Y.}~\bibnamefont {Ma}}, \ and\ \bibinfo
  {author} {\bibfnamefont {M.}~\bibnamefont {Rohlfing}},\ }\href {\doibase
  10.1021/ct2008999} {\bibfield  {journal} {\bibinfo  {journal} {J. Chem.
  Theory Comput.}\ }\textbf {\bibinfo {volume} {8}},\ \bibinfo {pages} {997}
  (\bibinfo {year} {2012})}\BibitemShut {NoStop}%
\bibitem [{\citenamefont {Rohlfing}(2012)}]{Rohlfing_2012}%
  \BibitemOpen
  \bibfield  {author} {\bibinfo {author} {\bibfnamefont {M.}~\bibnamefont
  {Rohlfing}},\ }\href {\doibase 10.1103/PhysRevLett.108.087402} {\bibfield
  {journal} {\bibinfo  {journal} {Phys. Rev. Lett.}\ }\textbf {\bibinfo
  {volume} {108}},\ \bibinfo {pages} {087402} (\bibinfo {year}
  {2012})}\BibitemShut {NoStop}%
\bibitem [{\citenamefont {Lettmann}\ and\ \citenamefont
  {Rohlfing}(2019)}]{Lettmann_2019}%
  \BibitemOpen
  \bibfield  {author} {\bibinfo {author} {\bibfnamefont {T.}~\bibnamefont
  {Lettmann}}\ and\ \bibinfo {author} {\bibfnamefont {M.}~\bibnamefont
  {Rohlfing}},\ }\href {\doibase 10.1021/acs.jctc.9b00223} {\bibfield
  {journal} {\bibinfo  {journal} {J. Chem. Theory Comput.}\ }\textbf {\bibinfo
  {volume} {15}},\ \bibinfo {pages} {4547} (\bibinfo {year}
  {2019})}\BibitemShut {NoStop}%
\bibitem [{\citenamefont {Bohm}\ and\ \citenamefont {Pines}(1951)}]{Bohm_1951}%
  \BibitemOpen
  \bibfield  {author} {\bibinfo {author} {\bibfnamefont {D.}~\bibnamefont
  {Bohm}}\ and\ \bibinfo {author} {\bibfnamefont {D.}~\bibnamefont {Pines}},\
  }\href {\doibase 10.1103/PhysRev.82.625} {\bibfield  {journal} {\bibinfo
  {journal} {Phys. Rev.}\ }\textbf {\bibinfo {volume} {82}},\ \bibinfo {pages}
  {625} (\bibinfo {year} {1951})}\BibitemShut {NoStop}%
\bibitem [{\citenamefont {Pines}\ and\ \citenamefont
  {Bohm}(1952)}]{Pines_1952}%
  \BibitemOpen
  \bibfield  {author} {\bibinfo {author} {\bibfnamefont {D.}~\bibnamefont
  {Pines}}\ and\ \bibinfo {author} {\bibfnamefont {D.}~\bibnamefont {Bohm}},\
  }\href {\doibase 10.1103/PhysRev.85.338} {\bibfield  {journal} {\bibinfo
  {journal} {Phys. Rev.}\ }\textbf {\bibinfo {volume} {85}},\ \bibinfo {pages}
  {338} (\bibinfo {year} {1952})}\BibitemShut {NoStop}%
\bibitem [{\citenamefont {Bohm}\ and\ \citenamefont {Pines}(1953)}]{Bohm_1953}%
  \BibitemOpen
  \bibfield  {author} {\bibinfo {author} {\bibfnamefont {D.}~\bibnamefont
  {Bohm}}\ and\ \bibinfo {author} {\bibfnamefont {D.}~\bibnamefont {Pines}},\
  }\href {\doibase 10.1103/PhysRev.92.609} {\bibfield  {journal} {\bibinfo
  {journal} {Phys. Rev.}\ }\textbf {\bibinfo {volume} {92}},\ \bibinfo {pages}
  {609} (\bibinfo {year} {1953})}\BibitemShut {NoStop}%
\bibitem [{\citenamefont {Ren}\ \emph {et~al.}(2012)\citenamefont {Ren},
  \citenamefont {Rinke}, \citenamefont {Joas},\ and\ \citenamefont
  {Scheffler}}]{Ren_2012}%
  \BibitemOpen
  \bibfield  {author} {\bibinfo {author} {\bibfnamefont {X.}~\bibnamefont
  {Ren}}, \bibinfo {author} {\bibfnamefont {P.}~\bibnamefont {Rinke}}, \bibinfo
  {author} {\bibfnamefont {C.}~\bibnamefont {Joas}}, \ and\ \bibinfo {author}
  {\bibfnamefont {M.}~\bibnamefont {Scheffler}},\ }\href {\doibase
  10.1007/s10853-012-6570-4} {\bibfield  {journal} {\bibinfo  {journal} {J.
  Mater. Sci.}\ }\textbf {\bibinfo {volume} {47}},\ \bibinfo {pages} {7447}
  (\bibinfo {year} {2012})}\BibitemShut {NoStop}%
\bibitem [{\citenamefont {Chen}\ \emph {et~al.}(2017)\citenamefont {Chen},
  \citenamefont {Voora}, \citenamefont {Agee}, \citenamefont {Balasubramani},\
  and\ \citenamefont {Furche}}]{Chen_2017}%
  \BibitemOpen
  \bibfield  {author} {\bibinfo {author} {\bibfnamefont {G.~P.}\ \bibnamefont
  {Chen}}, \bibinfo {author} {\bibfnamefont {V.~K.}\ \bibnamefont {Voora}},
  \bibinfo {author} {\bibfnamefont {M.~M.}\ \bibnamefont {Agee}}, \bibinfo
  {author} {\bibfnamefont {S.~G.}\ \bibnamefont {Balasubramani}}, \ and\
  \bibinfo {author} {\bibfnamefont {F.}~\bibnamefont {Furche}},\ }\href
  {\doibase 10.1146/annurev-physchem-040215-112308} {\bibfield  {journal}
  {\bibinfo  {journal} {Ann. Rev. Phys. Chem.}\ }\textbf {\bibinfo {volume}
  {68}},\ \bibinfo {pages} {421} (\bibinfo {year} {2017})}\BibitemShut
  {NoStop}%
\bibitem [{\citenamefont {Krylov}(2001)}]{Krylov_2001a}%
  \BibitemOpen
  \bibfield  {author} {\bibinfo {author} {\bibfnamefont {A.~I.}\ \bibnamefont
  {Krylov}},\ }\href {\doibase https://doi.org/10.1016/S0009-2614(01)00287-1}
  {\bibfield  {journal} {\bibinfo  {journal} {Chem. Phys. Lett.}\ }\textbf
  {\bibinfo {volume} {338}},\ \bibinfo {pages} {375 } (\bibinfo {year}
  {2001})}\BibitemShut {NoStop}%
\bibitem [{\citenamefont {Monino}\ and\ \citenamefont
  {Loos}(2021)}]{Monino_2021}%
  \BibitemOpen
  \bibfield  {author} {\bibinfo {author} {\bibfnamefont {E.}~\bibnamefont
  {Monino}}\ and\ \bibinfo {author} {\bibfnamefont {P.-F.}\ \bibnamefont
  {Loos}},\ }\href {\doibase 10.1021/acs.jctc.1c00074} {\bibfield  {journal}
  {\bibinfo  {journal} {J. Chem. Theory Comput.}\ }\textbf {\bibinfo {volume}
  {17}},\ \bibinfo {pages} {2852} (\bibinfo {year} {2021})}\BibitemShut
  {NoStop}%
\bibitem [{\citenamefont {Casanova}\ and\ \citenamefont
  {Krylov}(2020)}]{Casanova_2020}%
  \BibitemOpen
  \bibfield  {author} {\bibinfo {author} {\bibfnamefont {D.}~\bibnamefont
  {Casanova}}\ and\ \bibinfo {author} {\bibfnamefont {A.~I.}\ \bibnamefont
  {Krylov}},\ }\href {\doibase 10.1039/c9cp06507e} {\bibfield  {journal}
  {\bibinfo  {journal} {Phys. Chem. Chem. Phys.}\ }\textbf {\bibinfo {volume}
  {22}},\ \bibinfo {pages} {4326} (\bibinfo {year} {2020})}\BibitemShut
  {NoStop}%
\bibitem [{\citenamefont {Monino}\ \emph {et~al.}(2022)\citenamefont {Monino},
  \citenamefont {Boggio-Pasqua}, \citenamefont {Scemama}, \citenamefont
  {Jacquemin},\ and\ \citenamefont {Loos}}]{Monino_2022b}%
  \BibitemOpen
  \bibfield  {author} {\bibinfo {author} {\bibfnamefont {E.}~\bibnamefont
  {Monino}}, \bibinfo {author} {\bibfnamefont {M.}~\bibnamefont
  {Boggio-Pasqua}}, \bibinfo {author} {\bibfnamefont {A.}~\bibnamefont
  {Scemama}}, \bibinfo {author} {\bibfnamefont {D.}~\bibnamefont {Jacquemin}},
  \ and\ \bibinfo {author} {\bibfnamefont {P.-F.}\ \bibnamefont {Loos}},\
  }\href {\doibase 10.1021/acs.jpca.2c02480} {\bibfield  {journal} {\bibinfo
  {journal} {J. Phys. Chem. A}\ }\textbf {\bibinfo {volume} {126}},\ \bibinfo
  {pages} {4664} (\bibinfo {year} {2022})}\BibitemShut {NoStop}%
\bibitem [{\citenamefont {Bintrim}\ and\ \citenamefont
  {Berkelbach}(2021)}]{Bintrim_2021a}%
  \BibitemOpen
  \bibfield  {author} {\bibinfo {author} {\bibfnamefont {S.~J.}\ \bibnamefont
  {Bintrim}}\ and\ \bibinfo {author} {\bibfnamefont {T.~C.}\ \bibnamefont
  {Berkelbach}},\ }\href {\doibase 10.1063/5.0035141} {\bibfield  {journal}
  {\bibinfo  {journal} {J. Chem. Phys.}\ }\textbf {\bibinfo {volume} {154}},\
  \bibinfo {pages} {041101} (\bibinfo {year} {2021})}\BibitemShut {NoStop}%
\bibitem [{\citenamefont {Lange}\ and\ \citenamefont
  {Berkelbach}(2018)}]{Lange_2018}%
  \BibitemOpen
  \bibfield  {author} {\bibinfo {author} {\bibfnamefont {M.~F.}\ \bibnamefont
  {Lange}}\ and\ \bibinfo {author} {\bibfnamefont {T.~C.}\ \bibnamefont
  {Berkelbach}},\ }\href {\doibase 10.1021/acs.jctc.8b00455} {\bibfield
  {journal} {\bibinfo  {journal} {J. Chem. Theory. Comput.}\ }\textbf {\bibinfo
  {volume} {14}},\ \bibinfo {pages} {4224} (\bibinfo {year}
  {2018})}\BibitemShut {NoStop}%
\bibitem [{\citenamefont {Quintero-Monsebaiz}\ \emph
  {et~al.}(2022)\citenamefont {Quintero-Monsebaiz}, \citenamefont {Monino},
  \citenamefont {Marie},\ and\ \citenamefont {Loos}}]{Quintero_2022}%
  \BibitemOpen
  \bibfield  {author} {\bibinfo {author} {\bibfnamefont {R.}~\bibnamefont
  {Quintero-Monsebaiz}}, \bibinfo {author} {\bibfnamefont {E.}~\bibnamefont
  {Monino}}, \bibinfo {author} {\bibfnamefont {A.}~\bibnamefont {Marie}}, \
  and\ \bibinfo {author} {\bibfnamefont {P.-F.}\ \bibnamefont {Loos}},\ }\href
  {\doibase 10.1063/5.0130837} {\bibfield  {journal} {\bibinfo  {journal} {J.
  Chem. Phys.}\ }\textbf {\bibinfo {volume} {157}},\ \bibinfo {pages} {231102}
  (\bibinfo {year} {2022})}\BibitemShut {NoStop}%
\bibitem [{\citenamefont {T{\"o}lle}\ and\ \citenamefont
  {Kin-Lic~Chan}(2023)}]{Tolle_2023}%
  \BibitemOpen
  \bibfield  {author} {\bibinfo {author} {\bibfnamefont {J.}~\bibnamefont
  {T{\"o}lle}}\ and\ \bibinfo {author} {\bibfnamefont {G.}~\bibnamefont
  {Kin-Lic~Chan}},\ }\href {\doibase 10.1063/5.0139716} {\bibfield  {journal}
  {\bibinfo  {journal} {J. Chem. Phys.}\ }\textbf {\bibinfo {volume} {158}},\
  \bibinfo {pages} {124123} (\bibinfo {year} {2023})}\BibitemShut {NoStop}%
\bibitem [{\citenamefont {Bintrim}\ and\ \citenamefont
  {Berkelbach}(2022)}]{Bintrim_2022}%
  \BibitemOpen
  \bibfield  {author} {\bibinfo {author} {\bibfnamefont {S.~J.}\ \bibnamefont
  {Bintrim}}\ and\ \bibinfo {author} {\bibfnamefont {T.~C.}\ \bibnamefont
  {Berkelbach}},\ }\href {\doibase 10.1063/5.0074434} {\bibfield  {journal}
  {\bibinfo  {journal} {J. Chem. Phys.}\ }\textbf {\bibinfo {volume} {156}},\
  \bibinfo {pages} {044114} (\bibinfo {year} {2022})}\BibitemShut {NoStop}%
\bibitem [{\citenamefont {{van Setten}}\ \emph {et~al.}(2015)\citenamefont
  {{van Setten}}, \citenamefont {Caruso}, \citenamefont {Sharifzadeh},
  \citenamefont {Ren}, \citenamefont {Scheffler}, \citenamefont {Liu},
  \citenamefont {Lischner}, \citenamefont {Lin}, \citenamefont {Deslippe},
  \citenamefont {Louie}, \citenamefont {Yang}, \citenamefont {Weigend},
  \citenamefont {Neaton}, \citenamefont {Evers},\ and\ \citenamefont
  {Rinke}}]{vanSetten_2015}%
  \BibitemOpen
  \bibfield  {author} {\bibinfo {author} {\bibfnamefont {M.~J.}\ \bibnamefont
  {{van Setten}}}, \bibinfo {author} {\bibfnamefont {F.}~\bibnamefont
  {Caruso}}, \bibinfo {author} {\bibfnamefont {S.}~\bibnamefont {Sharifzadeh}},
  \bibinfo {author} {\bibfnamefont {X.}~\bibnamefont {Ren}}, \bibinfo {author}
  {\bibfnamefont {M.}~\bibnamefont {Scheffler}}, \bibinfo {author}
  {\bibfnamefont {F.}~\bibnamefont {Liu}}, \bibinfo {author} {\bibfnamefont
  {J.}~\bibnamefont {Lischner}}, \bibinfo {author} {\bibfnamefont
  {L.}~\bibnamefont {Lin}}, \bibinfo {author} {\bibfnamefont {J.~R.}\
  \bibnamefont {Deslippe}}, \bibinfo {author} {\bibfnamefont {S.~G.}\
  \bibnamefont {Louie}}, \bibinfo {author} {\bibfnamefont {C.}~\bibnamefont
  {Yang}}, \bibinfo {author} {\bibfnamefont {F.}~\bibnamefont {Weigend}},
  \bibinfo {author} {\bibfnamefont {J.~B.}\ \bibnamefont {Neaton}}, \bibinfo
  {author} {\bibfnamefont {F.}~\bibnamefont {Evers}}, \ and\ \bibinfo {author}
  {\bibfnamefont {P.}~\bibnamefont {Rinke}},\ }\href {\doibase
  10.1021/acs.jctc.5b00453} {\bibfield  {journal} {\bibinfo  {journal} {J.
  Chem. Theory Comput.}\ }\textbf {\bibinfo {volume} {11}},\ \bibinfo {pages}
  {5665} (\bibinfo {year} {2015})}\BibitemShut {NoStop}%
\bibitem [{\citenamefont {Riva}\ \emph {et~al.}(2022)\citenamefont {Riva},
  \citenamefont {Audinet}, \citenamefont {Vladaj}, \citenamefont {Romaniello},\
  and\ \citenamefont {Berger}}]{Riva_2022}%
  \BibitemOpen
  \bibfield  {author} {\bibinfo {author} {\bibfnamefont {G.}~\bibnamefont
  {Riva}}, \bibinfo {author} {\bibfnamefont {T.}~\bibnamefont {Audinet}},
  \bibinfo {author} {\bibfnamefont {M.}~\bibnamefont {Vladaj}}, \bibinfo
  {author} {\bibfnamefont {P.}~\bibnamefont {Romaniello}}, \ and\ \bibinfo
  {author} {\bibfnamefont {J.~A.}\ \bibnamefont {Berger}},\ }\href {\doibase
  10.21468/SciPostPhys.12.3.093} {\bibfield  {journal} {\bibinfo  {journal}
  {SciPost Phys.}\ }\textbf {\bibinfo {volume} {12}},\ \bibinfo {pages} {093}
  (\bibinfo {year} {2022})}\BibitemShut {NoStop}%
\bibitem [{\citenamefont {Rishi}, \citenamefont {Perera},\ and\ \citenamefont
  {Bartlett}(2020)}]{Rishi_2020}%
  \BibitemOpen
  \bibfield  {author} {\bibinfo {author} {\bibfnamefont {V.}~\bibnamefont
  {Rishi}}, \bibinfo {author} {\bibfnamefont {A.}~\bibnamefont {Perera}}, \
  and\ \bibinfo {author} {\bibfnamefont {R.~J.}\ \bibnamefont {Bartlett}},\
  }\href {\doibase 10.1063/5.0023862} {\bibfield  {journal} {\bibinfo
  {journal} {J. Chem. Phys.}\ }\textbf {\bibinfo {volume} {153}},\ \bibinfo
  {pages} {234101} (\bibinfo {year} {2020})}\BibitemShut {NoStop}%
\bibitem [{\citenamefont {Mattuck}(1992)}]{MattuckBook}%
  \BibitemOpen
  \bibfield  {author} {\bibinfo {author} {\bibfnamefont {R.~D.}\ \bibnamefont
  {Mattuck}},\ }\href@noop {} {\emph {\bibinfo {title} {A guide to {Feynman}
  diagrams in the many-body problem}}},\ \bibinfo {edition} {2nd}\ ed.,\ Dover
  books on physics and chemistry\ (\bibinfo  {publisher} {Dover Publications},\
  \bibinfo {address} {New York},\ \bibinfo {year} {1992})\BibitemShut {NoStop}%
\bibitem [{\citenamefont {Cederbaum}\ and\ \citenamefont
  {Domcke}(1977)}]{Cederbaum_1977}%
  \BibitemOpen
  \bibfield  {author} {\bibinfo {author} {\bibfnamefont {L.~S.}\ \bibnamefont
  {Cederbaum}}\ and\ \bibinfo {author} {\bibfnamefont {W.}~\bibnamefont
  {Domcke}},\ }in\ \href {\doibase 10.1002/9780470142554.ch4} {\emph {\bibinfo
  {booktitle} {Adv. Chem. Phys.}}}\ (\bibinfo  {publisher} {John Wiley {\&}
  Sons, Inc.},\ \bibinfo {year} {1977})\ pp.\ \bibinfo {pages}
  {205--344}\BibitemShut {NoStop}%
\bibitem [{\citenamefont {Oddershede}, \citenamefont {J{\o}rgensen},\ and\
  \citenamefont {Yeager}(1984)}]{Oddershede_1984}%
  \BibitemOpen
  \bibfield  {author} {\bibinfo {author} {\bibfnamefont {J.}~\bibnamefont
  {Oddershede}}, \bibinfo {author} {\bibfnamefont {P.}~\bibnamefont
  {J{\o}rgensen}}, \ and\ \bibinfo {author} {\bibfnamefont {D.~L.}\
  \bibnamefont {Yeager}},\ }\href {\doibase 10.1016/0167-7977(84)90003-0}
  {\bibfield  {journal} {\bibinfo  {journal} {Comp. Phys. Comm.}\ }\textbf
  {\bibinfo {volume} {2}},\ \bibinfo {pages} {33} (\bibinfo {year}
  {1984})}\BibitemShut {NoStop}%
\bibitem [{\citenamefont {Szabo}\ and\ \citenamefont
  {Ostlund}(1989)}]{SzaboBook}%
  \BibitemOpen
  \bibfield  {author} {\bibinfo {author} {\bibfnamefont {A.}~\bibnamefont
  {Szabo}}\ and\ \bibinfo {author} {\bibfnamefont {N.~S.}\ \bibnamefont
  {Ostlund}},\ }\href@noop {} {\emph {\bibinfo {title} {Modern quantum
  chemistry}}}\ (\bibinfo  {publisher} {McGraw-Hill},\ \bibinfo {address} {New
  York},\ \bibinfo {year} {1989})\BibitemShut {NoStop}%
\bibitem [{\citenamefont {Schirmer}(1982)}]{Schirmer_1982}%
  \BibitemOpen
  \bibfield  {author} {\bibinfo {author} {\bibfnamefont {J.}~\bibnamefont
  {Schirmer}},\ }\href {\doibase 10.1103/PhysRevA.26.2395} {\bibfield
  {journal} {\bibinfo  {journal} {Phys. Rev. A}\ }\textbf {\bibinfo {volume}
  {26}},\ \bibinfo {pages} {2395} (\bibinfo {year} {1982})}\BibitemShut
  {NoStop}%
\bibitem [{\citenamefont {Schirmer}, \citenamefont {Cederbaum},\ and\
  \citenamefont {Walter}(1983)}]{Schirmer_1983}%
  \BibitemOpen
  \bibfield  {author} {\bibinfo {author} {\bibfnamefont {J.}~\bibnamefont
  {Schirmer}}, \bibinfo {author} {\bibfnamefont {L.~S.}\ \bibnamefont
  {Cederbaum}}, \ and\ \bibinfo {author} {\bibfnamefont {O.}~\bibnamefont
  {Walter}},\ }\href {\doibase 10.1103/PhysRevA.28.1237} {\bibfield  {journal}
  {\bibinfo  {journal} {Phys. Rev. A}\ }\textbf {\bibinfo {volume} {28}},\
  \bibinfo {pages} {1237} (\bibinfo {year} {1983})}\BibitemShut {NoStop}%
\bibitem [{\citenamefont {Schirmer}\ and\ \citenamefont
  {Barth}(1984)}]{Schirmer_1984}%
  \BibitemOpen
  \bibfield  {author} {\bibinfo {author} {\bibfnamefont {J.}~\bibnamefont
  {Schirmer}}\ and\ \bibinfo {author} {\bibfnamefont {A.}~\bibnamefont
  {Barth}},\ }\href {\doibase 10.1007/BF01438358} {\bibfield  {journal}
  {\bibinfo  {journal} {Z. Phys. A}\ }\textbf {\bibinfo {volume} {317}},\
  \bibinfo {pages} {267} (\bibinfo {year} {1984})}\BibitemShut {NoStop}%
\bibitem [{\citenamefont {Dreuw}\ and\ \citenamefont
  {Wormit}(2015)}]{Dreuw_2015}%
  \BibitemOpen
  \bibfield  {author} {\bibinfo {author} {\bibfnamefont {A.}~\bibnamefont
  {Dreuw}}\ and\ \bibinfo {author} {\bibfnamefont {M.}~\bibnamefont {Wormit}},\
  }\href {\doibase 10.1002/wcms.1206} {\bibfield  {journal} {\bibinfo
  {journal} {Wiley Interdiscip. Rev. Comput. Mol. Sci.}\ }\textbf {\bibinfo
  {volume} {5}},\ \bibinfo {pages} {82} (\bibinfo {year} {2015})}\BibitemShut
  {NoStop}%
\bibitem [{\citenamefont {Gell-Mann}\ and\ \citenamefont
  {Brueckner}(1957)}]{Gell-Mann_1957}%
  \BibitemOpen
  \bibfield  {author} {\bibinfo {author} {\bibfnamefont {M.}~\bibnamefont
  {Gell-Mann}}\ and\ \bibinfo {author} {\bibfnamefont {K.~A.}\ \bibnamefont
  {Brueckner}},\ }\href {\doibase 10.1103/PhysRev.106.364} {\bibfield
  {journal} {\bibinfo  {journal} {Phys. Rev.}\ }\textbf {\bibinfo {volume}
  {106}},\ \bibinfo {pages} {364} (\bibinfo {year} {1957})}\BibitemShut
  {NoStop}%
\bibitem [{\citenamefont {Nozi\`eres}\ and\ \citenamefont
  {Pines}(1958)}]{Nozieres_1958}%
  \BibitemOpen
  \bibfield  {author} {\bibinfo {author} {\bibfnamefont {P.}~\bibnamefont
  {Nozi\`eres}}\ and\ \bibinfo {author} {\bibfnamefont {D.}~\bibnamefont
  {Pines}},\ }\href {\doibase 10.1103/PhysRev.111.442} {\bibfield  {journal}
  {\bibinfo  {journal} {Phys. Rev.}\ }\textbf {\bibinfo {volume} {111}},\
  \bibinfo {pages} {442} (\bibinfo {year} {1958})}\BibitemShut {NoStop}%
\bibitem [{\citenamefont {Foerster}, \citenamefont {Koval},\ and\ \citenamefont
  {S{\'a}nchez-Portal}(2011)}]{Foerster_2011}%
  \BibitemOpen
  \bibfield  {author} {\bibinfo {author} {\bibfnamefont {D.}~\bibnamefont
  {Foerster}}, \bibinfo {author} {\bibfnamefont {P.}~\bibnamefont {Koval}}, \
  and\ \bibinfo {author} {\bibfnamefont {D.}~\bibnamefont
  {S{\'a}nchez-Portal}},\ }\href {\doibase 10.1063/1.3624731} {\bibfield
  {journal} {\bibinfo  {journal} {J. Chem. Phys.}\ }\textbf {\bibinfo {volume}
  {135}},\ \bibinfo {pages} {074105} (\bibinfo {year} {2011})}\BibitemShut
  {NoStop}%
\bibitem [{\citenamefont {Liu}\ \emph {et~al.}(2016)\citenamefont {Liu},
  \citenamefont {Kaltak}, \citenamefont {Klime\ifmmode~\check{s}\else
  \v{s}\fi{}},\ and\ \citenamefont {Kresse}}]{Liu_2016}%
  \BibitemOpen
  \bibfield  {author} {\bibinfo {author} {\bibfnamefont {P.}~\bibnamefont
  {Liu}}, \bibinfo {author} {\bibfnamefont {M.}~\bibnamefont {Kaltak}},
  \bibinfo {author} {\bibfnamefont {J.~c.~v.}\ \bibnamefont
  {Klime\ifmmode~\check{s}\else \v{s}\fi{}}}, \ and\ \bibinfo {author}
  {\bibfnamefont {G.}~\bibnamefont {Kresse}},\ }\href {\doibase
  10.1103/PhysRevB.94.165109} {\bibfield  {journal} {\bibinfo  {journal} {Phys.
  Rev. B}\ }\textbf {\bibinfo {volume} {94}},\ \bibinfo {pages} {165109}
  (\bibinfo {year} {2016})}\BibitemShut {NoStop}%
\bibitem [{\citenamefont {F{\"o}rster}\ and\ \citenamefont
  {Visscher}(2021)}]{Forster_2021}%
  \BibitemOpen
  \bibfield  {author} {\bibinfo {author} {\bibfnamefont {A.}~\bibnamefont
  {F{\"o}rster}}\ and\ \bibinfo {author} {\bibfnamefont {L.}~\bibnamefont
  {Visscher}},\ }\href {\doibase 10.3389/fchem.2021.736591} {\bibfield
  {journal} {\bibinfo  {journal} {Front. Chem.}\ }\textbf {\bibinfo {volume}
  {9}},\ \bibinfo {pages} {736591} (\bibinfo {year} {2021})}\BibitemShut
  {NoStop}%
\bibitem [{\citenamefont {Duchemin}\ and\ \citenamefont
  {Blase}(2019)}]{Duchemin_2019}%
  \BibitemOpen
  \bibfield  {author} {\bibinfo {author} {\bibfnamefont {I.}~\bibnamefont
  {Duchemin}}\ and\ \bibinfo {author} {\bibfnamefont {X.}~\bibnamefont
  {Blase}},\ }\href {\doibase 10.1063/1.5090605} {\bibfield  {journal}
  {\bibinfo  {journal} {J. Chem. Phys.}\ }\textbf {\bibinfo {volume} {150}},\
  \bibinfo {pages} {174120} (\bibinfo {year} {2019})}\BibitemShut {NoStop}%
\bibitem [{\citenamefont {Duchemin}\ and\ \citenamefont
  {Blase}(2020)}]{Duchemin_2020}%
  \BibitemOpen
  \bibfield  {author} {\bibinfo {author} {\bibfnamefont {I.}~\bibnamefont
  {Duchemin}}\ and\ \bibinfo {author} {\bibfnamefont {X.}~\bibnamefont
  {Blase}},\ }\href {\doibase 10.1021/acs.jctc.9b01235} {\bibfield  {journal}
  {\bibinfo  {journal} {J. Chem. Theory Comput.}\ }\textbf {\bibinfo {volume}
  {16}},\ \bibinfo {pages} {1742} (\bibinfo {year} {2020})}\BibitemShut
  {NoStop}%
\bibitem [{\citenamefont {Duchemin}\ and\ \citenamefont
  {Blase}(2021)}]{Duchemin_2021}%
  \BibitemOpen
  \bibfield  {author} {\bibinfo {author} {\bibfnamefont {I.}~\bibnamefont
  {Duchemin}}\ and\ \bibinfo {author} {\bibfnamefont {X.}~\bibnamefont
  {Blase}},\ }\href {\doibase 10.1021/acs.jctc.1c00101} {\bibfield  {journal}
  {\bibinfo  {journal} {J. Chem. Theory Comput.}\ }\textbf {\bibinfo {volume}
  {17}},\ \bibinfo {pages} {2383} (\bibinfo {year} {2021})}\BibitemShut
  {NoStop}%
\bibitem [{\citenamefont {Romanova}\ and\ \citenamefont {Vl{\v
  c}ek}(2022)}]{Romanova_2022}%
  \BibitemOpen
  \bibfield  {author} {\bibinfo {author} {\bibfnamefont {M.}~\bibnamefont
  {Romanova}}\ and\ \bibinfo {author} {\bibfnamefont {V.}~\bibnamefont {Vl{\v
  c}ek}},\ }\href {\doibase https://doi.org/10.1038/s41524-022-00697-8}
  {\bibfield  {journal} {\bibinfo  {journal} {npj Comput Mater}\ }\textbf
  {\bibinfo {volume} {8}},\ \bibinfo {pages} {11} (\bibinfo {year}
  {2022})}\BibitemShut {NoStop}%
\bibitem [{\citenamefont {Weng}\ and\ \citenamefont {Vl{\v
  c}ek}(2021)}]{Weng_2021}%
  \BibitemOpen
  \bibfield  {author} {\bibinfo {author} {\bibfnamefont {G.}~\bibnamefont
  {Weng}}\ and\ \bibinfo {author} {\bibfnamefont {V.}~\bibnamefont {Vl{\v
  c}ek}},\ }\href {\doibase https://doi.org/10.1063/5.0058410} {\bibfield
  {journal} {\bibinfo  {journal} {J. Chem. Phys.}\ }\textbf {\bibinfo {volume}
  {155}},\ \bibinfo {pages} {054104} (\bibinfo {year} {2021})}\BibitemShut
  {NoStop}%
\bibitem [{\citenamefont {Romanova}\ and\ \citenamefont {Vl{\v
  c}ek}(2020)}]{Romanova_2020}%
  \BibitemOpen
  \bibfield  {author} {\bibinfo {author} {\bibfnamefont {M.}~\bibnamefont
  {Romanova}}\ and\ \bibinfo {author} {\bibfnamefont {V.}~\bibnamefont {Vl{\v
  c}ek}},\ }\href {\doibase https://doi.org/10.1063/5.0020430} {\bibfield
  {journal} {\bibinfo  {journal} {J. Chem. Phys.}\ }\textbf {\bibinfo {volume}
  {153}},\ \bibinfo {pages} {134103} (\bibinfo {year} {2020})}\BibitemShut
  {NoStop}%
\bibitem [{\citenamefont {Brooks}\ \emph {et~al.}(2020)\citenamefont {Brooks},
  \citenamefont {Weng}, \citenamefont {Taylor},\ and\ \citenamefont
  {Vlcek}}]{Brooks_2020}%
  \BibitemOpen
  \bibfield  {author} {\bibinfo {author} {\bibfnamefont {J.}~\bibnamefont
  {Brooks}}, \bibinfo {author} {\bibfnamefont {G.}~\bibnamefont {Weng}},
  \bibinfo {author} {\bibfnamefont {S.}~\bibnamefont {Taylor}}, \ and\ \bibinfo
  {author} {\bibfnamefont {V.}~\bibnamefont {Vlcek}},\ }\href {\doibase
  10.1088/1361-648X/ab6d8c} {\bibfield  {journal} {\bibinfo  {journal} {J.
  Phys.: Condens. Matter}\ }\textbf {\bibinfo {volume} {32}},\ \bibinfo {pages}
  {234001} (\bibinfo {year} {2020})}\BibitemShut {NoStop}%
\bibitem [{\citenamefont {Vl{\v c}ek}\ \emph {et~al.}(2017)\citenamefont {Vl{\v
  c}ek}, \citenamefont {Rabani}, \citenamefont {Neuhauser},\ and\ \citenamefont
  {Baer}}]{Vlcek_2017}%
  \BibitemOpen
  \bibfield  {author} {\bibinfo {author} {\bibfnamefont {V.}~\bibnamefont
  {Vl{\v c}ek}}, \bibinfo {author} {\bibfnamefont {E.}~\bibnamefont {Rabani}},
  \bibinfo {author} {\bibfnamefont {D.}~\bibnamefont {Neuhauser}}, \ and\
  \bibinfo {author} {\bibfnamefont {R.}~\bibnamefont {Baer}},\ }\href {\doibase
  10.1021/acs.jctc.7b00770} {\bibfield  {journal} {\bibinfo  {journal} {J.
  Chem. Theory Comput.}\ }\textbf {\bibinfo {volume} {13}},\ \bibinfo {pages}
  {4997} (\bibinfo {year} {2017})}\BibitemShut {NoStop}%
\bibitem [{\citenamefont {Cunningham}\ \emph {et~al.}(2018)\citenamefont
  {Cunningham}, \citenamefont {Gr\"uning}, \citenamefont {Azarhoosh},
  \citenamefont {Pashov},\ and\ \citenamefont {van
  Schilfgaarde}}]{Cunningham_2018}%
  \BibitemOpen
  \bibfield  {author} {\bibinfo {author} {\bibfnamefont {B.}~\bibnamefont
  {Cunningham}}, \bibinfo {author} {\bibfnamefont {M.}~\bibnamefont
  {Gr\"uning}}, \bibinfo {author} {\bibfnamefont {P.}~\bibnamefont
  {Azarhoosh}}, \bibinfo {author} {\bibfnamefont {D.}~\bibnamefont {Pashov}}, \
  and\ \bibinfo {author} {\bibfnamefont {M.}~\bibnamefont {van Schilfgaarde}},\
  }\href {\doibase 10.1103/PhysRevMaterials.2.034603} {\bibfield  {journal}
  {\bibinfo  {journal} {Phys. Rev. Mater.}\ }\textbf {\bibinfo {volume} {2}},\
  \bibinfo {pages} {034603} (\bibinfo {year} {2018})}\BibitemShut {NoStop}%
\bibitem [{\citenamefont {Dadkhah}\ \emph {et~al.}(2023)\citenamefont
  {Dadkhah}, \citenamefont {Lambrecht}, \citenamefont {Pashov},\ and\
  \citenamefont {van Schilfgaarde}}]{Dadkhah_2023}%
  \BibitemOpen
  \bibfield  {author} {\bibinfo {author} {\bibfnamefont {N.}~\bibnamefont
  {Dadkhah}}, \bibinfo {author} {\bibfnamefont {W.~R.~L.}\ \bibnamefont
  {Lambrecht}}, \bibinfo {author} {\bibfnamefont {D.}~\bibnamefont {Pashov}}, \
  and\ \bibinfo {author} {\bibfnamefont {M.}~\bibnamefont {van Schilfgaarde}},\
  }\href {\doibase 10.1103/PhysRevB.107.165201} {\bibfield  {journal} {\bibinfo
   {journal} {Phys. Rev. B}\ }\textbf {\bibinfo {volume} {107}},\ \bibinfo
  {pages} {165201} (\bibinfo {year} {2023})}\BibitemShut {NoStop}%
\bibitem [{\citenamefont {Grzeszczyk}\ \emph {et~al.}(2023)\citenamefont
  {Grzeszczyk}, \citenamefont {Acharya}, \citenamefont {Pashov}, \citenamefont
  {Chen}, \citenamefont {Vaklinova}, \citenamefont {van Schilfgaarde},
  \citenamefont {Watanabe}, \citenamefont {Taniguchi}, \citenamefont
  {Novoselov}, \citenamefont {Katsnelson},\ and\ \citenamefont
  {Koperski}}]{Grzeszczyk_2023}%
  \BibitemOpen
  \bibfield  {author} {\bibinfo {author} {\bibfnamefont {M.}~\bibnamefont
  {Grzeszczyk}}, \bibinfo {author} {\bibfnamefont {S.}~\bibnamefont {Acharya}},
  \bibinfo {author} {\bibfnamefont {D.}~\bibnamefont {Pashov}}, \bibinfo
  {author} {\bibfnamefont {Z.}~\bibnamefont {Chen}}, \bibinfo {author}
  {\bibfnamefont {K.}~\bibnamefont {Vaklinova}}, \bibinfo {author}
  {\bibfnamefont {M.}~\bibnamefont {van Schilfgaarde}}, \bibinfo {author}
  {\bibfnamefont {K.}~\bibnamefont {Watanabe}}, \bibinfo {author}
  {\bibfnamefont {T.}~\bibnamefont {Taniguchi}}, \bibinfo {author}
  {\bibfnamefont {K.~S.}\ \bibnamefont {Novoselov}}, \bibinfo {author}
  {\bibfnamefont {M.~I.}\ \bibnamefont {Katsnelson}}, \ and\ \bibinfo {author}
  {\bibfnamefont {M.}~\bibnamefont {Koperski}},\ }\href {\doibase
  https://doi.org/10.1002/adma.202209513} {\bibfield  {journal} {\bibinfo
  {journal} {Adv. Mater.}\ }\textbf {\bibinfo {volume} {35}},\ \bibinfo {pages}
  {2209513} (\bibinfo {year} {2023})}\BibitemShut {NoStop}%
\bibitem [{\citenamefont {Chen}\ and\ \citenamefont
  {Pasquarello}(2015)}]{Chen_2015}%
  \BibitemOpen
  \bibfield  {author} {\bibinfo {author} {\bibfnamefont {W.}~\bibnamefont
  {Chen}}\ and\ \bibinfo {author} {\bibfnamefont {A.}~\bibnamefont
  {Pasquarello}},\ }\href {\doibase 10.1103/PhysRevB.92.041115} {\bibfield
  {journal} {\bibinfo  {journal} {Phys. Rev. B}\ }\textbf {\bibinfo {volume}
  {92}},\ \bibinfo {pages} {041115} (\bibinfo {year} {2015})}\BibitemShut
  {NoStop}%
\bibitem [{\citenamefont {Caruso}\ \emph {et~al.}(2016)\citenamefont {Caruso},
  \citenamefont {Dauth}, \citenamefont {{van Setten}},\ and\ \citenamefont
  {Rinke}}]{Caruso_2016}%
  \BibitemOpen
  \bibfield  {author} {\bibinfo {author} {\bibfnamefont {F.}~\bibnamefont
  {Caruso}}, \bibinfo {author} {\bibfnamefont {M.}~\bibnamefont {Dauth}},
  \bibinfo {author} {\bibfnamefont {M.~J.}\ \bibnamefont {{van Setten}}}, \
  and\ \bibinfo {author} {\bibfnamefont {P.}~\bibnamefont {Rinke}},\ }\href
  {\doibase 10.1021/acs.jctc.6b00774} {\bibfield  {journal} {\bibinfo
  {journal} {J. Chem. Theory Comput.}\ }\textbf {\bibinfo {volume} {12}},\
  \bibinfo {pages} {5076} (\bibinfo {year} {2016})}\BibitemShut {NoStop}%
\bibitem [{\citenamefont {Dreuw}\ and\ \citenamefont
  {Head-Gordon}(2005)}]{Dreuw_2005}%
  \BibitemOpen
  \bibfield  {author} {\bibinfo {author} {\bibfnamefont {A.}~\bibnamefont
  {Dreuw}}\ and\ \bibinfo {author} {\bibfnamefont {M.}~\bibnamefont
  {Head-Gordon}},\ }\href {\doibase 10.1021/cr0505627} {\bibfield  {journal}
  {\bibinfo  {journal} {Chem. Rev.}\ }\textbf {\bibinfo {volume} {105}},\
  \bibinfo {pages} {4009} (\bibinfo {year} {2005})}\BibitemShut {NoStop}%
\bibitem [{\citenamefont {Schirmer}, \citenamefont {Trofimov},\ and\
  \citenamefont {Stelter}(1998)}]{Schirmer_1998}%
  \BibitemOpen
  \bibfield  {author} {\bibinfo {author} {\bibfnamefont {J.}~\bibnamefont
  {Schirmer}}, \bibinfo {author} {\bibfnamefont {A.~B.}\ \bibnamefont
  {Trofimov}}, \ and\ \bibinfo {author} {\bibfnamefont {G.}~\bibnamefont
  {Stelter}},\ }\href {\doibase 10.1063/1.477085} {\bibfield  {journal}
  {\bibinfo  {journal} {J. Chem. Phys.}\ }\textbf {\bibinfo {volume} {109}},\
  \bibinfo {pages} {4734} (\bibinfo {year} {1998})}\BibitemShut {NoStop}%
\bibitem [{\citenamefont {Ring}\ and\ \citenamefont
  {Schuck}(2004)}]{Schuck_Book}%
  \BibitemOpen
  \bibfield  {author} {\bibinfo {author} {\bibfnamefont {P.}~\bibnamefont
  {Ring}}\ and\ \bibinfo {author} {\bibfnamefont {P.}~\bibnamefont {Schuck}},\
  }\href@noop {} {\emph {\bibinfo {title} {The Nuclear Many-Body Problem}}}\
  (\bibinfo  {publisher} {Springer},\ \bibinfo {year} {2004})\BibitemShut
  {NoStop}%
\bibitem [{\citenamefont {{van Aggelen}}, \citenamefont {Yang},\ and\
  \citenamefont {Yang}(2013)}]{vanAggelen_2013}%
  \BibitemOpen
  \bibfield  {author} {\bibinfo {author} {\bibfnamefont {H.}~\bibnamefont {{van
  Aggelen}}}, \bibinfo {author} {\bibfnamefont {Y.}~\bibnamefont {Yang}}, \
  and\ \bibinfo {author} {\bibfnamefont {W.}~\bibnamefont {Yang}},\ }\href
  {\doibase 10.1103/PhysRevA.88.030501} {\bibfield  {journal} {\bibinfo
  {journal} {Phys. Rev. A}\ }\textbf {\bibinfo {volume} {88}},\ \bibinfo
  {pages} {030501} (\bibinfo {year} {2013})}\BibitemShut {NoStop}%
\bibitem [{\citenamefont {Peng}\ \emph {et~al.}(2013)\citenamefont {Peng},
  \citenamefont {Steinmann}, \citenamefont {{van Aggelen}},\ and\ \citenamefont
  {Yang}}]{Peng_2013}%
  \BibitemOpen
  \bibfield  {author} {\bibinfo {author} {\bibfnamefont {D.}~\bibnamefont
  {Peng}}, \bibinfo {author} {\bibfnamefont {S.~N.}\ \bibnamefont {Steinmann}},
  \bibinfo {author} {\bibfnamefont {H.}~\bibnamefont {{van Aggelen}}}, \ and\
  \bibinfo {author} {\bibfnamefont {W.}~\bibnamefont {Yang}},\ }\href {\doibase
  10.1063/1.4820556} {\bibfield  {journal} {\bibinfo  {journal} {J. Chem.
  Phys.}\ }\textbf {\bibinfo {volume} {139}},\ \bibinfo {pages} {104112}
  (\bibinfo {year} {2013})}\BibitemShut {NoStop}%
\bibitem [{\citenamefont {Scuseria}, \citenamefont {Henderson},\ and\
  \citenamefont {Bulik}(2013)}]{Scuseria_2013}%
  \BibitemOpen
  \bibfield  {author} {\bibinfo {author} {\bibfnamefont {G.~E.}\ \bibnamefont
  {Scuseria}}, \bibinfo {author} {\bibfnamefont {T.~M.}\ \bibnamefont
  {Henderson}}, \ and\ \bibinfo {author} {\bibfnamefont {I.~W.}\ \bibnamefont
  {Bulik}},\ }\href {\doibase 10.1063/1.4820557} {\bibfield  {journal}
  {\bibinfo  {journal} {J. Chem. Phys.}\ }\textbf {\bibinfo {volume} {139}},\
  \bibinfo {pages} {104113} (\bibinfo {year} {2013})}\BibitemShut {NoStop}%
\bibitem [{\citenamefont {Yang}\ \emph {et~al.}(2013)\citenamefont {Yang},
  \citenamefont {{van Aggelen}}, \citenamefont {Steinmann}, \citenamefont
  {Peng},\ and\ \citenamefont {Yang}}]{Yang_2013}%
  \BibitemOpen
  \bibfield  {author} {\bibinfo {author} {\bibfnamefont {Y.}~\bibnamefont
  {Yang}}, \bibinfo {author} {\bibfnamefont {H.}~\bibnamefont {{van Aggelen}}},
  \bibinfo {author} {\bibfnamefont {S.~N.}\ \bibnamefont {Steinmann}}, \bibinfo
  {author} {\bibfnamefont {D.}~\bibnamefont {Peng}}, \ and\ \bibinfo {author}
  {\bibfnamefont {W.}~\bibnamefont {Yang}},\ }\href {\doibase
  10.1063/1.4828728} {\bibfield  {journal} {\bibinfo  {journal} {J. Chem.
  Phys.}\ }\textbf {\bibinfo {volume} {139}},\ \bibinfo {pages} {174110}
  (\bibinfo {year} {2013})}\BibitemShut {NoStop}%
\bibitem [{\citenamefont {Yang}, \citenamefont {van Aggelen},\ and\
  \citenamefont {Yang}(2013)}]{Yang_2013b}%
  \BibitemOpen
  \bibfield  {author} {\bibinfo {author} {\bibfnamefont {Y.}~\bibnamefont
  {Yang}}, \bibinfo {author} {\bibfnamefont {H.}~\bibnamefont {van Aggelen}}, \
  and\ \bibinfo {author} {\bibfnamefont {W.}~\bibnamefont {Yang}},\ }\href
  {\doibase 10.1063/1.4834875} {\bibfield  {journal} {\bibinfo  {journal} {J.
  Chem. Phys.}\ }\textbf {\bibinfo {volume} {139}},\ \bibinfo {pages} {224105}
  (\bibinfo {year} {2013})}\BibitemShut {NoStop}%
\bibitem [{\citenamefont {van Aggelen}, \citenamefont {Yang},\ and\
  \citenamefont {Yang}(2014)}]{vanAggelen_2014}%
  \BibitemOpen
  \bibfield  {author} {\bibinfo {author} {\bibfnamefont {H.}~\bibnamefont {van
  Aggelen}}, \bibinfo {author} {\bibfnamefont {Y.}~\bibnamefont {Yang}}, \ and\
  \bibinfo {author} {\bibfnamefont {W.}~\bibnamefont {Yang}},\ }\href {\doibase
  10.1063/1.4865816} {\bibfield  {journal} {\bibinfo  {journal} {J. Chem.
  Phys.}\ }\textbf {\bibinfo {volume} {140}},\ \bibinfo {pages} {18A511}
  (\bibinfo {year} {2014})}\BibitemShut {NoStop}%
\bibitem [{\citenamefont {Yang}\ \emph {et~al.}(2014)\citenamefont {Yang},
  \citenamefont {Peng}, \citenamefont {Lu},\ and\ \citenamefont
  {Yang}}]{Yang_2014a}%
  \BibitemOpen
  \bibfield  {author} {\bibinfo {author} {\bibfnamefont {Y.}~\bibnamefont
  {Yang}}, \bibinfo {author} {\bibfnamefont {D.}~\bibnamefont {Peng}}, \bibinfo
  {author} {\bibfnamefont {J.}~\bibnamefont {Lu}}, \ and\ \bibinfo {author}
  {\bibfnamefont {W.}~\bibnamefont {Yang}},\ }\href {\doibase
  10.1063/1.4895792} {\bibfield  {journal} {\bibinfo  {journal} {J. Chem.
  Phys.}\ }\textbf {\bibinfo {volume} {141}},\ \bibinfo {pages} {124104}
  (\bibinfo {year} {2014})}\BibitemShut {NoStop}%
\bibitem [{\citenamefont {Zhang}\ and\ \citenamefont
  {Herbert}(2015)}]{Zhang_2015}%
  \BibitemOpen
  \bibfield  {author} {\bibinfo {author} {\bibfnamefont {X.}~\bibnamefont
  {Zhang}}\ and\ \bibinfo {author} {\bibfnamefont {J.~M.}\ \bibnamefont
  {Herbert}},\ }\href {\doibase 10.1063/1.4907376} {\bibfield  {journal}
  {\bibinfo  {journal} {J. Chem. Phys.}\ }\textbf {\bibinfo {volume} {142}},\
  \bibinfo {pages} {064109} (\bibinfo {year} {2015})}\BibitemShut {NoStop}%
\bibitem [{\citenamefont {Zhang}\ and\ \citenamefont
  {Yang}(2016)}]{Zhang_2016}%
  \BibitemOpen
  \bibfield  {author} {\bibinfo {author} {\bibfnamefont {D.}~\bibnamefont
  {Zhang}}\ and\ \bibinfo {author} {\bibfnamefont {W.}~\bibnamefont {Yang}},\
  }\href {\doibase 10.1063/1.4964501} {\bibfield  {journal} {\bibinfo
  {journal} {J. Chem. Phys.}\ }\textbf {\bibinfo {volume} {145}},\ \bibinfo
  {pages} {144105} (\bibinfo {year} {2016})}\BibitemShut {NoStop}%
\bibitem [{\citenamefont {Bannwarth}\ \emph {et~al.}(2020)\citenamefont
  {Bannwarth}, \citenamefont {Yu}, \citenamefont {Hohenstein},\ and\
  \citenamefont {Mart{\'\i}nez}}]{Bannwarth_2020}%
  \BibitemOpen
  \bibfield  {author} {\bibinfo {author} {\bibfnamefont {C.}~\bibnamefont
  {Bannwarth}}, \bibinfo {author} {\bibfnamefont {J.~K.}\ \bibnamefont {Yu}},
  \bibinfo {author} {\bibfnamefont {E.~G.}\ \bibnamefont {Hohenstein}}, \ and\
  \bibinfo {author} {\bibfnamefont {T.~J.}\ \bibnamefont {Mart{\'\i}nez}},\
  }\href {\doibase 10.1063/5.0003985} {\bibfield  {journal} {\bibinfo
  {journal} {J. Chem. Phys.}\ }\textbf {\bibinfo {volume} {153}},\ \bibinfo
  {pages} {024110} (\bibinfo {year} {2020})}\BibitemShut {NoStop}%
\bibitem [{\citenamefont {Rebolini}(2014)}]{Rebolini_PhD}%
  \BibitemOpen
  \bibfield  {author} {\bibinfo {author} {\bibfnamefont {E.}~\bibnamefont
  {Rebolini}},\ }\emph {\bibinfo {title} {Range-Separated Density-Functional
  Theory for Molecular Excitation Energies}},\ \href
  {https://tel.archives-ouvertes.fr/tel-01027522} {Ph.D. thesis},\ \bibinfo
  {school} {Universit{\'e} Pierre et Marie Curie --- Paris VI} (\bibinfo {year}
  {2014})\BibitemShut {NoStop}%
\bibitem [{\citenamefont {Schirmer}\ and\ \citenamefont
  {Trofimov}(2004)}]{Schirmer_2004}%
  \BibitemOpen
  \bibfield  {author} {\bibinfo {author} {\bibfnamefont {J.}~\bibnamefont
  {Schirmer}}\ and\ \bibinfo {author} {\bibfnamefont {A.~B.}\ \bibnamefont
  {Trofimov}},\ }\href {\doibase 10.1063/1.1752875} {\bibfield  {journal}
  {\bibinfo  {journal} {J. Chem. Phys.}\ }\textbf {\bibinfo {volume} {120}},\
  \bibinfo {pages} {11449} (\bibinfo {year} {2004})}\BibitemShut {NoStop}%
\bibitem [{\citenamefont {Wormit}\ \emph {et~al.}(2014)\citenamefont {Wormit},
  \citenamefont {Rehn}, \citenamefont {Harbach}, \citenamefont {Wenzel},
  \citenamefont {Krauter}, \citenamefont {Epifanovsky},\ and\ \citenamefont
  {Dreuw}}]{Wormit_2014}%
  \BibitemOpen
  \bibfield  {author} {\bibinfo {author} {\bibfnamefont {M.}~\bibnamefont
  {Wormit}}, \bibinfo {author} {\bibfnamefont {D.~R.}\ \bibnamefont {Rehn}},
  \bibinfo {author} {\bibfnamefont {P.~H.}\ \bibnamefont {Harbach}}, \bibinfo
  {author} {\bibfnamefont {J.}~\bibnamefont {Wenzel}}, \bibinfo {author}
  {\bibfnamefont {C.~M.}\ \bibnamefont {Krauter}}, \bibinfo {author}
  {\bibfnamefont {E.}~\bibnamefont {Epifanovsky}}, \ and\ \bibinfo {author}
  {\bibfnamefont {A.}~\bibnamefont {Dreuw}},\ }\href {\doibase
  10.1080/00268976.2013.859313} {\bibfield  {journal} {\bibinfo  {journal}
  {Mol. Phys.}\ }\textbf {\bibinfo {volume} {112}},\ \bibinfo {pages} {774}
  (\bibinfo {year} {2014})}\BibitemShut {NoStop}%
\bibitem [{\citenamefont {Wormit}(2009)}]{Wormit_PhD}%
  \BibitemOpen
  \bibfield  {author} {\bibinfo {author} {\bibfnamefont {M.}~\bibnamefont
  {Wormit}},\ }\emph {\bibinfo {title} {Development and Application of Reliable
  Methods for the Calculation of Excited States: From Light-Harvesting
  Complexes to Medium-Sized Molecules}},\ \href@noop {} {Ph.D. thesis},\
  \bibinfo  {school} {Frankfurt University} (\bibinfo {year}
  {2009})\BibitemShut {NoStop}%
\bibitem [{\citenamefont {Albrecht}\ \emph {et~al.}(1998)\citenamefont
  {Albrecht}, \citenamefont {Reining}, \citenamefont {Del~Sole},\ and\
  \citenamefont {Onida}}]{Albrecht_1998}%
  \BibitemOpen
  \bibfield  {author} {\bibinfo {author} {\bibfnamefont {S.}~\bibnamefont
  {Albrecht}}, \bibinfo {author} {\bibfnamefont {L.}~\bibnamefont {Reining}},
  \bibinfo {author} {\bibfnamefont {R.}~\bibnamefont {Del~Sole}}, \ and\
  \bibinfo {author} {\bibfnamefont {G.}~\bibnamefont {Onida}},\ }\href
  {\doibase 10.1103/PhysRevLett.80.4510} {\bibfield  {journal} {\bibinfo
  {journal} {Phys. Rev. Lett.}\ }\textbf {\bibinfo {volume} {80}},\ \bibinfo
  {pages} {4510} (\bibinfo {year} {1998})}\BibitemShut {NoStop}%
\bibitem [{\citenamefont {Benedict}, \citenamefont {Shirley},\ and\
  \citenamefont {Bohn}(1998)}]{Benedict_1998}%
  \BibitemOpen
  \bibfield  {author} {\bibinfo {author} {\bibfnamefont {L.~X.}\ \bibnamefont
  {Benedict}}, \bibinfo {author} {\bibfnamefont {E.~L.}\ \bibnamefont
  {Shirley}}, \ and\ \bibinfo {author} {\bibfnamefont {R.~B.}\ \bibnamefont
  {Bohn}},\ }\href {\doibase 10.1103/PhysRevLett.80.4514} {\bibfield  {journal}
  {\bibinfo  {journal} {Phys. Rev. Lett.}\ }\textbf {\bibinfo {volume} {80}},\
  \bibinfo {pages} {4514} (\bibinfo {year} {1998})}\BibitemShut {NoStop}%
\bibitem [{\citenamefont {Marini}\ and\ \citenamefont
  {Del~Sole}(2003)}]{Marini_2003}%
  \BibitemOpen
  \bibfield  {author} {\bibinfo {author} {\bibfnamefont {A.}~\bibnamefont
  {Marini}}\ and\ \bibinfo {author} {\bibfnamefont {R.}~\bibnamefont
  {Del~Sole}},\ }\href {\doibase 10.1103/PhysRevLett.91.176402} {\bibfield
  {journal} {\bibinfo  {journal} {Phys. Rev. Lett.}\ }\textbf {\bibinfo
  {volume} {91}},\ \bibinfo {pages} {176402} (\bibinfo {year}
  {2003})}\BibitemShut {NoStop}%
\bibitem [{\citenamefont {Hanke}\ and\ \citenamefont
  {Sham}(1980)}]{Hanke_1980}%
  \BibitemOpen
  \bibfield  {author} {\bibinfo {author} {\bibfnamefont {W.}~\bibnamefont
  {Hanke}}\ and\ \bibinfo {author} {\bibfnamefont {L.~J.}\ \bibnamefont
  {Sham}},\ }\href {\doibase 10.1103/PhysRevB.21.4656} {\bibfield  {journal}
  {\bibinfo  {journal} {Phys. Rev. B}\ }\textbf {\bibinfo {volume} {21}},\
  \bibinfo {pages} {4656} (\bibinfo {year} {1980})}\BibitemShut {NoStop}%
\bibitem [{\citenamefont {Yamada}\ \emph {et~al.}(2022)\citenamefont {Yamada},
  \citenamefont {Noguchi}, \citenamefont {Ishii}, \citenamefont {Hirose},
  \citenamefont {Sugino},\ and\ \citenamefont {Ohno}}]{Yamada_2022}%
  \BibitemOpen
  \bibfield  {author} {\bibinfo {author} {\bibfnamefont {S.}~\bibnamefont
  {Yamada}}, \bibinfo {author} {\bibfnamefont {Y.}~\bibnamefont {Noguchi}},
  \bibinfo {author} {\bibfnamefont {K.}~\bibnamefont {Ishii}}, \bibinfo
  {author} {\bibfnamefont {D.}~\bibnamefont {Hirose}}, \bibinfo {author}
  {\bibfnamefont {O.}~\bibnamefont {Sugino}}, \ and\ \bibinfo {author}
  {\bibfnamefont {K.}~\bibnamefont {Ohno}},\ }\href {\doibase
  10.1103/PhysRevB.106.045113} {\bibfield  {journal} {\bibinfo  {journal}
  {Phys. Rev. B}\ }\textbf {\bibinfo {volume} {106}},\ \bibinfo {pages}
  {045113} (\bibinfo {year} {2022})}\BibitemShut {NoStop}%
\bibitem [{\citenamefont {Schreiber}\ \emph {et~al.}(2008)\citenamefont
  {Schreiber}, \citenamefont {Silva-Junior}, \citenamefont {Sauer},\ and\
  \citenamefont {Thiel}}]{Schreiber_2008}%
  \BibitemOpen
  \bibfield  {author} {\bibinfo {author} {\bibfnamefont {M.}~\bibnamefont
  {Schreiber}}, \bibinfo {author} {\bibfnamefont {M.~R.}\ \bibnamefont
  {Silva-Junior}}, \bibinfo {author} {\bibfnamefont {S.~P.~A.}\ \bibnamefont
  {Sauer}}, \ and\ \bibinfo {author} {\bibfnamefont {W.}~\bibnamefont
  {Thiel}},\ }\href@noop {} {\bibfield  {journal} {\bibinfo  {journal} {J.
  Chem. Phys.}\ }\textbf {\bibinfo {volume} {128}},\ \bibinfo {pages} {134110}
  (\bibinfo {year} {2008})}\BibitemShut {NoStop}%
\bibitem [{\citenamefont {Silva-Junior}\ \emph
  {et~al.}(2010{\natexlab{a}})\citenamefont {Silva-Junior}, \citenamefont
  {Schreiber}, \citenamefont {Sauer},\ and\ \citenamefont
  {Thiel}}]{Silva-Junior_2010}%
  \BibitemOpen
  \bibfield  {author} {\bibinfo {author} {\bibfnamefont {M.~R.}\ \bibnamefont
  {Silva-Junior}}, \bibinfo {author} {\bibfnamefont {M.}~\bibnamefont
  {Schreiber}}, \bibinfo {author} {\bibfnamefont {S.~P.~A.}\ \bibnamefont
  {Sauer}}, \ and\ \bibinfo {author} {\bibfnamefont {W.}~\bibnamefont
  {Thiel}},\ }\href {\doibase 10.1063/1.3499598} {\bibfield  {journal}
  {\bibinfo  {journal} {J. Chem. Phys.}\ }\textbf {\bibinfo {volume} {133}},\
  \bibinfo {pages} {174318} (\bibinfo {year} {2010}{\natexlab{a}})}\BibitemShut
  {NoStop}%
\bibitem [{\citenamefont {Silva-Junior}\ \emph
  {et~al.}(2010{\natexlab{b}})\citenamefont {Silva-Junior}, \citenamefont
  {Sauer}, \citenamefont {Schreiber},\ and\ \citenamefont
  {Thiel}}]{Silva-Junior_2010b}%
  \BibitemOpen
  \bibfield  {author} {\bibinfo {author} {\bibfnamefont {M.~R.}\ \bibnamefont
  {Silva-Junior}}, \bibinfo {author} {\bibfnamefont {S.~P.}\ \bibnamefont
  {Sauer}}, \bibinfo {author} {\bibfnamefont {M.}~\bibnamefont {Schreiber}}, \
  and\ \bibinfo {author} {\bibfnamefont {W.}~\bibnamefont {Thiel}},\ }\href
  {\doibase 10.1080/00268970903549047} {\bibfield  {journal} {\bibinfo
  {journal} {Mol. Phys.}\ }\textbf {\bibinfo {volume} {108}},\ \bibinfo {pages}
  {453} (\bibinfo {year} {2010}{\natexlab{b}})}\BibitemShut {NoStop}%
\bibitem [{\citenamefont {Silva-Junior}\ \emph
  {et~al.}(2010{\natexlab{c}})\citenamefont {Silva-Junior}, \citenamefont
  {Schreiber}, \citenamefont {Sauer},\ and\ \citenamefont
  {Thiel}}]{Silva-Junior_2010c}%
  \BibitemOpen
  \bibfield  {author} {\bibinfo {author} {\bibfnamefont {M.~R.}\ \bibnamefont
  {Silva-Junior}}, \bibinfo {author} {\bibfnamefont {M.}~\bibnamefont
  {Schreiber}}, \bibinfo {author} {\bibfnamefont {S.~P.~A.}\ \bibnamefont
  {Sauer}}, \ and\ \bibinfo {author} {\bibfnamefont {W.}~\bibnamefont
  {Thiel}},\ }\href@noop {} {\bibfield  {journal} {\bibinfo  {journal} {J.
  Chem. Phys.}\ }\textbf {\bibinfo {volume} {133}},\ \bibinfo {pages} {174318}
  (\bibinfo {year} {2010}{\natexlab{c}})}\BibitemShut {NoStop}%
\bibitem [{\citenamefont {Loos}\ \emph
  {et~al.}(2020{\natexlab{b}})\citenamefont {Loos}, \citenamefont {Pradines},
  \citenamefont {Scemama}, \citenamefont {Giner},\ and\ \citenamefont
  {Toulouse}}]{Loos_2020}%
  \BibitemOpen
  \bibfield  {author} {\bibinfo {author} {\bibfnamefont {P.-F.}\ \bibnamefont
  {Loos}}, \bibinfo {author} {\bibfnamefont {B.}~\bibnamefont {Pradines}},
  \bibinfo {author} {\bibfnamefont {A.}~\bibnamefont {Scemama}}, \bibinfo
  {author} {\bibfnamefont {E.}~\bibnamefont {Giner}}, \ and\ \bibinfo {author}
  {\bibfnamefont {J.}~\bibnamefont {Toulouse}},\ }\href {\doibase
  10.1021/acs.jctc.9b01067} {\bibfield  {journal} {\bibinfo  {journal} {J.
  Chem. Theory Comput.}\ }\textbf {\bibinfo {volume} {16}},\ \bibinfo {pages}
  {1018} (\bibinfo {year} {2020}{\natexlab{b}})}\BibitemShut {NoStop}%
\bibitem [{\citenamefont {Head-Gordon}\ \emph {et~al.}(1994)\citenamefont
  {Head-Gordon}, \citenamefont {Rico}, \citenamefont {Oumi},\ and\
  \citenamefont {Lee}}]{Head-Gordon_1994}%
  \BibitemOpen
  \bibfield  {author} {\bibinfo {author} {\bibfnamefont {M.}~\bibnamefont
  {Head-Gordon}}, \bibinfo {author} {\bibfnamefont {R.~J.}\ \bibnamefont
  {Rico}}, \bibinfo {author} {\bibfnamefont {M.}~\bibnamefont {Oumi}}, \ and\
  \bibinfo {author} {\bibfnamefont {T.~J.}\ \bibnamefont {Lee}},\ }\href
  {\doibase 10.1016/0009-2614(94)00070-0} {\bibfield  {journal} {\bibinfo
  {journal} {Chem. Phys. Lett.}\ }\textbf {\bibinfo {volume} {219}},\ \bibinfo
  {pages} {21} (\bibinfo {year} {1994})}\BibitemShut {NoStop}%
\bibitem [{\citenamefont {Head-Gordon}, \citenamefont {Maurice},\ and\
  \citenamefont {Oumi}(1995)}]{Head-Gordon_1995}%
  \BibitemOpen
  \bibfield  {author} {\bibinfo {author} {\bibfnamefont {M.}~\bibnamefont
  {Head-Gordon}}, \bibinfo {author} {\bibfnamefont {D.}~\bibnamefont
  {Maurice}}, \ and\ \bibinfo {author} {\bibfnamefont {M.}~\bibnamefont
  {Oumi}},\ }\href {\doibase 10.1016/0009-2614(95)01111-L} {\bibfield
  {journal} {\bibinfo  {journal} {Chem. Phys. Lett.}\ }\textbf {\bibinfo
  {volume} {246}},\ \bibinfo {pages} {114} (\bibinfo {year}
  {1995})}\BibitemShut {NoStop}%
\bibitem [{\citenamefont {Trofimov}\ and\ \citenamefont
  {Schirmer}(1997)}]{Trofimov_1997}%
  \BibitemOpen
  \bibfield  {author} {\bibinfo {author} {\bibfnamefont {A.}~\bibnamefont
  {Trofimov}}\ and\ \bibinfo {author} {\bibfnamefont {J.}~\bibnamefont
  {Schirmer}},\ }\href {\doibase https://doi.org/10.1016/S0301-0104(96)00303-5}
  {\bibfield  {journal} {\bibinfo  {journal} {Chem. Phys.}\ }\textbf {\bibinfo
  {volume} {214}},\ \bibinfo {pages} {153} (\bibinfo {year}
  {1997})}\BibitemShut {NoStop}%
\bibitem [{\citenamefont {Christiansen}, \citenamefont {Koch},\ and\
  \citenamefont {J{\o}rgensen}(1995{\natexlab{a}})}]{Christiansen_1995a}%
  \BibitemOpen
  \bibfield  {author} {\bibinfo {author} {\bibfnamefont {O.}~\bibnamefont
  {Christiansen}}, \bibinfo {author} {\bibfnamefont {H.}~\bibnamefont {Koch}},
  \ and\ \bibinfo {author} {\bibfnamefont {P.}~\bibnamefont {J{\o}rgensen}},\
  }\href {\doibase http://dx.doi.org/10.1016/0009-2614(95)00841-Q} {\bibfield
  {journal} {\bibinfo  {journal} {Chem. Phys. Lett.}\ }\textbf {\bibinfo
  {volume} {243}},\ \bibinfo {pages} {409} (\bibinfo {year}
  {1995}{\natexlab{a}})}\BibitemShut {NoStop}%
\bibitem [{\citenamefont {Purvis~III}\ and\ \citenamefont
  {Bartlett}(1982)}]{Purvis_1982}%
  \BibitemOpen
  \bibfield  {author} {\bibinfo {author} {\bibfnamefont {G.~P.}\ \bibnamefont
  {Purvis~III}}\ and\ \bibinfo {author} {\bibfnamefont {R.~J.}\ \bibnamefont
  {Bartlett}},\ }\href {\doibase 10.1063/1.443164} {\bibfield  {journal}
  {\bibinfo  {journal} {J. Chem. Phys.}\ }\textbf {\bibinfo {volume} {76}},\
  \bibinfo {pages} {1910} (\bibinfo {year} {1982})}\BibitemShut {NoStop}%
\bibitem [{\citenamefont {Stanton}\ and\ \citenamefont
  {Bartlett}(1993)}]{Stanton_1993}%
  \BibitemOpen
  \bibfield  {author} {\bibinfo {author} {\bibfnamefont {J.~F.}\ \bibnamefont
  {Stanton}}\ and\ \bibinfo {author} {\bibfnamefont {R.~J.}\ \bibnamefont
  {Bartlett}},\ }\href {\doibase 10.1063/1.464746} {\bibfield  {journal}
  {\bibinfo  {journal} {J. Chem. Phys.}\ }\textbf {\bibinfo {volume} {98}},\
  \bibinfo {pages} {7029} (\bibinfo {year} {1993})}\BibitemShut {NoStop}%
\bibitem [{\citenamefont {Loos}(2019)}]{QuAcK}%
  \BibitemOpen
  \bibfield  {author} {\bibinfo {author} {\bibfnamefont {P.~F.}\ \bibnamefont
  {Loos}},\ }\href {\doibase 10.5281/zenodo.3745928} {\enquote {\bibinfo
  {title} {{{QuAcK: a software for emerging quantum electronic structure
  methods}}},}\ } (\bibinfo {year} {2019}),\ \bibinfo {note}
  {\url{https://github.com/pfloos/QuAcK}}\BibitemShut {NoStop}%
\bibitem [{\citenamefont {{\v C}{\'\i}{\v z}ek}\ and\ \citenamefont
  {Paldus}(1967)}]{Cizek_1967}%
  \BibitemOpen
  \bibfield  {author} {\bibinfo {author} {\bibfnamefont {J.}~\bibnamefont {{\v
  C}{\'\i}{\v z}ek}}\ and\ \bibinfo {author} {\bibfnamefont {J.}~\bibnamefont
  {Paldus}},\ }\href {\doibase 10.1063/1.1701562} {\bibfield  {journal}
  {\bibinfo  {journal} {J. Chem. Phys.}\ }\textbf {\bibinfo {volume} {47}},\
  \bibinfo {pages} {3976} (\bibinfo {year} {1967})}\BibitemShut {NoStop}%
\bibitem [{\citenamefont {Loos}\ \emph {et~al.}(2018)\citenamefont {Loos},
  \citenamefont {Scemama}, \citenamefont {Blondel}, \citenamefont {Garniron},
  \citenamefont {Caffarel},\ and\ \citenamefont {Jacquemin}}]{Loos_2018a}%
  \BibitemOpen
  \bibfield  {author} {\bibinfo {author} {\bibfnamefont {P.~F.}\ \bibnamefont
  {Loos}}, \bibinfo {author} {\bibfnamefont {A.}~\bibnamefont {Scemama}},
  \bibinfo {author} {\bibfnamefont {A.}~\bibnamefont {Blondel}}, \bibinfo
  {author} {\bibfnamefont {Y.}~\bibnamefont {Garniron}}, \bibinfo {author}
  {\bibfnamefont {M.}~\bibnamefont {Caffarel}}, \ and\ \bibinfo {author}
  {\bibfnamefont {D.}~\bibnamefont {Jacquemin}},\ }\href {\doibase
  10.1021/acs.jctc.8b00406} {\bibfield  {journal} {\bibinfo  {journal} {J.
  Chem. Theory Comput.}\ }\textbf {\bibinfo {volume} {14}},\ \bibinfo {pages}
  {4360} (\bibinfo {year} {2018})}\BibitemShut {NoStop}%
\bibitem [{\citenamefont {V{\'e}ril}\ \emph {et~al.}()\citenamefont
  {V{\'e}ril}, \citenamefont {Scemama}, \citenamefont {Caffarel}, \citenamefont
  {Lipparini}, \citenamefont {Boggio-Pasqua}, \citenamefont {Jacquemin},\ and\
  \citenamefont {Loos}}]{Veril_2021}%
  \BibitemOpen
  \bibfield  {author} {\bibinfo {author} {\bibfnamefont {M.}~\bibnamefont
  {V{\'e}ril}}, \bibinfo {author} {\bibfnamefont {A.}~\bibnamefont {Scemama}},
  \bibinfo {author} {\bibfnamefont {M.}~\bibnamefont {Caffarel}}, \bibinfo
  {author} {\bibfnamefont {F.}~\bibnamefont {Lipparini}}, \bibinfo {author}
  {\bibfnamefont {M.}~\bibnamefont {Boggio-Pasqua}}, \bibinfo {author}
  {\bibfnamefont {D.}~\bibnamefont {Jacquemin}}, \ and\ \bibinfo {author}
  {\bibfnamefont {P.-F.}\ \bibnamefont {Loos}},\ }\href {\doibase
  https://doi.org/10.1002/wcms.1517} {\bibfield  {journal} {\bibinfo  {journal}
  {WIREs Comput. Mol. Sci.}\ }\textbf {\bibinfo {volume} {11}},\ \bibinfo
  {pages} {e1517}}\BibitemShut {NoStop}%
\bibitem [{\citenamefont {Endo}\ \emph {et~al.}(2009)\citenamefont {Endo},
  \citenamefont {Ogasawara}, \citenamefont {Takahashi}, \citenamefont
  {Yokoyama}, \citenamefont {Kato},\ and\ \citenamefont {Adachi}}]{Endo_2009}%
  \BibitemOpen
  \bibfield  {author} {\bibinfo {author} {\bibfnamefont {A.}~\bibnamefont
  {Endo}}, \bibinfo {author} {\bibfnamefont {M.}~\bibnamefont {Ogasawara}},
  \bibinfo {author} {\bibfnamefont {A.}~\bibnamefont {Takahashi}}, \bibinfo
  {author} {\bibfnamefont {D.}~\bibnamefont {Yokoyama}}, \bibinfo {author}
  {\bibfnamefont {Y.}~\bibnamefont {Kato}}, \ and\ \bibinfo {author}
  {\bibfnamefont {C.}~\bibnamefont {Adachi}},\ }\href {\doibase
  10.1002/adma.200900983} {\bibfield  {journal} {\bibinfo  {journal} {Adv.
  Mater.}\ }\textbf {\bibinfo {volume} {21}},\ \bibinfo {pages} {4802}
  (\bibinfo {year} {2009})}\BibitemShut {NoStop}%
\bibitem [{\citenamefont {Uoyama}\ \emph {et~al.}(2012)\citenamefont {Uoyama},
  \citenamefont {Goushi}, \citenamefont {Shizu}, \citenamefont {Nomura},\ and\
  \citenamefont {Adachi}}]{Uoyama_2012}%
  \BibitemOpen
  \bibfield  {author} {\bibinfo {author} {\bibfnamefont {H.}~\bibnamefont
  {Uoyama}}, \bibinfo {author} {\bibfnamefont {K.}~\bibnamefont {Goushi}},
  \bibinfo {author} {\bibfnamefont {K.}~\bibnamefont {Shizu}}, \bibinfo
  {author} {\bibfnamefont {H.}~\bibnamefont {Nomura}}, \ and\ \bibinfo {author}
  {\bibfnamefont {C.}~\bibnamefont {Adachi}},\ }\href {\doibase
  10.1038/nature11687} {\bibfield  {journal} {\bibinfo  {journal} {Nature}\
  }\textbf {\bibinfo {volume} {492}},\ \bibinfo {pages} {234} (\bibinfo {year}
  {2012})}\BibitemShut {NoStop}%
\bibitem [{\citenamefont {Baldo}\ \emph {et~al.}(1998)\citenamefont {Baldo},
  \citenamefont {O{\textquotesingle}Brien}, \citenamefont {You}, \citenamefont
  {Shoustikov}, \citenamefont {Sibley}, \citenamefont {Thompson},\ and\
  \citenamefont {Forrest}}]{Baldo_1998}%
  \BibitemOpen
  \bibfield  {author} {\bibinfo {author} {\bibfnamefont {M.~A.}\ \bibnamefont
  {Baldo}}, \bibinfo {author} {\bibfnamefont {D.~F.}\ \bibnamefont
  {O{\textquotesingle}Brien}}, \bibinfo {author} {\bibfnamefont
  {Y.}~\bibnamefont {You}}, \bibinfo {author} {\bibfnamefont {A.}~\bibnamefont
  {Shoustikov}}, \bibinfo {author} {\bibfnamefont {S.}~\bibnamefont {Sibley}},
  \bibinfo {author} {\bibfnamefont {M.~E.}\ \bibnamefont {Thompson}}, \ and\
  \bibinfo {author} {\bibfnamefont {S.~R.}\ \bibnamefont {Forrest}},\ }\href
  {\doibase 10.1038/25954} {\bibfield  {journal} {\bibinfo  {journal} {Nature}\
  }\textbf {\bibinfo {volume} {395}},\ \bibinfo {pages} {151} (\bibinfo {year}
  {1998})}\BibitemShut {NoStop}%
\bibitem [{\citenamefont {Adachi}\ \emph {et~al.}(2001)\citenamefont {Adachi},
  \citenamefont {Baldo}, \citenamefont {Thompson},\ and\ \citenamefont
  {Forrest}}]{Adachi_2001}%
  \BibitemOpen
  \bibfield  {author} {\bibinfo {author} {\bibfnamefont {C.}~\bibnamefont
  {Adachi}}, \bibinfo {author} {\bibfnamefont {M.~A.}\ \bibnamefont {Baldo}},
  \bibinfo {author} {\bibfnamefont {M.~E.}\ \bibnamefont {Thompson}}, \ and\
  \bibinfo {author} {\bibfnamefont {S.~R.}\ \bibnamefont {Forrest}},\ }\href
  {\doibase 10.1063/1.1409582} {\bibfield  {journal} {\bibinfo  {journal} {J.
  Appl. Phys.}\ }\textbf {\bibinfo {volume} {90}},\ \bibinfo {pages} {5048}
  (\bibinfo {year} {2001})}\BibitemShut {NoStop}%
\bibitem [{\citenamefont {de~Silva}(2019)}]{deSilva_2019}%
  \BibitemOpen
  \bibfield  {author} {\bibinfo {author} {\bibfnamefont {P.}~\bibnamefont
  {de~Silva}},\ }\href {\doibase 10.1021/acs.jpclett.9b02333} {\bibfield
  {journal} {\bibinfo  {journal} {J. Phys. Chem. Lett.}\ }\textbf {\bibinfo
  {volume} {10}},\ \bibinfo {pages} {5674} (\bibinfo {year}
  {2019})}\BibitemShut {NoStop}%
\bibitem [{\citenamefont {Sanz-Rodrigo}\ \emph {et~al.}(2021)\citenamefont
  {Sanz-Rodrigo}, \citenamefont {Ricci}, \citenamefont {Olivier},\ and\
  \citenamefont {Sancho-Garc{\'\i}a}}]{Sanz-Rodrigo_2021}%
  \BibitemOpen
  \bibfield  {author} {\bibinfo {author} {\bibfnamefont {J.}~\bibnamefont
  {Sanz-Rodrigo}}, \bibinfo {author} {\bibfnamefont {G.}~\bibnamefont {Ricci}},
  \bibinfo {author} {\bibfnamefont {Y.}~\bibnamefont {Olivier}}, \ and\
  \bibinfo {author} {\bibfnamefont {J.~C.}\ \bibnamefont
  {Sancho-Garc{\'\i}a}},\ }\href {\doibase 10.1021/acs.jpca.0c08029} {\bibfield
   {journal} {\bibinfo  {journal} {J. Phys. Chem. A}\ }\textbf {\bibinfo
  {volume} {125}},\ \bibinfo {pages} {513} (\bibinfo {year}
  {2021})}\BibitemShut {NoStop}%
\bibitem [{\citenamefont {Ricci}\ \emph {et~al.}(2021)\citenamefont {Ricci},
  \citenamefont {San-Fabi{\'a}n}, \citenamefont {Olivier},\ and\ \citenamefont
  {Sancho-Garc{\'\i}a}}]{Ricci_2021}%
  \BibitemOpen
  \bibfield  {author} {\bibinfo {author} {\bibfnamefont {G.}~\bibnamefont
  {Ricci}}, \bibinfo {author} {\bibfnamefont {E.}~\bibnamefont
  {San-Fabi{\'a}n}}, \bibinfo {author} {\bibfnamefont {Y.}~\bibnamefont
  {Olivier}}, \ and\ \bibinfo {author} {\bibfnamefont {J.~C.}\ \bibnamefont
  {Sancho-Garc{\'\i}a}},\ }\href {\doibase
  https://doi.org/10.1002/cphc.202000926} {\bibfield  {journal} {\bibinfo
  {journal} {ChemPhysChem}\ }\textbf {\bibinfo {volume} {22}},\ \bibinfo
  {pages} {553} (\bibinfo {year} {2021})}\BibitemShut {NoStop}%
\bibitem [{\citenamefont {Olivier}\ \emph {et~al.}(2017)\citenamefont
  {Olivier}, \citenamefont {Yurash}, \citenamefont {Muccioli}, \citenamefont
  {D'Avino}, \citenamefont {Mikhnenko}, \citenamefont {Sancho-Garc\'{\i}a},
  \citenamefont {Adachi}, \citenamefont {Nguyen},\ and\ \citenamefont
  {Beljonne}}]{Olivier_2017}%
  \BibitemOpen
  \bibfield  {author} {\bibinfo {author} {\bibfnamefont {Y.}~\bibnamefont
  {Olivier}}, \bibinfo {author} {\bibfnamefont {B.}~\bibnamefont {Yurash}},
  \bibinfo {author} {\bibfnamefont {L.}~\bibnamefont {Muccioli}}, \bibinfo
  {author} {\bibfnamefont {G.}~\bibnamefont {D'Avino}}, \bibinfo {author}
  {\bibfnamefont {O.}~\bibnamefont {Mikhnenko}}, \bibinfo {author}
  {\bibfnamefont {J.~C.}\ \bibnamefont {Sancho-Garc\'{\i}a}}, \bibinfo {author}
  {\bibfnamefont {C.}~\bibnamefont {Adachi}}, \bibinfo {author} {\bibfnamefont
  {T.-Q.}\ \bibnamefont {Nguyen}}, \ and\ \bibinfo {author} {\bibfnamefont
  {D.}~\bibnamefont {Beljonne}},\ }\href {\doibase
  10.1103/PhysRevMaterials.1.075602} {\bibfield  {journal} {\bibinfo  {journal}
  {Phys. Rev. Mater.}\ }\textbf {\bibinfo {volume} {1}},\ \bibinfo {pages}
  {075602} (\bibinfo {year} {2017})}\BibitemShut {NoStop}%
\bibitem [{\citenamefont {Olivier}\ \emph {et~al.}(2018)\citenamefont
  {Olivier}, \citenamefont {Sancho-Garcia}, \citenamefont {Muccioli},
  \citenamefont {D'Avino},\ and\ \citenamefont {Beljonne}}]{Olivier_2018}%
  \BibitemOpen
  \bibfield  {author} {\bibinfo {author} {\bibfnamefont {Y.}~\bibnamefont
  {Olivier}}, \bibinfo {author} {\bibfnamefont {J.-C.}\ \bibnamefont
  {Sancho-Garcia}}, \bibinfo {author} {\bibfnamefont {L.}~\bibnamefont
  {Muccioli}}, \bibinfo {author} {\bibfnamefont {G.}~\bibnamefont {D'Avino}}, \
  and\ \bibinfo {author} {\bibfnamefont {D.}~\bibnamefont {Beljonne}},\ }\href
  {\doibase 10.1021/acs.jpclett.8b02327} {\bibfield  {journal} {\bibinfo
  {journal} {J. Chem. Theory Comput.}\ }\textbf {\bibinfo {volume} {9}},\
  \bibinfo {pages} {6149} (\bibinfo {year} {2018})}\BibitemShut {NoStop}%
\bibitem [{\citenamefont {Sancho-Garc{\'{\i}}a}\ \emph
  {et~al.}(2022)\citenamefont {Sancho-Garc{\'{\i}}a}, \citenamefont
  {Br{\'{e}}mond}, \citenamefont {Ricci}, \citenamefont
  {P{\'{e}}rez-Jim{\'{e}}nez}, \citenamefont {Olivier},\ and\ \citenamefont
  {Adamo}}]{SanchoGarca_2022}%
  \BibitemOpen
  \bibfield  {author} {\bibinfo {author} {\bibfnamefont {J.~C.}\ \bibnamefont
  {Sancho-Garc{\'{\i}}a}}, \bibinfo {author} {\bibfnamefont {E.}~\bibnamefont
  {Br{\'{e}}mond}}, \bibinfo {author} {\bibfnamefont {G.}~\bibnamefont
  {Ricci}}, \bibinfo {author} {\bibfnamefont {A.~J.}\ \bibnamefont
  {P{\'{e}}rez-Jim{\'{e}}nez}}, \bibinfo {author} {\bibfnamefont
  {Y.}~\bibnamefont {Olivier}}, \ and\ \bibinfo {author} {\bibfnamefont
  {C.}~\bibnamefont {Adamo}},\ }\href {\doibase 10.1063/5.0076545} {\bibfield
  {journal} {\bibinfo  {journal} {J. Chem. Phys.}\ }\textbf {\bibinfo {volume}
  {156}},\ \bibinfo {pages} {034105} (\bibinfo {year} {2022})}\BibitemShut
  {NoStop}%
\bibitem [{\citenamefont {Curtis}\ \emph {et~al.}(2023)\citenamefont {Curtis},
  \citenamefont {Adeyiga}, \citenamefont {Suleiman},\ and\ \citenamefont
  {Odoh}}]{Curtis_2023}%
  \BibitemOpen
  \bibfield  {author} {\bibinfo {author} {\bibfnamefont {K.}~\bibnamefont
  {Curtis}}, \bibinfo {author} {\bibfnamefont {O.}~\bibnamefont {Adeyiga}},
  \bibinfo {author} {\bibfnamefont {O.}~\bibnamefont {Suleiman}}, \ and\
  \bibinfo {author} {\bibfnamefont {S.~O.}\ \bibnamefont {Odoh}},\ }\href
  {\doibase 10.1063/5.0133727} {\bibfield  {journal} {\bibinfo  {journal} {J.
  Chem. Phys.}\ }\textbf {\bibinfo {volume} {158}},\ \bibinfo {pages} {024116}
  (\bibinfo {year} {2023})}\BibitemShut {NoStop}%
\bibitem [{\citenamefont {Trofimov}, \citenamefont {Stelter},\ and\
  \citenamefont {Schirmer}(2002)}]{Trofimov_2002}%
  \BibitemOpen
  \bibfield  {author} {\bibinfo {author} {\bibfnamefont {A.~B.}\ \bibnamefont
  {Trofimov}}, \bibinfo {author} {\bibfnamefont {G.}~\bibnamefont {Stelter}}, \
  and\ \bibinfo {author} {\bibfnamefont {J.}~\bibnamefont {Schirmer}},\ }\href
  {\doibase 10.1063/1.1504708} {\bibfield  {journal} {\bibinfo  {journal} {J.
  Chem. Phys.}\ }\textbf {\bibinfo {volume} {117}},\ \bibinfo {pages} {6402}
  (\bibinfo {year} {2002})}\BibitemShut {NoStop}%
\bibitem [{\citenamefont {Harbach}, \citenamefont {Wormit},\ and\ \citenamefont
  {Dreuw}(2014)}]{Harbach_2014}%
  \BibitemOpen
  \bibfield  {author} {\bibinfo {author} {\bibfnamefont {P.~H.~P.}\
  \bibnamefont {Harbach}}, \bibinfo {author} {\bibfnamefont {M.}~\bibnamefont
  {Wormit}}, \ and\ \bibinfo {author} {\bibfnamefont {A.}~\bibnamefont
  {Dreuw}},\ }\href {\doibase 10.1063/1.4892418} {\bibfield  {journal}
  {\bibinfo  {journal} {J. Chem. Phys.}\ }\textbf {\bibinfo {volume} {141}},\
  \bibinfo {pages} {064113} (\bibinfo {year} {2014})}\BibitemShut {NoStop}%
\bibitem [{\citenamefont {Christiansen}, \citenamefont {Koch},\ and\
  \citenamefont {J{\o}rgensen}(1995{\natexlab{b}})}]{Christiansen_1995b}%
  \BibitemOpen
  \bibfield  {author} {\bibinfo {author} {\bibfnamefont {O.}~\bibnamefont
  {Christiansen}}, \bibinfo {author} {\bibfnamefont {H.}~\bibnamefont {Koch}},
  \ and\ \bibinfo {author} {\bibfnamefont {P.}~\bibnamefont {J{\o}rgensen}},\
  }\href {\doibase http://dx.doi.org/10.1063/1.470315} {\bibfield  {journal}
  {\bibinfo  {journal} {J. Chem. Phys.}\ }\textbf {\bibinfo {volume} {103}},\
  \bibinfo {pages} {7429} (\bibinfo {year} {1995}{\natexlab{b}})}\BibitemShut
  {NoStop}%
\bibitem [{\citenamefont {Koch}\ \emph {et~al.}(1997)\citenamefont {Koch},
  \citenamefont {Christiansen}, \citenamefont {Jorgensen}, \citenamefont
  {Sanchez~de Mer{\'a}s},\ and\ \citenamefont {Helgaker}}]{Koch_1997}%
  \BibitemOpen
  \bibfield  {author} {\bibinfo {author} {\bibfnamefont {H.}~\bibnamefont
  {Koch}}, \bibinfo {author} {\bibfnamefont {O.}~\bibnamefont {Christiansen}},
  \bibinfo {author} {\bibfnamefont {P.}~\bibnamefont {Jorgensen}}, \bibinfo
  {author} {\bibfnamefont {A.~M.}\ \bibnamefont {Sanchez~de Mer{\'a}s}}, \ and\
  \bibinfo {author} {\bibfnamefont {T.}~\bibnamefont {Helgaker}},\ }\href
  {\doibase http://dx.doi.org/10.1063/1.473322} {\bibfield  {journal} {\bibinfo
   {journal} {J. Chem. Phys.}\ }\textbf {\bibinfo {volume} {106}},\ \bibinfo
  {pages} {1808} (\bibinfo {year} {1997})}\BibitemShut {NoStop}%
\bibitem [{\citenamefont {Loos}\ and\ \citenamefont
  {Jacquemin}(2020)}]{Loos_2020b}%
  \BibitemOpen
  \bibfield  {author} {\bibinfo {author} {\bibfnamefont {P.-F.}\ \bibnamefont
  {Loos}}\ and\ \bibinfo {author} {\bibfnamefont {D.}~\bibnamefont
  {Jacquemin}},\ }\href {\doibase 10.1021/acs.jpclett.9b03652} {\bibfield
  {journal} {\bibinfo  {journal} {J. Phys. Chem. Lett.}\ }\textbf {\bibinfo
  {volume} {11}},\ \bibinfo {pages} {974} (\bibinfo {year} {2020})}\BibitemShut
  {NoStop}%
\end{thebibliography}%

\end{document}